\documentclass[%
 reprint,
superscriptaddress,
 amsmath,amssymb,
 aps,
]{revtex4-2}

\usepackage{graphicx}
\usepackage{dcolumn}
\usepackage{bm}
\usepackage{hyperref}
\hypersetup{
  colorlinks   = true, 
  urlcolor     = blue, 
  linkcolor    = blue, 
  citecolor   = blue 
}

\usepackage[normalem]{ulem}
\usepackage{xcolor}
\usepackage{mathdots}
\usepackage{comment}
\usepackage{verbatim}
\usepackage{bbm}
\usepackage{amsthm}

\newcommand{\abs}[1]{\lvert #1 \rvert}
\newcommand{\ket}[1]{\lvert #1 \rangle}
\newcommand{\bra}[1]{\langle #1 \rvert}

\newcommand{\braket}[2]{\langle #1 \vert #2 \rangle}

\newcommand{\tr}{\operatorname{tr}}

\newcommand{\medcirc}{\mathord{\raisebox{-1pt}{\scalebox{1.4}{$\circ$}}}}

\usepackage{textcomp}

\usepackage[smallerops]{newtxmath}

\begin{document}

\preprint{APS/123-QED}

\title{Infinite-Level Hierarchy of Solvable Quantum Circuits}

\author{Michael A. Rampp}
\affiliation{Max Planck Institute for the Physics of Complex Systems, 01187 Dresden, Germany}

\author{Suhail A. Rather}
\affiliation{Max Planck Institute for the Physics of Complex Systems, 01187 Dresden, Germany}
\affiliation{Dahlem Center for Complex Quantum Systems, Freie Universit\"at Berlin, 14195 Berlin, Germany}

\author{Pieter W. Claeys}
\affiliation{Max Planck Institute for the Physics of Complex Systems, 01187 Dresden, Germany}
\affiliation{School of Physics, Trinity College Dublin, Dublin 2, Ireland}

\date{\today}

\begin{abstract}
Dual-unitary circuits have emerged as a paradigm of exactly solvable yet non-integrable quantum dynamics. 
Recently, a generalization of dual unitarity attempting to extend the phenomenology of exactly solvable circuits has been introduced through a hierarchy of conditions, with dual unitarity as the first level. 
However, beyond the second level the proposed generalized dual-unitary hierarchy ceases to be solvable in the whole spacetime. 
We present an infinite hierarchy of solvability conditions remedying this problem.
These new conditions can be combined with the generalized dual-unitary hierarchy to obtain circuits for which correlation functions and entanglement dynamics can be analyzed exactly in the whole spacetime. 
We show that this novel hierarchy possesses non-trivial solutions at every level. 
Our results demonstrate that dual unitarity can be systematically extended while preserving solvability, opening up investigations of exactly solvable non-integrable systems with more general properties.
\end{abstract}

\maketitle


\begin{figure*}[t]
    \centering
    \includegraphics[width = 0.95\textwidth]{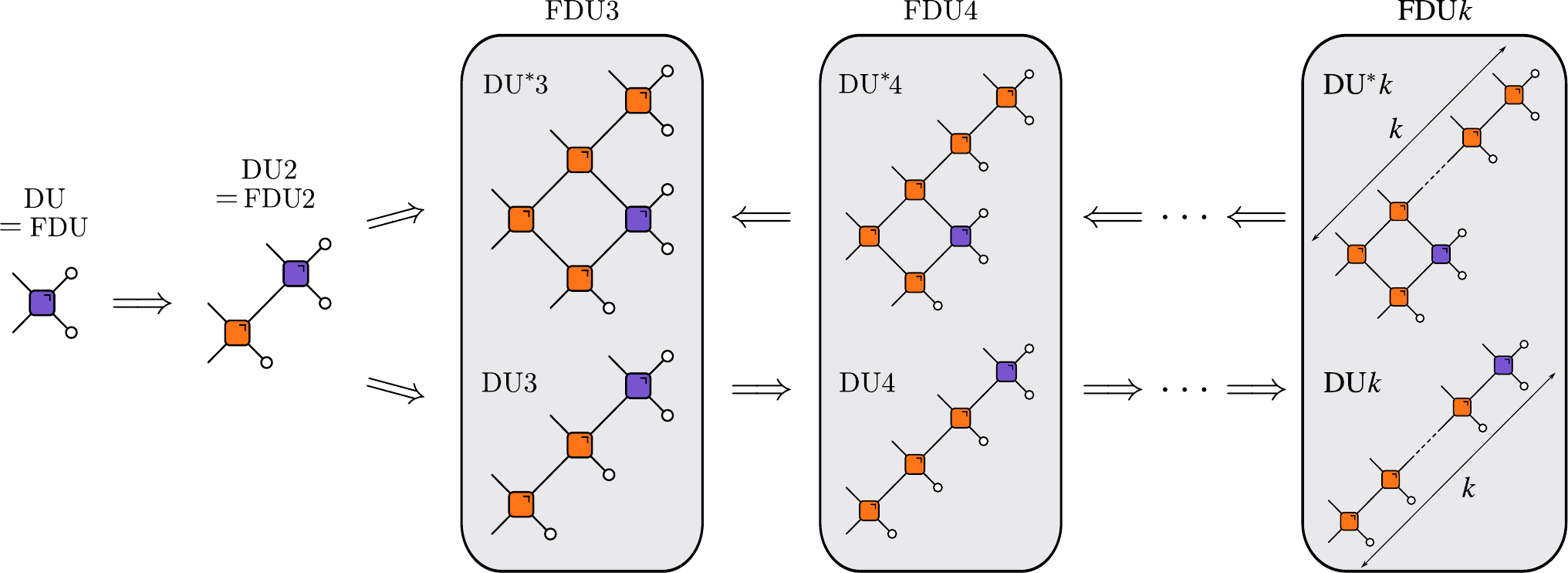}
    \caption{Illustration of the full dual-unitary hierarchy and the logical inclusion of the conditions. The solvability conditions are illustrated in the folded picture. Each condition allow for the systematic removal of a single gate, where the gate depicted in dark purple is removed on the right-hand side of the associated solvability condition. Dual unitarity (DU) is the strongest and most constrained condition, followed by DU2. The complementary dual-unitarity conditions (DU$^*k\geq3$) we introduce in this paper are logically independent from the generalized dual-unitarity conditions (DU$k\geq3$). The DU$^*k$ conditions become stronger with increasing $k$, while the DU$k$ conditions weaken. The FDU$k$ conditions are defined by combining DU$^*k$ with the associated DU$k$ condition, yielding solvability in the whole spacetime.}
    \label{fig:conditions_overview}
\end{figure*}

\section{Introduction}

Recently, there has been a flurry of activity surrounding dual-unitary circuits, where many quantities of interest in many-body dynamics can be computed exactly~\cite{Akila2016,Bertini2018,Gopalakrishnan2019,Bertini2019,Bertini2019a,Claeys2020,Claeys2021,Bertini2026}. Dual-unitary circuits require neither integrability nor randomness, opening a window into the emergence of quantum chaos and thermalization in generic structured dynamics. The power of dual-unitary circuits is based on a space-time duality, which is imposed by requiring the building block of a circuit ---the two-site unitary gate--- to be unitary in the spatial direction, in addition to the familiar temporal unitarity. A gate satisfying this condition is called a dual-unitary gate. It is a natural question to ask if dual unitarity can be extended to produce a wider variety of phenomena while maintaining the exact solvability. Recently, there have been several proposed extensions of dual unitarity, expanding the range of non-integrable yet solvable models~\cite{Jonay2021,Milbradt2023,Sommers2023,Mestyan2024,Yu2024,Rampp2025,Breach2025,Rampp2025a,Pickering2026}. One of these proposals, by Yu et al.~\cite{Yu2024}, is aimed at constructing an infinite hierarchy of conditions, with dual-unitarity at the bottom logically implying all other conditions. However, in the proposed hierarchy only the second level, i.e. the so-called DU2 condition as the first generalization of dual unitarity, maintains complete solvability of correlation functions and entanglement dynamics. For the higher levels of the hierarchy, the solvability is restricted to a particular region of spacetime close to the edge of the causal light cone~\cite{Yu2024,Rampp2024}. The generalized dual-unitary hierarchy has been used to investigate entanglement and operator dynamics~\cite{Foligno2024,Rampp2024,Sommers2024,Liu2025} and it has been applied to the quantum Floquet-East model~\cite{Bertini2024a}.

In this paper, we present an extension of the generalized dual-unitary hierarchy of Yu et al. that remedies this problem. We supplement the hierarchy with additional conditions providing solvability of correlation functions and entanglement dynamics in the region of spacetime \emph{not covered} by generalized dual-unitarity, thereby yielding the solution for the whole spacetime. In this way, we obtain an infinite hierarchy of inequivalent classes of exactly solvable quantum circuits. We call these additional conditions \emph{complementary dual unitarity} (DU$^*$), because they provide solvability in a region of spacetime exactly complementary to generalized dual unitarity while maintaining consistency of both. We call the combination of both \emph{full dual unitarity} (FDU). This can be summarized by the symbolic equation
\begin{align}
    \mathrm{DU}k\,+\,\mathrm{DU}^*k = \mathrm{FDU}k, \nonumber
\end{align} 
where DU$k$ refers to generalized dual unitarity of level $k$.
The hierarchy is depicted in Fig.~\ref{fig:conditions_overview}.
We show that the hierarchy of full dual-unitary circuits possesses non-trivial solutions at every level by using an infinitely large family of circuits defined on exotic spacetime lattices previously introduced by the authors~\cite{Rampp2025a}. This family of solutions also constitutes the first non-trivial examples of gates satisfying the generalized dual-unitarity equations for levels greater than three.

The outline of the paper is as follows. We begin by recalling the hierarchy of generalized dual-unitary circuits and its breakdown of solvability. In Sec.~\ref{sec:hierarchy_corr} we first introduce complementary dual-unitarity of level three (DU$^*3$) and show how it can be used to compute dynamical correlation functions. Then, we generalize this approach to higher levels of the hierarchy. We find that correlation functions are generically non-zero in a finite region of spacetime below a finite threshold velocity. Crucially, these correlations can be efficiently evaluated through a sequence of low-dimensional quantum channels. Furthermore, in Sec.~\ref{sec:hierarchy_elt} we show how the full dual-unitary hierarchy can be used to gain insight into entanglement and operator dynamics via the entanglement line tension. We find that the entanglement line tension takes a piecewise linear form with up to five kinks. Finally, in Sec.~\ref{sec:solutions} we discuss solutions to the full dual-unitarity equations. We present non-trivial solutions at every level of the hierarchy using spacetime lattice constructions and discuss their phenomenology. We also numerically analyze the tangent space of these solutions and conclude that there exist further solutions beyond the analytical ones presented here.

\subsection{Generalized dual-unitary circuits}

We consider unitary circuits in a brickwork geometry~\cite{Fisher2023}. We graphically denote two-site unitary gates $U$ in tensor network notation as
\begin{align}
    \bra{ab}U\ket{cd} = \vcenter{\hbox{\includegraphics[width=0.06\textwidth]{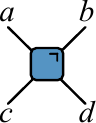}}}, \quad
    \bra{ab}U^{\dagger}\ket{cd} = \vcenter{\hbox{\includegraphics[width=0.06\textwidth]{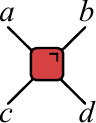}}}\,,
\end{align}
with indices $a,b,c,d \in \{1,\dots, q\}$ labeling basis states in a local $q$-dimension Hilbert space.
Unitarity is expressed in tensor network notation as
\begin{align}
    \vcenter{\hbox{\includegraphics[width=0.053\textwidth]{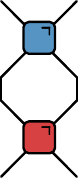}}}\,\,=\,\,\vcenter{\hbox{\includegraphics[width=0.053\textwidth]{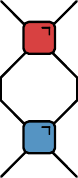}}}=\,\,\vcenter{\hbox{\includegraphics[width=0.035\textwidth]{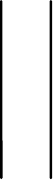}}}\,\,,
\end{align}
where we have made the indices implicit.
The two-site gates are applied to a chain of $L$ qudits (of dimension $q$) in a brickwork pattern, yielding the evolution operator after a finite amount of time steps, i.e., layers, as
\begin{align}\label{eq:brickwork}
  \mathcal{U}(t) = \!\!\vcenter{\hbox{\includegraphics[width=0.38\textwidth]{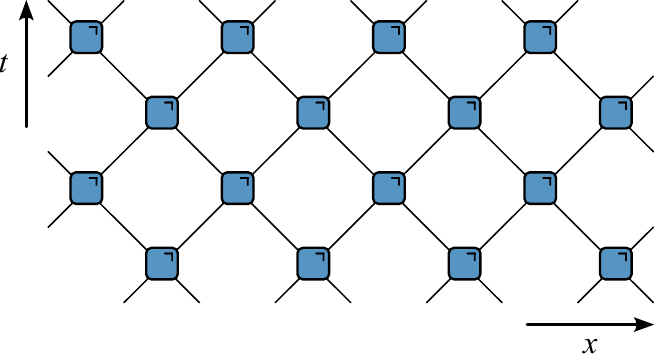}}}\,\,,
\end{align}
illustrated here for $t=4$. We are interested in the properties of the bulk system in the regime $t\ll L$, where the boundaries do not influence the dynamics of operators in the bulk.

Dual-unitary (DU) gates are a subclass of unitary gates that satisfy additional unitarity conditions in the spatial direction~\cite{Bertini2019,Gopalakrishnan2019}:
\begin{align}
    \vcenter{\hbox{\includegraphics[width=0.067\textwidth]{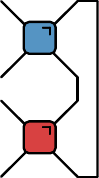}}}\,\,=\,\,\vcenter{\hbox{\includegraphics[width=0.022\textwidth]{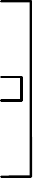}}}\,\,,\qquad\vcenter{\hbox{\includegraphics[width=0.067\textwidth]{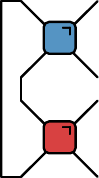}}}\,\,=\,\,\vcenter{\hbox{\includegraphics[width=0.022\textwidth]{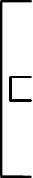}}}\,\,.
\end{align}
Circuits composed of DU gates lead to dynamics where many features can be analyzed exactly despite the absence of an extensive number of conserved quantities and without averaging over randomized realizations (see Ref.~\cite{Bertini2026} for a recent review).

It is often convenient to work directly in the Heisenberg picture and introduce the folded gate
\begin{align}
    \vcenter{\hbox{\includegraphics[width=0.061\textwidth]{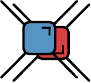}}} \,\,\equiv\,\,\vcenter{\hbox{\includegraphics[width=0.053\textwidth]{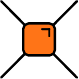}}}\,\,,
\end{align}
representing the superoperator $U\otimes U^*$. The normalized identity matrix in the folded representation becomes the vector
\begin{align}\label{eq:folded_identity}
    \vcenter{\hbox{\includegraphics[width=0.014\textwidth]{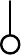}}} \,\equiv\, \frac{1}{\sqrt{q}}\,\, \vcenter{\hbox{\includegraphics[width=0.0166\textwidth]{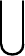}}}\,\,.
\end{align}
Unitarity is then represented as
\begin{align}
    \vcenter{\hbox{\includegraphics[width=0.063\textwidth]{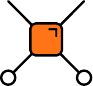}}}\,\, =\,\,  \vcenter{\hbox{\includegraphics[width=0.053\textwidth]{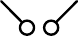}}}\,\,,\qquad
    \vcenter{\hbox{\includegraphics[width=0.063\textwidth]{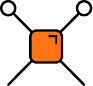}}}\,\, =\,\, \vcenter{\hbox{\includegraphics[width=0.053\textwidth]{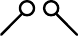}}}\,\,,
    \label{eq:unitarity_folded}
\end{align}
and dual-unitarity is represented as
\begin{align}
    \vcenter{\hbox{\includegraphics[width=0.058\textwidth]{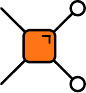}}}\,\, =\,\,  \vcenter{\hbox{\includegraphics[width=0.024\textwidth]{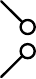}}}\,\,,\qquad
    \vcenter{\hbox{\includegraphics[width=0.058\textwidth]{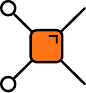}}}\,\, =\,\, \vcenter{\hbox{\includegraphics[width=0.024\textwidth]{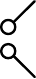}}}\,\,.
    \label{eq:dual_unitarity_folded}
\end{align}

Dual unitarity enables the exact evaluation of dynamical correlation functions of local operators, thus giving valuable information about the relaxation dynamics of a many-body system~\cite{Bertini2019}. For concreteness, we consider a traceless one-site operator $\sigma$ on an arbitrary even site of the lattice which we designate as $x=0$. Note that we here only consider translationally invariant circuits. After time evolution for $t$ steps, we take the expectation value w.r.t. the infinite-temperature (maximally mixed) state:
\begin{equation}
    C(x,t) = \langle \sigma(0,t)\rho(x,0)\rangle = \frac{1}{q^L} \tr\left[\mathcal{U}(t)^\dagger\sigma(0)\mathcal{U}(t)\rho(x)\right].
\end{equation}
This correlation function can be represented as a two-dimensional tensor network using unitarity~\cite{Bertini2026}. For $(t-x)\in2\mathbb{Z}$, we obtain the representation
\begin{equation} \label{eq:corr_tn}
    C(x,t) =\,\,\vcenter{\hbox{\includegraphics[width=0.3\textwidth]{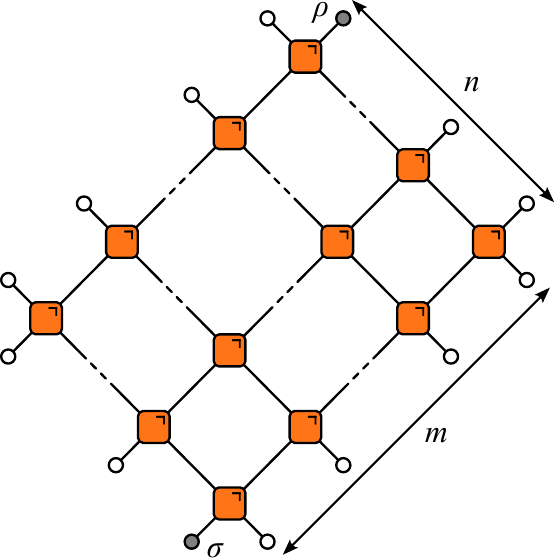}}}\,\,,
\end{equation}
whose side lengths are given by the light-cone coordinates
\begin{align}\label{eq:lc_coords}
    m = \frac{t+x}{2}, \quad n = \frac{t-x+2}{2},
\end{align}
and where we denote the vectorized operators graphically as
\begin{align}
    \vcenter{\hbox{\includegraphics[height=0.03\textheight]{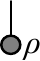}}} \,\equiv\, \frac{1}{\sqrt{q}}\,\, \vcenter{\hbox{\includegraphics[height=0.03\textheight]{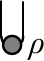}}}\,\equiv \ket{\rho}.
\end{align}
When $\rho$ equals the identity operator, this returns Eq.~\eqref{eq:folded_identity}.
For $(t-x)\in2\mathbb{Z}+1$, the tensor network reads
\begin{equation}
    C(x,t) =\,\,\vcenter{\hbox{\includegraphics[width=0.3\textwidth]{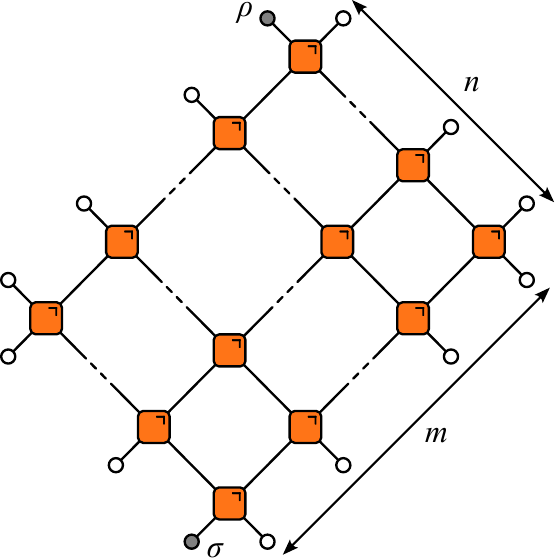}}}\,\,,
\end{equation}
with side lengths
\begin{align}
    m = \frac{t+x+1}{2}, \quad n = \frac{t-x+1}{2}.
\end{align}
On the edge of the light cone, $x=t$, the tensor network reduces to the one-dimensional contraction~\cite{Bertini2019}
\begin{align} \label{eq:corr_lc}
    C(t,t) = \,\,\vcenter{\hbox{\includegraphics[width=0.2\textwidth]{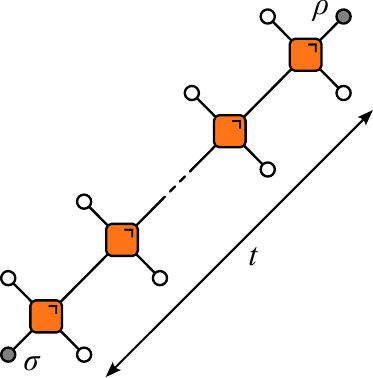}}}\,\, = \bra{\rho}\mathcal{M}_+^t\ket{\sigma},
\end{align}
generated by the light-cone channel $\mathcal{M}_+$, whose matrix elements follow from Eq.~\eqref{eq:corr_lc} as
\begin{align}\label{eq:channel_v1}
    (\mathcal{M}_{+})_{a,b} = \,\,\vcenter{\hbox{\includegraphics[height=0.13\columnwidth]{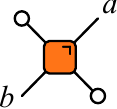}}}\,\,.
\end{align}
Note that $a,b \in \{1, \dots, q^2\}$ now label states in a doubled (folded) Hilbert space.
Equation \eqref{eq:corr_lc} can be efficiently evaluated at late times in any circuit as it consists of the repeated application of a low-dimensional channel. Inside the light cone, however, both sides of the tensor network Eq.~\eqref{eq:corr_tn} grow without bounds, making such an evaluation exponentially hard. Dual unitarity circumvents this by enabling the exact contraction of these expressions. Repeatedly applying Eqs.~\eqref{eq:dual_unitarity_folded} to Eq.~\eqref{eq:corr_tn} yields~\cite{Bertini2019}:
\begin{align}
    &\vcenter{\hbox{\includegraphics[width=0.3\textwidth]{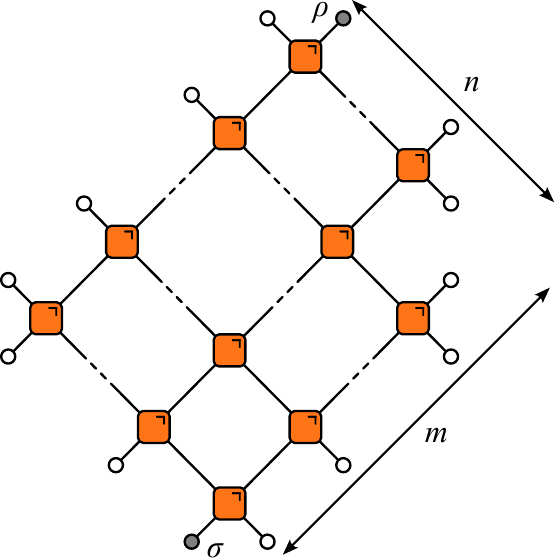}}} 
    \,\,=\,\,\,\vcenter{\hbox{\includegraphics[width=0.05\textwidth]{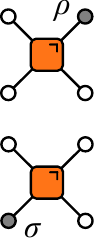}}} \nonumber\\
    &\qquad\qquad\qquad\qquad\qquad\qquad\propto \tr[\sigma]\tr[\rho] = 0,
\end{align}
which vanishes exactly because of the tracelessness of the operators.

Reference~\cite{Yu2024} introduced the DU2 conditions to achieve non-vanishing correlations inside the light cone while preserving solvability. These conditions read
\begin{align}\label{eq:DU2}
    \vcenter{\hbox{\includegraphics[height=0.185\columnwidth]{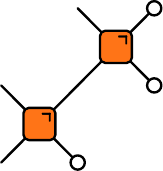}}}\,\,=\,\,\vcenter{\hbox{\includegraphics[height=0.185\columnwidth]{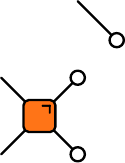}}}\,, \quad\vcenter{\hbox{\includegraphics[height=0.185\columnwidth]{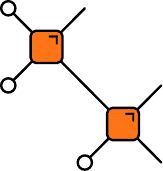}}}\,\,=\,\,\vcenter{\hbox{\includegraphics[height=0.185\columnwidth]{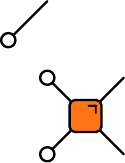}}}\,\,.
\end{align}
Dual unitary gates automatically satisfy the DU2 conditions, but non-dual-unitary solutions also exist. DU2 is therefore a generalization of dual unitarity, which additionally allows for non-vanishing correlations for the worldline $x=0$. These correlations are represented by a square-shaped tensor network which can be simplified through the repeated application of Eq.~\eqref{eq:DU2} to
\begin{align}
    C(0,t)=\vcenter{\hbox{\includegraphics[height=0.6\columnwidth]{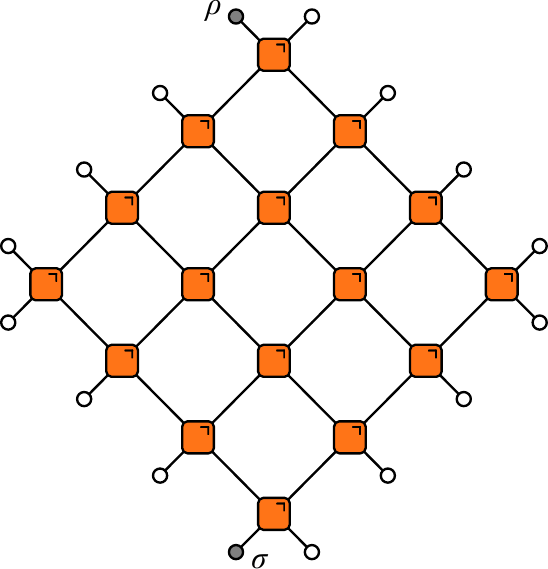}}}=\vcenter{\hbox{\includegraphics[height=0.6\columnwidth]{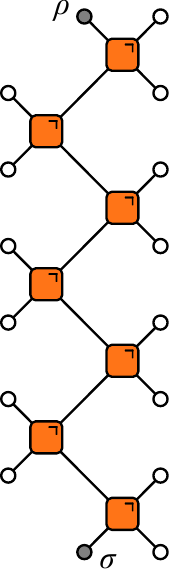}}}\,\,, \label{eq:corr_v0_du2}
\end{align}
now generated by the channel satisfying
\begin{equation}\label{eq:channel_v0}
    (\mathcal{M}_{1})_{a,b} = \,\,\vcenter{\hbox{\includegraphics[height=0.2\columnwidth]{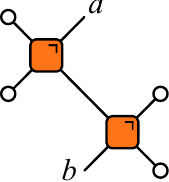}}}\,\,.
\end{equation}
No additional non-vanishing correlation functions are however possible: Away from the edges of the causal light cone, $|x|=t$, and this worldline, $x=0$, the correlations still vanish in the remaining interior of the light cone.

\begin{figure}[t]
    \centering
    \includegraphics[width = 0.35\textwidth]{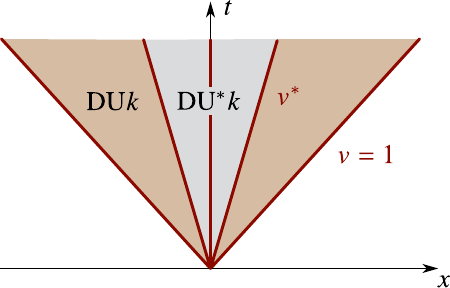}
    \caption{Diagram illustrating the domain of solvability and rays of information flow for FDU$k$ gates. The velocity separating the domains where DU$k$ and DU$^*k$ lead to solvability is given by $v^*=(k-2)/k$ ($k\geq3$). The correlation functions vanish exactly inside the DU$k$ domain, and may be non-vanishing inside the DU$^*k$ domain.}
    \label{fig:FDUk_solvability}
\end{figure}

To obtain more general behavior, the authors of Ref.~\cite{Yu2024} further introduced the DU3 conditions
\begin{align}\label{eq:DU3}
    \vcenter{\hbox{\includegraphics[height=0.27\columnwidth]{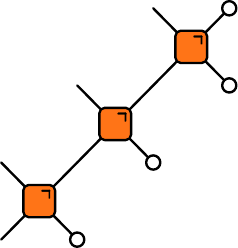}}}\!\!\!\!=\!\!\!\!\vcenter{\hbox{\includegraphics[height=0.27\columnwidth]{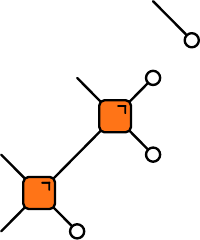}}} ,\,\,\vcenter{\hbox{\includegraphics[height=0.27\columnwidth]{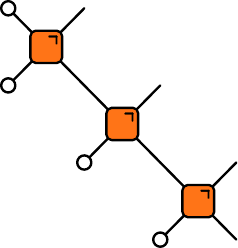}}}\!\!\!\!=\!\!\!\!\vcenter{\hbox{\includegraphics[height=0.27\columnwidth]{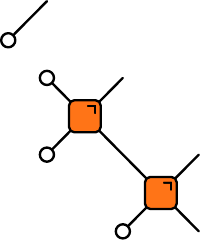}}}\,.
\end{align}
Again, DU2 gates automatically satisfy DU3. The problem with the DU3 condition is that correlation functions are no longer solvable in the whole spacetime, but only in a finite region corresponding to a restricted range of velocities. If the light-cone coordinates satisfy $m=2n-1$, corresponding to a ray of velocity $v=1/3$, the correlations can be evaluated exactly as
\begin{align}
\vcenter{\hbox{\includegraphics[height=0.7\columnwidth]{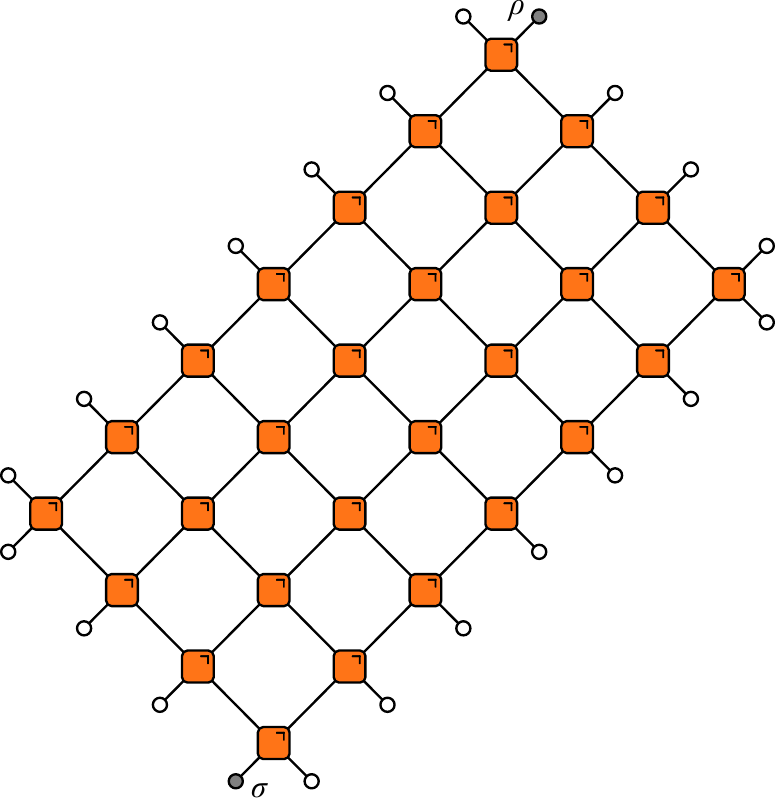}}}\!\!\!= \!\!\!\!\!\vcenter{\hbox{\includegraphics[height=0.7\columnwidth]{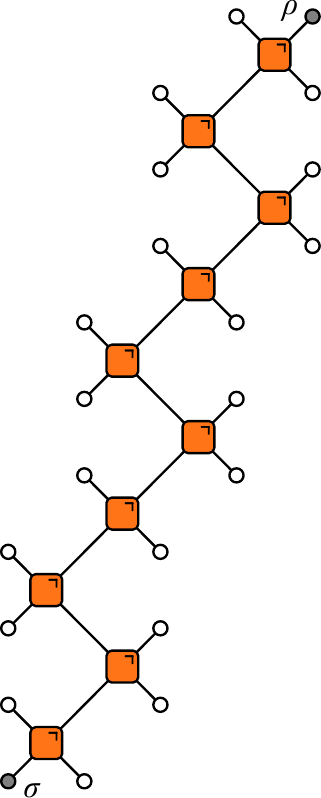}}}\!\!, \label{eq:corr_v13_fdu3}
\end{align}
with generating quantum channel satisfying
\begin{align}\label{eq:channel_v13}
    (\mathcal{M}_{2})_{a,b} = \,\,\vcenter{\hbox{\includegraphics[height=0.23\columnwidth]{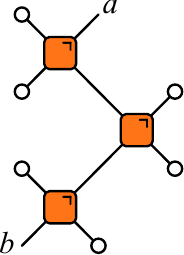}}}\,\,.
\end{align}
Note that this channel can be decomposed as $\mathcal{M}_2 = \mathcal{M}_1 \mathcal{M}_+$. 

Above this velocity, for $m>2n-1$, the DU3 conditions force the correlations to vanish. Here, applying the DU3 condition repeatedly disconnects diagrams of the form
\begin{align}
    \vcenter{\hbox{\includegraphics[height=0.2\textwidth]{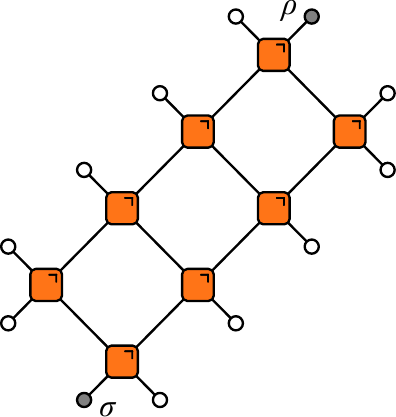}}}= \!\!\!\!\vcenter{\hbox{\includegraphics[height=0.2\textwidth]{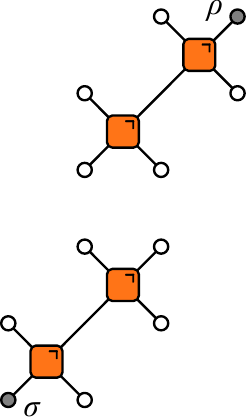}}}\!\!\!\! = 0,
\end{align}
which again factorizes and evaluates to zero because of the tracelessness of the operators $\sigma,\,\rho$. However, for $m<2n-1$ the DU3 conditions alone are not sufficient to simplify the diagram in such a way that it can be efficiently evaluated. The velocity $v=1/3$ acts as a threshold below which the dynamics remains inacessible.

Further generalization to the $k$-th level, DU$k$, only reduces the solvable area. The results are analogous to the DU3 condition with non-trivial solvable correlations along the threshold velocity $v^*_k=(k-2)/k$. Above this threshold the correlations vanish, while they are not accessible below.

\section{Dynamical correlation functions}
\label{sec:hierarchy_corr}

In this section we introduce \emph{full} dual unitarity of the third level, FDU3, and use it to derive exact expressions for dynamical correlation functions. We next generalize this approach to the full-dual-unitarity condition of arbitrary level $k$ and show that dynamical correlations remain exactly solvable for any level.

\subsection{Full dual unitarity of the third level}

We first define \emph{complementary} dual unitarity of the third level (DU$^*3$) via the condition
\begin{align}
    \vcenter{\hbox{\includegraphics[height=0.17\textwidth]{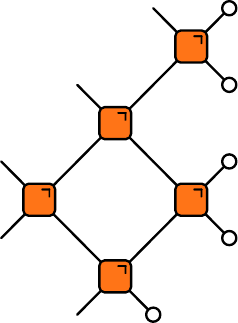}}}\,\, = \,\,\vcenter{\hbox{\includegraphics[height=0.17\textwidth]{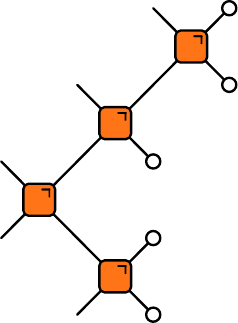}}}\,\,. \label{eq:cond_fdu3}
\end{align}
Furthermore, we require that not only the unitary gate $U$ satisfies Eq.~\eqref{eq:cond_fdu3}, but also its left-right flipped version $SUS$ (where $S$ is the swap gate), its transposed $U^T$, and $SU^TS$. These conditions correspond to mirrored and rotated versions of Eq.~\eqref{eq:cond_fdu3}. Rather than a single constraint, this hence corresponds to a set of different and generally inequivalent conditions.
Equation~\eqref{eq:cond_fdu3} has been proposed previously in the closing remarks of Ref.~\cite{Yu2024}, but the authors did not pursue their investigation further because of the difficulty of finding solutions to this equation. Indeed, Eq.~\eqref{eq:cond_fdu3} is a non-linear tensor equation on a high-dimensional space, making it hard to address via conventional methods or numerical searches. We show in Sec.~\ref{sec:solutions} that certain geometric constructions based on DU gates on smaller Hilbert spaces constitute solutions to Eq.~\eqref{eq:cond_fdu3}, motivating us to further investigate its properties and define the extension to the full dual-unitary hierarchy.

To treat the region which is inaccessible to DU3 we turn to the DU$^*3$ condition. Consider correlation functions along $v=0$.
First, we note that the DU$^*3$ condition also implies the following ``corner removal'' condition on four gates
\begin{align}
\vcenter{\hbox{\includegraphics[height=0.13\textwidth]{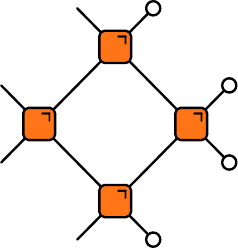}}}\,\, = \,\, \vcenter{\hbox{\includegraphics[height=0.13\textwidth]{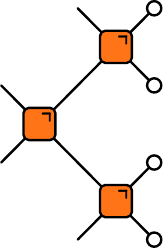}}}\,\,. \label{eq:corner}
\end{align}
This follows from contracting the FDU3 condition with a single identity state~\eqref{eq:folded_identity} from the top and using unitarity as
\begin{align}
\vcenter{\hbox{\includegraphics[height=0.17\textwidth]{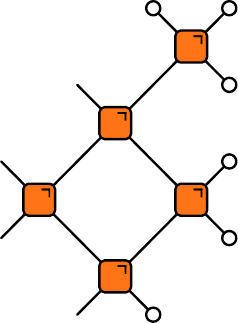}}}\,\, = \,\, \vcenter{\hbox{\includegraphics[height=0.17\textwidth]{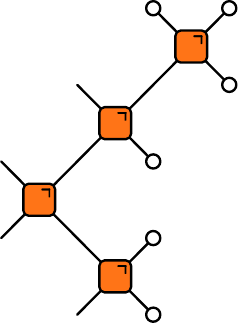}}}\,\,. \label{eq:corner_deriv}
\end{align}
By repeatedly acting with the corner removal and the DU$^*3$ condition, we obtain
\begin{align}
C(0,t)=\,\,\vcenter{\hbox{\includegraphics[height=0.6\columnwidth]{figs/corr_fdu3_v0_rhs.pdf}}}\,\,. \label{eq:corr_v0_fdu3}
\end{align}
This is the same result as for DU2 gates, Eq.~\eqref{eq:corr_v0_du2}, even though the FDU3 gates do not satisfy the DU2 condition. As an alternative interpretation, we also see from this result that FDU3 circuits give rise to the same influence matrix along $v=0$ as DU2 circuits~\cite{Foligno2024}.

Next, we consider rays in the intermediate range $0<m\leq2n-2$, corresponding to correlation functions inside the causal light cone with velocities $v \leq 1/3$. For simplicity, we focus on the case $(t-x)\in2\mathbb{Z}$. We also set $n\geq3$. We have to distinguish two cases. 
First we consider the isolated ray $m=2n-2$. In this case the correlator can be expressed as
\begin{align}
\vcenter{\hbox{\includegraphics[height=0.66\columnwidth]{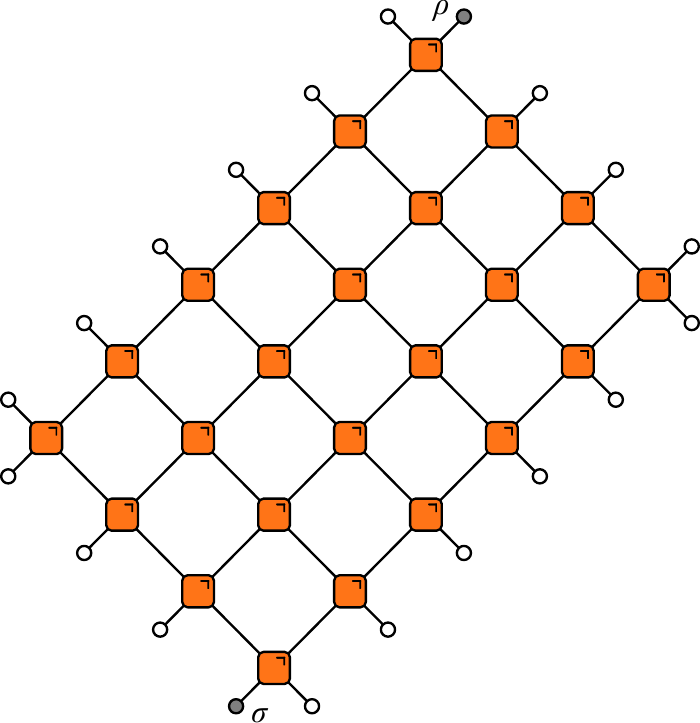}}}=\!\!\!\!\!\!\!\!\vcenter{\hbox{\includegraphics[height=0.66\columnwidth]{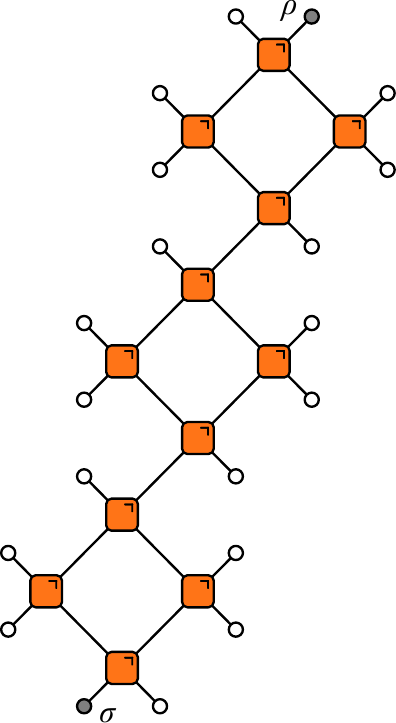}}}, \label{eq:corr_v13_fdu3_large}
\end{align}
through the repeated application of the DU$^*3$ condition starting from the left and right corners.
This result can again be evaluated as a one-dimensional contraction, generated by the channel defined from
\begin{equation}
    (\mathcal{M}'_{2})_{a,b} = \vcenter{\hbox{\includegraphics[height=0.24\columnwidth]{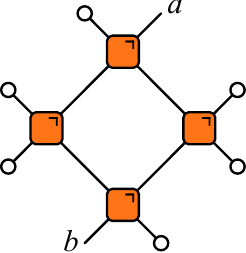}}}\,\,, \label{eq:channel_v13_large}
\end{equation}
such that 
\begin{equation}
    C = \bra{\rho}(\mathcal{M}_2')^{n-1}\ket{\sigma}, \quad m=2n-2.
\end{equation}
In contrast to the cases we have seen before, the reduced diagram is wider than a single gate in certain places. If the reduced diagram always has width one, this means that the only terms in the operator wave function contributing to the correlation functions are those with support on only one site. In contrast, for $m=2n-2$ operator histories involving two sites at certain time steps may also contribute. 

Second, we consider the region $0<m\leq2n-3$. We again employ the following algorithm: we apply the DU$^*3$ condition starting from the right corner until no simplifications are possible. Then we turn to the left corner and do the same. In this way the full diagram can be reduced to the form
\begin{align}
\vcenter{\hbox{\includegraphics[height=0.65\columnwidth]{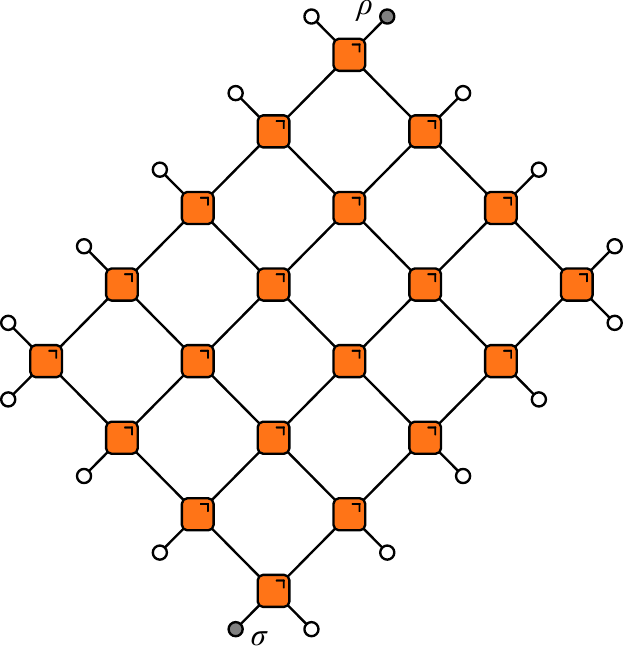}}}\,\, = \!\!\!\!\!\!\!\!\vcenter{\hbox{\includegraphics[height=0.65\columnwidth]{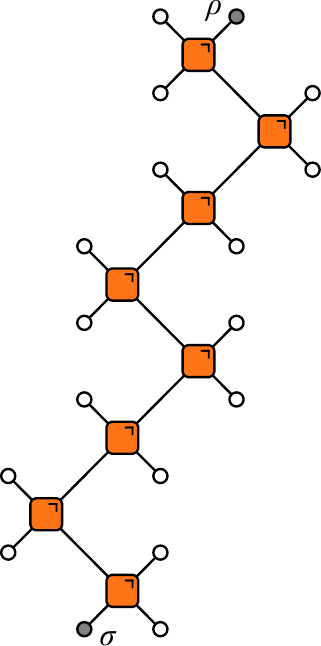}}}\!. \label{eq:corr_inter_fdu3}
\end{align}
This expression can again be efficiently evaluated through a combination of the channels $\mathcal{M}_1$ and $\mathcal{M}_2$ generating the correlations along $v=0$ and $v=1/3$, Eqs.~\eqref{eq:channel_v0} and~\eqref{eq:channel_v13}, respectively, as
\begin{equation}
    C = \bra{\rho}\mathcal{M}_2^{m-n+1}\mathcal{M}_1^{2n-m-2}\ket{\sigma}, \quad m\leq2n-3. \label{eq:corr_inter_fdu3_channels}
\end{equation}
Importantly, this expression is generically nonvanishing in the entire domain. In contrast with previous solvable circuit dynamics, where nonvanishing correlation functions are restricted to isolated rays, correlations functions here can be nontrivially supported in extended regions of spacetime. 

It turns out that the reduction given by Eq.~\eqref{eq:corr_inter_fdu3} is not unique but that the correlation function can equivalently be written as
\begin{align}
    C = \bra{\rho}\mathcal{M}_1^{2n-m-3}\mathcal{M}_2^{m-n+1}\mathcal{M}_1\ket{\sigma}.
\end{align}
The consistency of these two expressions leads to a set of commutation condition on the channels $\mathcal{M}_1,\,\mathcal{M}_2$. The simplest such condition is
\begin{equation}
    \mathcal{M}_1\left[\mathcal{M}_1, \mathcal{M}_2 \right] = 0, \label{eq:consistency}
\end{equation}
namely that $\mathcal{M}_1$ and $\mathcal{M}_2$ commute in a subspace defined by $\mathcal{M}_1$.

The description in terms of quantum channels enables us to extract the asymptotic behavior of the correlation function. The channels $\mathcal{M}_1,\,\mathcal{M}_2$ are non-expanding.
Therefore, they asymptotically project the initial operator on their respective largest subleading eigenvectors $\ket{v_1},\,\ket{v_{2}}$ with eigenvalues $\lambda_1,\,\lambda_{2}$ that are smaller than one in modulus for generic (ergodic) circuits. The asymptotic late-time behavior of the correlation function follows as
\begin{align}
    C(x=vt,t) &\approx \braket{\rho}{v_2}\braket{v_2}{v_1}\braket{v_1}{\sigma}\,\lambda_1^{2n-m-2}\lambda_{2}^{m-n+1}\nonumber\\
    &\propto \left(\lambda_1^{\frac{1-3v}{2}}\lambda_{2}^v\right)^t,
\end{align}
again with $m,n$ defined as in Eq.~\eqref{eq:lc_coords}. Note that even though $\mathcal{M}_2=\mathcal{M}_1\mathcal{M}_+$, generally $\lambda_2\neq\lambda_1\lambda_+$. 
Defining a decay exponent through $C(vt,t)\propto \exp(-\gamma(v)t)$ yields for $\gamma(v)$ the linear interpolation between the decay exponents along $v=0$ and $v=1/3$,
\begin{equation}
    \gamma(v) = \frac{1-3v}{2}\log\abs{\lambda_1}^{-1} + v\log\abs{\lambda_{2}}^{-1}.
\end{equation}
This shows an important distinction to generic many-body dynamics, where the decay exponent is expected to have a finite curvature as a function of velocity, to be contrasted with the observed (piecewise) linear behavior.

\subsection{Higher levels}

\begin{figure*}[t]
    \centering
    \includegraphics[width = .95\textwidth]{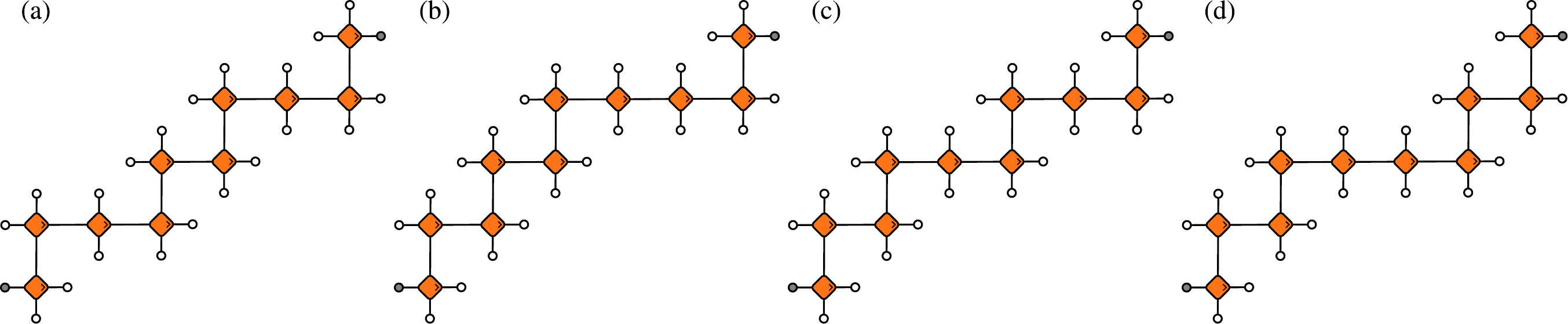}
    \caption{Illustration of the possible reduced forms of a diagram in the thin path domain of a FDU4 circuit. The unreduced diagram has side lengths $m=6$ and $n=5$. All possible fully reduced forms are equivalent and of the form (a)-(d), up to rotation by 180 degrees.}
    \label{fig:FDU4_paths}
\end{figure*}

Let us now define the DU$^*k$ condition in generality. We say that a unitary gate $U$ satisfies DU$^*k$ if $U$ itself, as well as $SUS$, $U^T$, and $SU^TS$, satisfy the condition
\begin{align}
    \vcenter{\hbox{\includegraphics[width=0.2\textwidth]{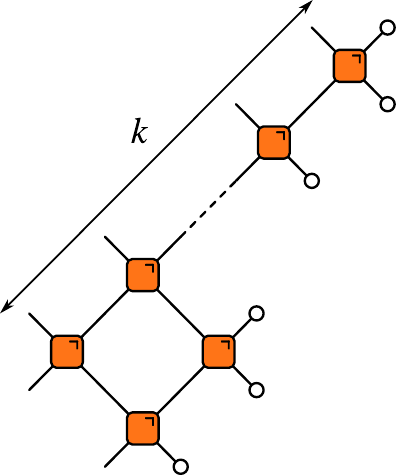}}}\!\!\!\!\! = \!\vcenter{\hbox{\includegraphics[width=0.2\textwidth]{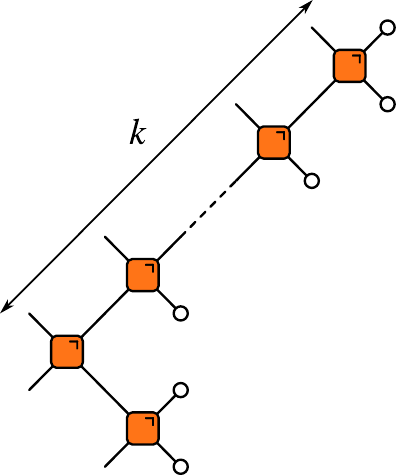}}}\,\,, \label{eq:cond_dustark}
\end{align}
where $k\geq3$. These conditions again correspond to generalized ``corner removal'' conditions. Using unitarity, it is easy to see that a gate satisfying DU$^*(k+1)$ implies that it also satisfies DU$^*k$. Similarly, if a gate satisfies DU or DU2 it automatically satisfies DU$^*k$ for all $k\geq3$. Thus, we call a gate a DU$^*k$ gate if it satisfies DU$^*k$ but does not satisfy DU, DU2, or any DU$^*\ell$ condition with $\ell>k$. The logical inclusion of the DU$^*k$ conditions suggests that the domain in spacetime in which the dynamics can be solved exactly grows with increasing $k$. Indeed, we will show in the next section that dynamical correlation functions can be solved exactly on rays $x=vt$ in spacetime with $\abs{v}\leq v^*=(k-2)/k$. This threshold velocity coincides with the threshold below which DU$k$ circuits cease to be solvable. The area of solvability in spacetime of DU$^*k$ circuits is hence complementary to DU$k$ circuits, as depicted in Fig.~\ref{fig:FDUk_solvability}.

For the discussion of the correlator in FDU$k$ circuits, we distinguish three domains in spacetime. For clarity, we always consider $m\geq n$ and the even component where $(t-x)\in2\mathbb{Z}$. 
\begin{itemize}
\item Thin path domain, $m\leq (k-1)n - (2k-3)$, corresponding to an extended region in spacetime with $v \leq v^*$. In this range, the correlator can be expressed purely as a sequence of the channels $\mathcal{M}_1,\,\mathcal{M}_+$. Each sequence corresponds to a particular path in spacetime connecting the initial and the final operator. Remarkably, different paths corresponding to different sequences of quantum channels can be obtained, but all give the same result.
\item Thick path domain, $(k-1)n-(2k-4)\leq m\leq (k-1)n-2$, corresponding to isolated rays with $v = v^*$. In this range, the reduced correlator is given by more complicated channels, generalizing Eq.~\eqref{eq:corr_v13_fdu3_large}. The diagram no longer has the ``skeleton'' form, but is wider in certain places.
\item DU$k$ domain, $m\geq(k-1)n-1$, corresponding to an extended region in space time with $v \geq v^*$. In this range, the correlator can be evaluated using the DU$k$ condition only, yielding known results. 
\end{itemize}
Starting in this section, we rotate the tensor networks by 45 degrees for convenience. We also drop the operator labels if they are not stricly necessary. In this section, we focus on presenting results and relegate the derivations to App.~\ref{app:corr_higher}.

We first consider the thin path domain, in which the correlator can be expressed through a sequence of the quantum channels $\mathcal{M}_1$, $\mathcal{M}_+$ alone. This sequence can be related to a path in spacetime. It is useful to introduce the quantum channel
\begin{equation} \label{eq:channel_thin}
    \mathcal{M}_{w} \equiv \mathcal{M}_1\mathcal{M}_+^{w-1}=\,\,\vcenter{\hbox{\includegraphics[height=0.3\columnwidth]{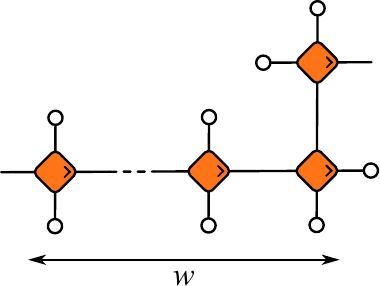}}}\,\,.
\end{equation}
This channel corresponds to a step of width $w$ and height $h=1$. In a FDU$k$ circuit, the maximum possible step width is given by $k-1$. Now, any path that connects the initial and final operator and contains only steps of height $h=1$ and $w\leq k-1$ corresponds to a possible reduced form of the diagram Eq.~\eqref{eq:corr_tn}.  
Let us call $a_j$ the number of steps of width $j$. The total height and width of the diagram yield the consistency conditions
\begin{align}
    n = 1 + \sum_{j=1}^{k-1}a_j, \quad
    m = \sum_{j=1}^{k-1}ja_j. \label{eq:step_constraints}
\end{align}
The constraint that the gate just above the initial operator cannot be removed implies $a_1\geq1$. Any solution of Eq.~\eqref{eq:step_constraints} with $a_1\geq1$ and $a_{j>1}\geq0$ together with a valid ordering of the steps corresponds to a sequence of quantum channels which can be related to the correlator as 
\begin{equation} \label{eq:corr_fduk}
    C = \bra{\rho}\mathcal{M}_{w_N}\dots\mathcal{M}_{w_2}\mathcal{M}_1\ket{\sigma}.
\end{equation} 
An example of a diagram with multiple distinct reduced forms is given in Fig.~\ref{fig:FDU4_paths}. Again, the equivalence of difference paths can be translated to a set of algebraic conditions on the quantum channels. 

Next, we consider the thick path domain, corresponding to the range $(k-1)n-(2k-4)\leq m\leq (k-1)n-2$. In this range, the diagrams do not reduce to the skeleton form any more, leading to the appearance of a more general set of quantum channels. We introduce the channels
\begin{align}
    \mathcal{M}_+^{k-j-2}\mathcal{M}_{j+1}' = \,\,\vcenter{\hbox{\includegraphics[height=0.3\columnwidth]{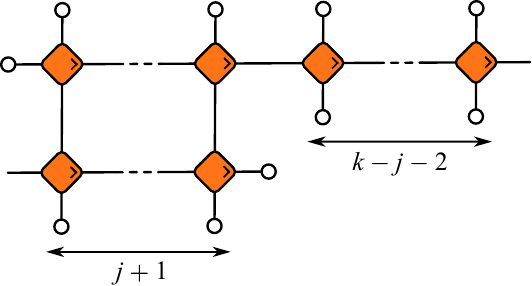}}}\,\,.
\end{align}
Then, we consider the range $m=(k-1)n-(2k-3)+j$ with $j=1,\dots,k-2$, in which case the correlator reads
\begin{align} \label{eq:thick_path1}
    C= \bra{\rho}\mathcal{M}_{j+1}'\left(\mathcal{M}_+^{k-j-2}\mathcal{M}_{j+1}'\right)^{n-2}\ket{\sigma}. 
\end{align}
When $j$ grows larger than $k-2$, this expression has to be modified slightly. For $k-1\leq j\leq2k-1$ we find
\begin{align} \label{eq:thick_path2}
    C = \bra{\rho}\left(\mathcal{M}_+^{j-(k-2)}\mathcal{M}_{2k-3-j}'\right)^{n-1}\mathcal{M}_+^{j-(k-2)}\ket{\sigma}. 
\end{align}

Finally, we touch on the remaining DU$k$ domain with $m\geq(k-1)n-1$. In this range, the DU$k$ condition alone is sufficient to simplify the correlator completely~\cite{Yu2024}. The calculation is analogous to that presented above for the DU3 condition. For $m=(k-1)n-1$ we obtain
\begin{equation}
    C = \bra{\rho}\mathcal{M}_+^{k-2}\mathcal{M}_{k-1}^{n-1}\ket{\sigma},
\end{equation}
and for $m\geq(k-1)n$ the correlator vanishes identically, except when $n=1$.

\section{Entanglement line tension}
\label{sec:hierarchy_elt}

Dual unitarity generally allows for an exact characterization of not just the dynamics of correlation functions, but also of entanglement. This result directly extends to the presented hierarchy. To obtain insight into the dynamics itself, independent from any particular initial state, we consider the operator entanglement of the time evolution operator itself. The bipartition of the indices of the time evolution operator is described by an entanglement cut of slope $v=x/t$ connecting the initial and final time slices. In the scaling limit, the R\'{e}nyi-$\alpha$ operator entanglement takes the form~\cite{Jonay2018,Zhou2019}
\begin{equation}
    S^{(\alpha)} (x=vt,t) \approx \log (q)\, \mathcal{E}_{\alpha}(v)\,t. \label{eq:ELT_def}
\end{equation}
The scaling function $\mathcal{E}_{\alpha}(v)$, also known as the entanglement line tension (ELT), contains all information about entanglement and operator dynamics at large scales~\cite{Jonay2018}. In particular, the bipartite entanglement entropy after a quench from a translationally invariant state reads
\begin{equation}
    S^{(\alpha)}_A(t) = \log (q)\, \mathcal{E}_\alpha(0)\, t + S^{(\alpha)}_A(0).
\end{equation}
The ELT at $v=0$ therefore yields the rate of entanglement growth, which is also known as the entanglement velocity $v_E^{(\alpha)}$  in this setting. Furthermore, the butterfly velocity $v_B$ defining a system-specific emergent causal light cone can be extracted from the ELT through the self-consistent relation
\begin{equation}
    v_B = \mathcal{E}_2(v_B).
\end{equation}

To express the ELT in terms of tensor network diagrams in brickwork circuits, it is convenient to introduce a generalization of the folded representation to higher numbers of replicas
\begin{align}
(U \otimes U^*)^{\otimes\alpha} =\, \vcenter{\hbox{\includegraphics[width=0.11\columnwidth]{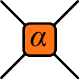}}}\,, \label{fig:folded_gate}
\end{align}
along with the permutation states extending Eq.~\eqref{eq:folded_identity} to multiple replicas:
\begin{align}\label{eq:permutation_states}
\vcenter{\hbox{\includegraphics[height=0.025\textheight]{figs/circle.pdf}}} \,= \frac{1}{q^\frac{\alpha}{2}}\, \overbrace{\vcenter{\hbox{\includegraphics[height=0.025\textheight]{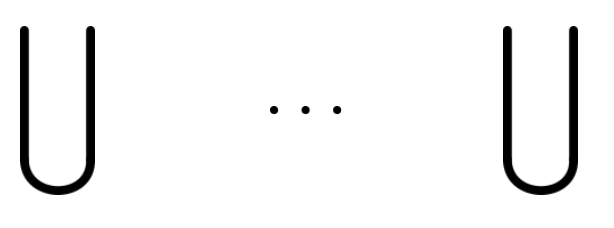}}}}^{2\alpha}\,, \qquad \vcenter{\hbox{\includegraphics[height=0.025\textheight]{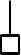}}}\, = \frac{1}{q^\frac{\alpha}{2}}\,\overbrace{\vcenter{\hbox{\includegraphics[height=0.025\textheight]{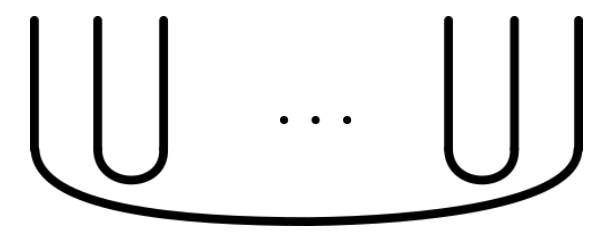}}}}^{2\alpha}\,. \,\,
\end{align}
Unitarity is then graphically expressed as
\begin{align}
    \label{eq:unitarity_folded_circ}
    \vcenter{\hbox{\includegraphics[width=0.063\textwidth]{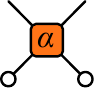}}}\,\, &=&\,\,  \vcenter{\hbox{\includegraphics[width=0.053\textwidth]{figs/folded_eig_2}}}\,\,,\qquad
    \vcenter{\hbox{\includegraphics[width=0.063\textwidth]{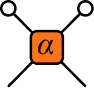}}}\,\, &=&\,\, \vcenter{\hbox{\includegraphics[width=0.053\textwidth]{figs/folded_eig_4}}}\,\,,\\
    \label{eq:unitarity_folded_sq}
    \vcenter{\hbox{\includegraphics[width=0.063\textwidth]{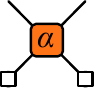}}}\,\, &=&\,\,  \vcenter{\hbox{\includegraphics[width=0.053\textwidth]{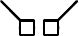}}}\,\,,\qquad
    \vcenter{\hbox{\includegraphics[width=0.063\textwidth]{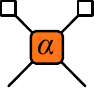}}}\,\, &=&\,\, \vcenter{\hbox{\includegraphics[width=0.053\textwidth]{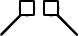}}}\,\,.
\end{align}
The operator entanglement can then be written as
\begin{equation}
     S_{\alpha} (x,t) \equiv \frac{1}{1-\alpha} \log  Z_\alpha(m,n),
\end{equation} 
where $Z_\alpha$ is defined as
\begin{align}
    Z_\alpha(m,n) \equiv \tr_A\left[ \left(\operatorname{tr}_{\bar{A}} \ket{\mathcal{U}(t)}\bra{\mathcal{U}(t)}\right)^\alpha \right], \label{eq:z_alpha}
\end{align}
and graphically represented as
\begin{align} \label{eq:elt_tn}
    Z_\alpha(m,n) = \,\,\vcenter{\hbox{\includegraphics[height = .64\columnwidth]{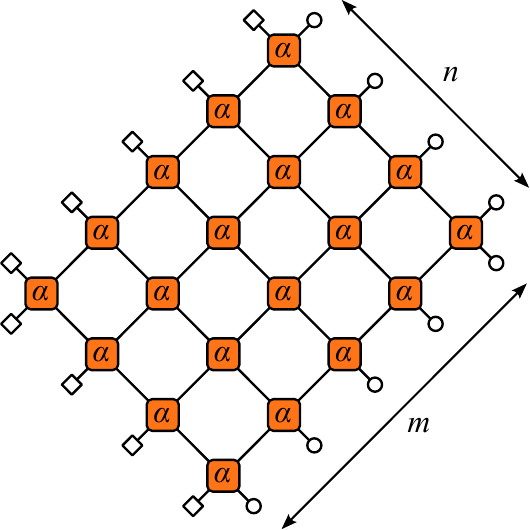}}}\,\,.
\end{align}
The size of this tensor network is set by the coordinates of the entanglement cut as
\begin{equation}
    n = \frac{t-x -(x \,\mathrm{mod}\, 2)}{2}, \quad m = \frac{t+x -(x\, \mathrm{mod}\, 2)}{2}.
\end{equation}

\subsection{Full dual unitarity of the third level}

The diagram Eq.~\eqref{eq:elt_tn} has a similar form to the correlation function Eq.~\eqref{eq:corr_tn}, but for higher number of replicas. Therefore, it can be simplified in a similar manner. In this section, we show how this is done in FDU3 circuits and discuss the physical implications of the result.

First, we focus on the range $n\leq m\leq2n+1$ where the DU3 condition alone is insufficient to solve the problem. We show that the diagrams can be reduced to a \emph{skeleton} form, a one-dimensional subset of the original two-dimensional diagram that corresponds to a path in spacetime, analogous to correlation functions. As for correlation functions, each diagram has multiple reduced forms corresponding to the different possible paths. However, the boundary conditions for $Z_\alpha(m,n)$ enable additional simplifications for a subset of paths, leading to a simple form of the ELT.

\begin{figure}[t]
    \centering
    \includegraphics[width = 0.45\textwidth]{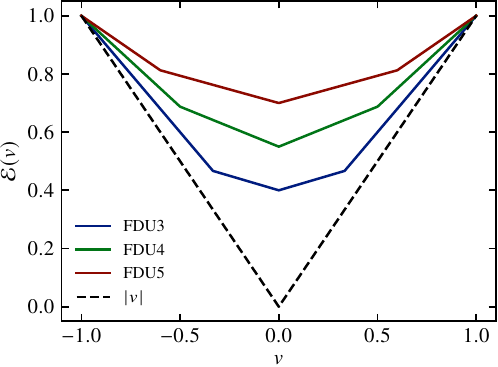}
    \caption{Diagram illustrating possible forms of the ELT for FDU$k$ circuits.}
    \label{fig:FDUk_ELT}
\end{figure}

Analogous to the computation for correlation functions, applying the DU$^*3$ condition to the diagram $Z_\alpha(m,n)$ yields a lower boundary described by a sequence of steps $(w_1,w_2,\dots,w_N)$ with individual widths $1\leq w_{j>1}\leq2$. The constraint that the gate just below the gate in the top right corner cannot be removed is absent for $Z_\alpha(m,n)$ because there is no inserted operator. Hence, all steps have height $h=1$ and therefore the number of steps is $N=n$, leading to a simplified expression (again rotated by 90 degrees for convenience):
\begin{align}
    \vcenter{\hbox{\includegraphics[width=0.55\columnwidth]{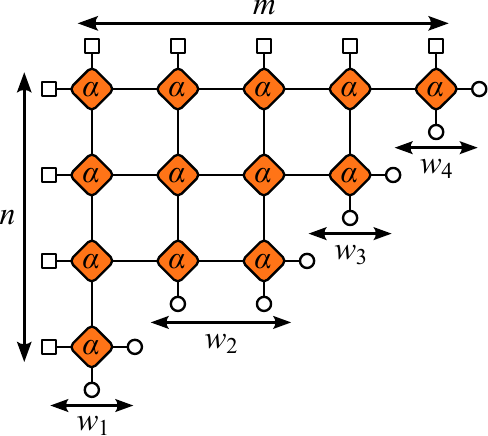}}}\,\,.
\end{align}
From the total width of the diagram being $m$ it follows that
\begin{equation}
    \sum_{j=1}^n w_j=m.
\end{equation}
Whenever we simplify the bottom boundary in such a way that the lowest step has a width $w_1\leq3$, then the top boundary can be reduced such that the diagram becomes a skeleton diagram. This is possible as long as $m\leq2n+1$.
Skeleton diagrams for $Z_\alpha$ can be expressed in terms of low-dimensional transfer matrices as
\begin{align}
    Z_\alpha(m,n) &= \,\,\vcenter{\hbox{\includegraphics[width=0.55\columnwidth]{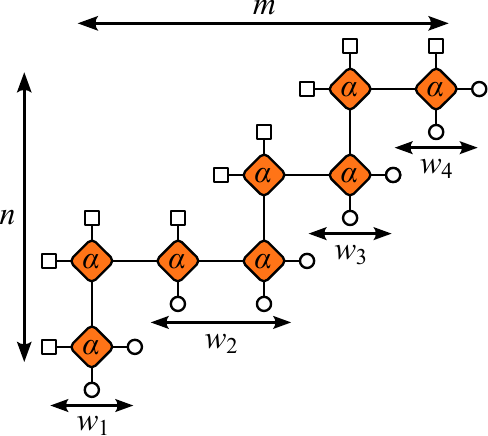}}}\,\,,
\end{align}
expressed in terms of transfer matrices given by
\begin{align}
   \mathcal{M}_{+}^{(\alpha)} = \,\,\vcenter{\hbox{\includegraphics[width=0.11\columnwidth]{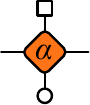}}} \,,\quad \mathcal{M}_{w}^{(\alpha)} =\,\,\vcenter{\hbox{\includegraphics[width=0.38\columnwidth]{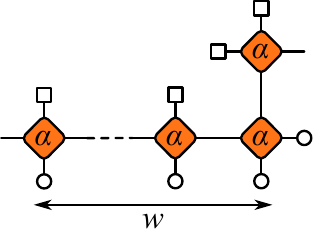}}}\,\,, 
\end{align}
and $\alpha$ is the number of replicas. In contrast to the quantum channels introduced in the previous section, these multi-replica objects are not trace preserving, which is why we refer to them as transfer matrices.

We have seen that there are multiple paths that give the same result. However, some of the paths yield diagrams that can be simplified further using the DU3 condition. If $w_n=2$, then the topmost boundary has three adjacent gates, one of which can be removed using DU3 from the left, thereby disconnecting part of the diagram:
\begin{align}
    \vcenter{\hbox{\includegraphics[width=0.46\columnwidth]{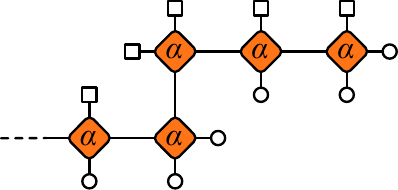}}}=\vcenter{\hbox{\includegraphics[width=0.46\columnwidth]{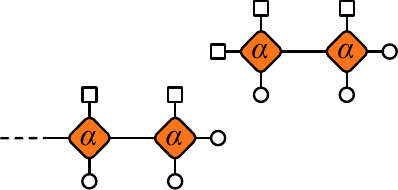}}}\,\,.
\end{align}
The next lower step can be disconnected in the same manner if it also has width $w_{n-1}=2$. This is continued until a step of width one is encountered. Analogous simplifications can be performed starting from the bottom of the diagram. We introduce the quantity
\begin{align}
    B_{2}^{(\alpha)} \equiv q^3\left(\,\,\vcenter{\hbox{\includegraphics[width=0.22\columnwidth]{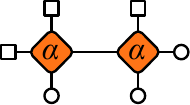}}}\,\,\right)^{1/(\alpha-1)}. \label{eq:B2_def}
\end{align}
Let us now consider paths where all steps of width one are adjacent to each other. Let $a_1$ be the number of steps of width one and $a_2$ the number of steps of width two. Using the DU3 property the diagram factorizes into a large connected part and a product of small disconnected parts whose value is given by Eq.~\eqref{eq:B2_def} each:

\begin{widetext}
\begin{equation}
Z_\alpha(m,n) = \,\,\vcenter{\hbox{\includegraphics[width=0.25\columnwidth]{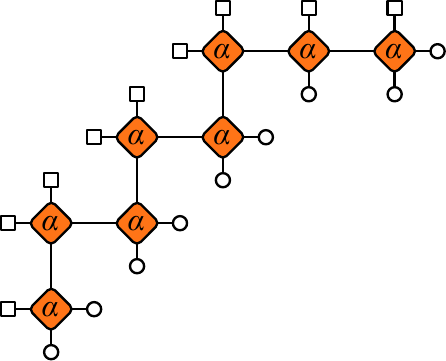}}}\!\!\!\!\!\!\!\!= \vcenter{\hbox{\includegraphics[width=0.25\columnwidth]{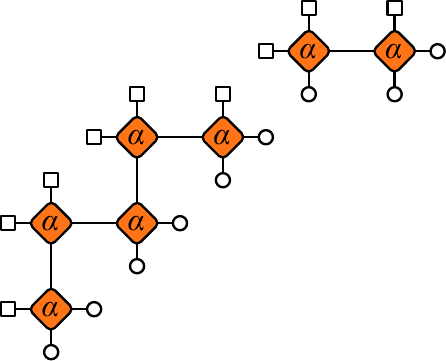}}}\!\!\!\!\!\!\!\!\!\!\!\!=\bra{\medcirc}\mathcal{M}_{+}^{(\alpha)}\left(\mathcal{M}^{(\alpha)}_{1}\right)^{a_1-1}\ket{\Box}\left(\frac{B_{2}^{(\alpha)}}{q^3}\right)^{(\alpha-1)a_2}.
\end{equation}
\end{widetext}
where we introduced the notation $\bra{\medcirc}$ and $\ket{\Box}$ for the states~\eqref{eq:permutation_states}.
Using the constraints $m=a_1+2a_2$ and $n=a_1+a_2$ yields
\begin{equation}
    Z_\alpha(m,n) = \bra{\medcirc}\mathcal{M}_{+}^{(\alpha)}\left(\mathcal{M}^{(\alpha)}_{1}\right)^{2n-m-1}\ket{\Box}\left(\frac{B_{2}^{(\alpha)}}{q^{3}}\right)^{(\alpha-1)(m-n)}.
\end{equation}

The ELT is extracted by considering the asymptotic decay of $Z_\alpha$ for $t\rightarrow\infty$. First, we set $m=n$ and find that the entanglement velocity can be expressed through the leading eigenvalue $\lambda^{(\alpha)}_1$ of $\mathcal{M}_{1}^{(\alpha)}$ as
\begin{equation}
    v_E^{(\alpha)} = \mathcal{E}_\alpha(0) = \frac{1}{2(\alpha-1)}\frac{\log\abs{\lambda^{(\alpha)}_1}^{-1}}{\log q}. \label{eq:ve}
\end{equation}
In general, this equation depends on the R\'{e}nyi index $\alpha$ as opposed to DU and DU2 circuits. However, so far no FDU3 gate exbiting a non-flat spectrum of entanglement velocities is known.
For general velocities $0\leq v\leq1/3$, we find
\begin{equation}
    \mathcal{E}_\alpha(v) = (1-3v)v_E^{(\alpha)} - v \frac{\log B_{2}^{(\alpha)}}{\log q} +3v.
\end{equation}
For $v>1/3$ the diagram factorizes because of DU3 and we recover the known result~\cite{Rampp2024}:
\begin{equation}
    \mathcal{E}_\alpha(v) = 1 - (1-v)\frac{\log B_{2}^{(\alpha)}}{2\log q}, \quad v>1/3.
\end{equation}
We observe that continuity at $v=1/3$ is automatically satisfied. This shows that the DU3 and DU$^*3$ conditions are compatible with each other and do not lead to inconsistencies along $v=1/3$.

The ELT in FDU3 circuits is therefore piecewise linear (see Fig.~\ref{fig:FDUk_ELT}), in contrast to generic many-body systems. This is a reflection of the fact that FDU3 circuits transport information along discrete directions in spacetime only (see Ref.~\cite{Rampp2025a} for a discussion of this connection).

\subsection{Higher levels}

In FDU$k\geq4$ circuits, the above picture is generalized to steps of maximal width $w\leq k-1$ as before, yielding
\begin{align}
    Z_\alpha(m,n) = \bra{\medcirc}\left(\mathcal{M}^{(\alpha)}_{+}\right)^{w_n}\mathcal{M}^{(\alpha)}_{w_{n-1}}\dots\mathcal{M}^{(\alpha)}_{w_1}\ket{\Box},
\end{align}
in the domain $m\leq(k-1)n+1$. Again, paths which have steps of maximal width at the beginning or end can be partially disconnected using the DU$k$ condition.
To describe the disconnected parts, we introduce
\begin{align}
    B_{k-1}^{(\alpha)} \equiv q^{k}\left(\,\,\vcenter{\hbox{\includegraphics[width=0.32\columnwidth]{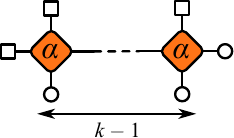}}}\,\,\right)^{1/(\alpha-1)}.
\end{align}

Considering only extremal paths that consist only of steps of width one and steps of maximal width $k-1$ is sufficient to extract the ELT from the scaling limit. We give the details of this construction in App.~\ref{app:elt_higher}. Note that the entanglement velocity is again given by Eq.~\eqref{eq:ve} just as in FDU3 circuits, because the DU$^*3$ condition is valid for all FDU$k$ circuits for $v=0$.
The ELT follows as
\begin{equation} \label{eq:elt_fduk}
    \mathcal{E}_\alpha(v) = \left(1-\frac{k}{k-2}v\right)v_E^{(\alpha)} - \frac{v}{k-2} \frac{\log B^{(\alpha)}_{k-1}}{\log q} +kv,
\end{equation}
in the range $0\leq v\leq\frac{k-2}{k}$.  In the regime where DU$k$ is valid, $v>\frac{k-2}{k}$ the ELT reads
\begin{align}
    \mathcal{E}_\alpha(v) = 1 - (1-v)\frac{\log B_{k-1}^{(\alpha)}}{2\log q}.
\end{align}
Again, continuity at $v_*$ is automatically satisfied.
Overall, we find a piecewise linear ELT with at most five kinks at the velocities $v=0,\pm \frac{k-2}{k},\pm1$, as shown in Fig.~\ref{fig:FDUk_ELT}. These kinks correspond to the possible directions where information propagates in FDU$k$ circuits (see again Ref.~\cite{Rampp2025a}).

\section{Solutions to the full dual-unitary hierarchy}
\label{sec:solutions}

While the FDU$k$ conditions enable to derive simple solutions to dynamical quantities of interest, finding gates that satisfy the FDU$k$ conditions is difficult, as they are non-linear high-dimensional tensor equations. In this section, we show that the FDU$k$ conditions possess non-trivial solutions that are distinct from dual-unitary dynamics for any $k\geq3$. We show that certain infinite families of spacetime lattice constructions introduced by us in Ref.~\cite{Rampp2025a} and labeled by an integer $k$ satisfy the FDU$k$ conditions. We also discuss the existence of FDU$k$ in smaller Hilbert space dimensions and beyond constructions based on spacetime lattices. We report numerical results that the tangent space of the analytical families Eqs~\eqref{eq:4pyr_chm} and \eqref{eq:4rock_chm} contains gates not described by those families in Sec.~\ref{sec:tangent}. In App.~\ref{app:clifford}, we present Clifford FDU3 gates that go beyond our analytical constructions.

\subsection{FDU$k$ gates from spacetime lattices}

Spacetime lattices of DU gates giving rise to exactly solvable dynamics, as introduced and discussed in Ref.~\cite{Rampp2025a}, are closely related to the full dual-unitary hierarchy. The dynamics of these models can be understood through their lattice structure, imposing constraints on the information flow and forming a trivial spacetime knot (for details see Ref.~\cite{Rampp2025a}), but as we show now certain spacetime lattice circuits can alternatively also be seen as manifestations of full dual-unitary circuits.

Before turning to more complex examples of spacetime lattices, we note that even dual-unitary brickwork circuits can be viewed through the lens of full dual unitarity. As apparent from Eq.~\eqref{eq:brickwork}, a brickwork arrangement of dual-unitary gates can be associated to a square lattice in spacetime.
Alternatively, we can choose a brickwork lattice of composite unitary gates acting on a composite $q^k$-dimensional local Hilbert space (for $k\geq3$),
\begin{align}
U_k = \vcenter{\hbox{\includegraphics[height=0.45\columnwidth]{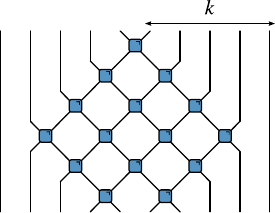}}}\,\,, \label{eq:transf_du}
\end{align}
for which the corresponding brickwork circuit leads to a spacetime lattice of dual-unitary gates with a seemingly more complex geometry.
This unitary gate $U_k$ is obtained by first forming a $(k-1)$ by $(k-1)$ square of DU gates. From this the bottommost gate is removed and one ancilla qudit is added to the left and right each.
While the individual gates~\eqref{eq:transf_du} do not satisfy the dual-unitary conditions~\eqref{eq:dual_unitarity_folded}, the full brickwork circuit directly reproduces the brickwork circuit of dual-unitary gates~\eqref{eq:brickwork}.
On the level of the spacetime lattice, this corresponds to applying a coordinate transformation breaking the square space-time symmetry.
The new coordinates of the $q^k$-dimensional circuit are asymptotically given by 
\begin{equation}
    \xi \sim \frac{2(k-2)^2}{k}x, \quad \tau\sim2(k-2)t,
\end{equation}
where $x,\,t$ are the coordinates of the underlying DU circuit. The transformed gate Eq.~\eqref{eq:transf_du} does not satisfy dual unitarity anymore, but the overall circuit is equivalent and thus still solvable. The local conditions expressing this solvability are precisely the FDU$k$ conditions.

In Ref.~\cite{Rampp2025a} the following family of gates was shown to lead to solvable dynamics:
\begin{align}
V_k = \,\,\vcenter{\hbox{\includegraphics[height=0.45\columnwidth]{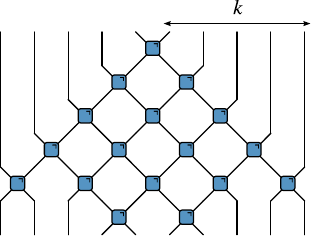}}}\,\,. \label{eq:fam1}
\end{align}
This gate can be seen as a local dressing of the coordinate-transformed dual-unitary gate introduced in Eq.~\eqref{eq:transf_du} by adding gates at the bottom, connecting the outermost legs to the ancilla qudits. The local dressing on the outermost bond breaks the dual unitarity of the dynamics, but keeps the solvability intact. For $k=3$ we obtain the gate
\begin{align}
V_3 = \,\,\vcenter{\hbox{\includegraphics[height=0.25\columnwidth]{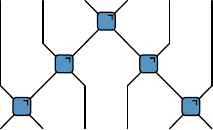}}}\,\,,
\end{align}
which satisfies the DU2 condition. For $k\geq4$, $V_k$ satisfies FDU$(k-1)$, as we show diagrammatically in App.~\ref{app:family_proofs}. In particular, for $k=4$ we obtain the pyramid gate
\begin{align}
V_4 = \,\,\vcenter{\hbox{\includegraphics[height=0.35\columnwidth]{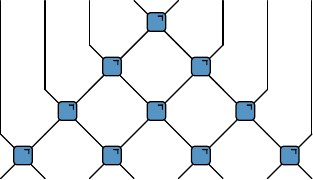}}}\,\,.
\end{align}

For $V_k$, the correlations are non-trivial along $\abs{v}=(k-3)/(k-1)$ and $\abs{v}=1$. The ELT has kinks at $\abs{v}=(k-3)/(k-1)$ and yields the entanglement velocity $v_E=(k-2)/k$~\cite{Rampp2025a}. We demonstrate this explicitly for $V_4$ in App.~\ref{app:channels_v4} using the channel description introduced in the previous sections.

The following family of spacetime lattice gates also leads to solvable dynamics,
\begin{align}
W_k = \,\,\vcenter{\hbox{\includegraphics[height=0.5\columnwidth]{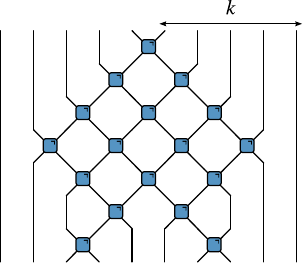}}}\,\,. \label{eq:fam2}
\end{align}
It can be established by similar means as for $V_k$ that $W_k$ with $k\geq4$ satisfies FDU$(k-1)$.

The above constructions have correlations propagating along four discrete directions in spacetime which distinguishes them from previous examples of solvable models. However, the FDU$k$ condition \emph{a priori} allows five distinct directions. Therefore, examples of gates that exhaust that possibility are of interest. An example is provided by the following gate
\begin{align}
U_{\rm{rays}} = \,\,\vcenter{\hbox{\includegraphics[height=0.35\columnwidth]{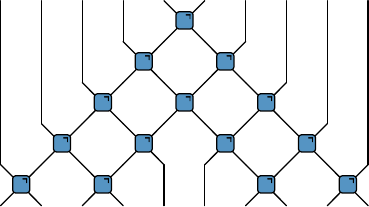}}}\,\,,\nonumber
\end{align}
which was introduced in Ref.~\cite{Rampp2025a}. It satisfies FDU3 and an analysis of the channels reveals the expected propagation of correlations along $v=0$ and $\abs{v}=1/3,1$.

\subsection{FDU$k$ gates in smaller Hilbert spaces}

The above constructions yield examples of FDU$k$ gates for any $k$. However, the size of the local Hilbert space grows exponentially with $k$, as $q^k$ for the coordinate transformed dual-unitaries and $q^{k+1}$ for the other constructions. This makes implementation on digital quantum devices impractical for large $k$. It also raises the question of the minimal local Hilbert space dimension in which FDU$k$ gates exist. While we do not answer this question in full, we present gates living in significantly smaller Hilbert spaces. These are obtained from the spacetime lattice approach by generalizing it to biunitary connections. For an introduction to biunitary connections and their applications to quantum dynamics, we refer to Refs.~\cite{Reutter2019,Claeys2024}. 

For any spacetime lattice gate where the local dimension is an even power of $q$, i.e. $q^{2j}$, there exists a construction that is defined on a local $q^j$-dimensional Hilbert space, thereby compressing the Hilbert space exponentially. The compressed gate is formed from  complex Hadamard matrices (CHM). A matrix $H\in\mathbb{C}^{q\times q}$ is called a CHM if it is proportional to a unitary matrix, $H^\dagger H=HH^\dagger=q\mathbbm{1}$, and its entries are all phases, $\abs{H_{ab}}=1$~\cite{Tadej2006}. CHMs have various applications in quantum information theory~\cite{Werner2001,Englert2001,Wojcik2003,Klappenecker2004}. We represent a CHM graphically as
\begin{align}
    H_{ab} = \,\,\vcenter{\hbox{\includegraphics[width=0.024\textwidth]{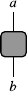}}}\,\,,
\end{align}
and we will consider gates constructed from multiple CHMs, `glued' together by delta tensors delta tensors with an arbitrary number of legs, graphically represented as
\begin{align}
    \vcenter{\hbox{\includegraphics[height = .16\columnwidth]{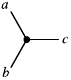}}} = \delta_{abc}\,,
\end{align}
here depicted for three legs but more generally defined for any number of legs.

For the above constructions, we obtain a set of gates acting on a local $q^2$-dimension Hilbert space:
\begin{subequations}
    \begin{align}
        (\tilde{U}_{4})_{abcd,efgh}&= \,\,\vcenter{\hbox{\includegraphics[width=0.3\columnwidth]{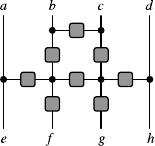}}}\,\,,\\
        (\tilde{V}_{4})_{abcd,efgh}&= \,\,\vcenter{\hbox{\includegraphics[width=0.3\columnwidth]{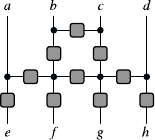}}}\,\,,\label{eq:4pyr_chm}\\ 
        (\tilde{W}_{4})_{abcd,efgh}&= \,\,\vcenter{\hbox{\includegraphics[width=0.3\columnwidth]{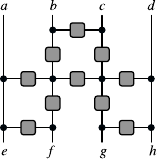}}}\,\,.\label{eq:4rock_chm}
    \end{align}
\end{subequations}
Here $\tilde{U}_4$ can be seen as a coordinate transformation of the (generalized) self-dual kicked Ising model on the square lattice~\cite{Akila2016,Gutkin2020,Claeys2022a}. It can be shown that $\tilde{U}_4$ satisfies FDU4, while $\tilde{V}_4$ and $\tilde{W}_4$ satisfy FDU3. Importantly, this establishes the existence of FDU3 and FDU4 gates for ququads.

The existence of FDU$k$ gates for qubits remains an open question. We note in passing that sheared dual-unitary gates~\cite{Sommers2024,Rampp2024} satisfy a weaker version of the FDU3 condition in one half of spacetime only. For sheared dual-unitaries, only $U$ and $SU^TS$ satisfy Eq.~\eqref{eq:cond_fdu3}.

\subsection{Tangent space analysis}
\label{sec:tangent}

To gain more insight into the structure of the solution space of the generalized dual-unitarity conditions we numerically determine the dimension of the tangent space at known solutions of the conditions. This enables us to determine the dimension of the solution space not covered by the analytic constructions. We follow the procedure outlined in Ref.~\cite{Prosen2021}. We numerically construct the Jacobian $J$ at a known exact solution of the condition under question. The dimension of the tangent space is then given by the number of vanishing singular values. In practice, the singular values never vanish exactly, but the tangent space can be determined reliably if there is a large enough gap in the singular value spectrum.

We restrict ourselves to the analytic families Eqs.~\eqref{eq:4pyr_chm} and~\eqref{eq:4rock_chm} for $q=4$ and the DU3, DU$^*3$, and FDU3 conditions because of the computational cost associated with constructing the tensor equations. In fact, the number of equations associated to the DU$^*3$ condition is $4q^{10}$ ($2q^8$ for DU$3$) which is much larger than the number of parameters ($q^4$ real parameters for a unitary two-site gate). Thus, the information contained in the equations is likely highly redundant. To reduce the computational cost, we therefore randomly choose a fraction $p_{*3}$ ($p_3$) of components and construct the Jacobian for this fraction of equations. If the fraction is large enough, we expect to reproduce the tangent space of the full set of equations. We verify this by plugging the gates obtained by deforming the original solution along the directions of the tangent space back into the solvability conditions. We find that the resulting gates satisfy the conditions to numerical precision, showing that the sampled equations do not overestimate the tangent space. We also observe that the tangent space remains insensitive to increasing the fraction. For the gate~\eqref{eq:4pyr_chm} we use $p_{*3}=10^{-4}$ and $p_3=6 \times 10^{-3}$, and for the gate~\eqref{eq:4rock_chm} we use $p_{*3}=5 \times 10^{-4}$ and $p_3=3 \times 10^{-2}$.

For the gate $\tilde{V}_4$ from Eq.~\eqref{eq:4pyr_chm} we find the dimensions $D_{\mathrm{DU}3}=41$, $D_{\mathrm{DU}^*3}=45$, and $D_{\mathrm{FDU}3}=38$ for the tangent spaces associated to the DU3, DU$^*3$, and FDU3 condition respectively. For the gate $\tilde{W}_4$ from Eq.~\eqref{eq:4rock_chm} we find $D_{\mathrm{DU}3}=47$, $D_{\mathrm{DU}^*3}=45$, and $D_{\mathrm{FDU}3}=40$. To compare with the number of parameters of the analytical families, we need to take the $SU(q)\times SU(q)$ gauge symmetry of the FDU3 condition into account. Given a solution $U$ to the FDU3 conditions, the gauge transformed gate $(u\otimes v)U(v^\dagger\otimes u^\dagger)$ where $u,v$ are local unitaries in $SU(q)$ also satisfies the FDU3 conditions. This follows because the local rotations cancel on the internal legs of the condition. Therefore, we need to add $2\,\textrm{dim}(SU(4))=30$ parameters from the gauge freedom. Having fixed the gauge, there remain two internal phases, four local phases (the other four can be absorbed into the gauge transformations), and one global phase, yielding $37$ parameters both for the 4-pyramid and 4-rocket. We conclude that there is a one-parameter family of transformations preserving FDU3 not included in the analytical construction from Eq.~\eqref{eq:4pyr_chm} and a three-parameter family for the construction from Eq.~\eqref{eq:4rock_chm}.

\section{Conclusion and outlook}

In this paper, we have proposed and investigated an infinite hierarchy of conditions leading to exactly solvable non-integrable quantum dynamics, yielding a novel generalized dual-unitary hierarchy that we term the full dual unitary hierarchy. In contrast to the previously proposed generalized dual unitary hierarchy of Ref.~\cite{Yu2024}, we add additional conditions that provide solvability in all of spacetime. We have investigated dynamical correlation functions and the entanglement line tension, showing that they can be analyzed by considering certain low-dimensional quantum channels. Information in full dual-unitary circuits flows along at most five directions in spacetime, yielding a piecewise linear entanglement line tension. 

A natural next step is to investigate the dynamics of FDU$k$ circuits after quenches from states with low entanglement. We expect that the concept of a solvable state~\cite{Piroli2020} that is compatible with the solvability condition can be extended to FDU$k$ circuits. It would be of interest to find analytical solutions for indicators of quantum chaos or ergodicity, such as the spectral form factor~\cite{Haake2018,Thouless1977,Altshuler1986} or deep thermalization~\cite{Ho2022,Cotler2023}.

We have shown that the full dual-unitary hierarchy possesses non-trivial solutions for any level $k$, showing dynamics distinct from dual unitarity. The complete characterization of the solution space, however, remains an open problem. In particular, the minimal local Hilbert space dimension for which FDU$k$ gates exist is unknown.

In principle, there are many different conditions similar to the DU$^*k$ and DU$k$ conditions yielding solvability of correlation functions and entanglement dynamics in some part of spacetime. For example, the condition
\begin{align}
    \vcenter{\hbox{\includegraphics[width=0.18\textwidth]{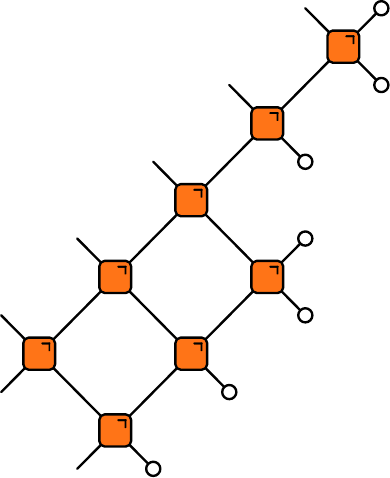}}}\!\! = \,\,\vcenter{\hbox{\includegraphics[width=0.18\textwidth]{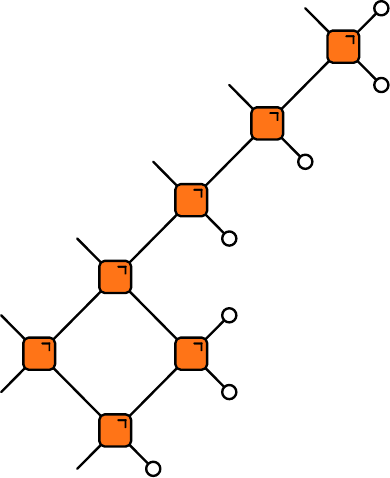}}}, \label{eq:cond_alt}
\end{align}
leads to solvability in the range $1/3\leq v\leq1/2$. It would be desirable to have an organizing principle bringing order to the wealth of conditions. Moreover, there are no generally applicable techniques to solve the resulting non-linear tensor equations. Thus, finding gates that satisfy such conditions poses a major difficulty.

It would also be interesting to further explore the connections between spacetime lattice constructions and solvability conditions. Is it possible to find a condition on a local unit cell that expresses the solvability of any solvable spacetime lattice circuit? And conversely, given a solvability condition, is it always possible to find a solution in terms of a spacetime lattice unit cell? Such a connection would provide a surprising duality between the geometry of spacetime and the form of interactions.

\begin{acknowledgments}
We wish to acknowledge useful discussions with Cecilia De Fazio.
\end{acknowledgments}

\appendix

\begin{widetext}

\section{Operator Schmidt decomposition of FDU$k$ gates}

We note that any DU$^*k$ gate must have a rank deficient operator Schmidt decomposition~\cite{Nielsen2003}. To show this, we first assume the contrary, which implies that the space-time dual gate possesses an inverse (depicted in dark green) satisfying
\begin{align}
    \vcenter{\hbox{\includegraphics[height=0.045\paperwidth]{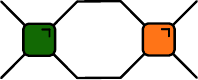}}}\,\, = \,\,\vcenter{\hbox{\includegraphics[width=0.035\textwidth,angle=90]{figs/unitarity2_v2.pdf}}}\,\,.
\end{align}
Acting repeatedly with this inverse on the DU$^*k$ condition, we obtain
\begin{align}
    \vcenter{\hbox{\includegraphics[height=0.17\paperwidth]{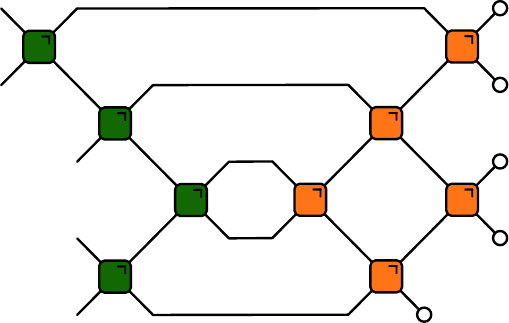}}}\,\, = \,\,\vcenter{\hbox{\includegraphics[height=0.17\paperwidth]{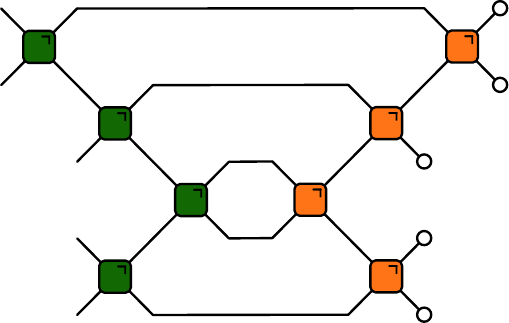}}}\,\,\implies\,\,\,\,
    \vcenter{\hbox{\includegraphics[height=0.17\paperwidth]{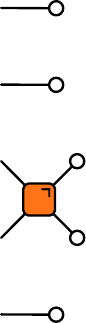}}}\,\, = \,\,\vcenter{\hbox{\includegraphics[height=0.17\paperwidth]{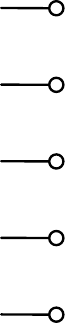}}}\,\,.
\end{align}
This implies that the gate is DU, which is in contradiction to the assumption that it is a DU$^*k$ gate. Thus, the inverse of the dual cannot exist, implying rank deficiency.

\section{Derivations on FDU$k$ circuits with $k\geq4$}
\label{app:derivations_higher}

\subsection{Dynamical correlation functions}
\label{app:corr_higher}

\begin{figure}[t]
    \centering
    \includegraphics[width=0.45\textwidth]{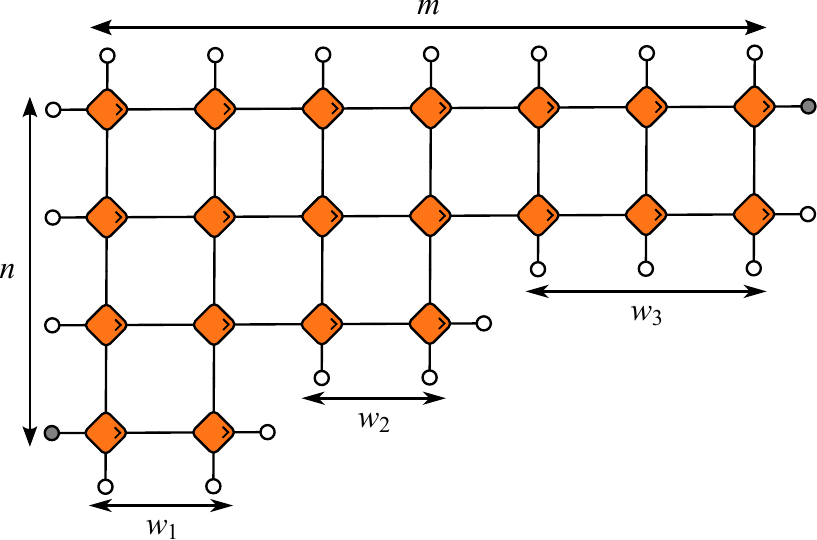}
    \caption{Applying the DU$^*k$ condition to a tensor network diagram representing a correlation function leads to the appearance of steps on the boundary of the diagram. These steps have height $h=1$ and are characterized by their width $w_j$. For DU$^*k$, the steps can have a width of at most $k-1$.}
    \label{fig:boundary_shape}
\end{figure}

\subsubsection{Thin path domain}

Let us first analyze the correlator in the thin path domain. The crucial insight is the following: when reducing the diagram using the DU$^*k$ condition or any of the DU$^*\ell$ conditions with $k<\ell$, the boundaries of the diagram acquires a particular shape. For concreteness, we start the reduction from the bottom right corner. The resulting boundary contains steps that have at most width $w=k-1$ and always have height $h=1$, with the exception of the lowest part of the boundary, which may be wider than $k-1$, because it might not be possible to reduce it further. Additionally, the DU$^*k$ condition does not enable the removal of the gate below the top right corner. This is illustrated in Fig.~\ref{fig:boundary_shape}. Analogous considerations hold when reducing from the top left corner.

Assume now that the diagram can be reduced to a thin path. The path is then determined by the shape of the boundary and the restriction on the width and height of its steps directly translates to restrictions on the path. The number of steps of width $j$ -- we call this $a_j$ -- is constrained by the total height and width of the diagram, yielding Eq.~\eqref{eq:step_constraints}.
Let us determine the range of $m,\,n$ for which the reduction to a thin path is possible, i.e., Eq.~\eqref{eq:step_constraints} has a solution. For given $n$, what is the maximal $m$ such that a solution exists? In this extreme case, the thin path has $a_1=1$ and $a_{k-1}$ maximal. All other $a_j$ vanish. Using $n=1+a_{k-1}$ yields $m_{\rm{max}}=(k-1)n-(2k-3)$.  

To find the mathematical expression for the reduced diagram corresponding to a thin path, we associate a step of width $w$ to the quantum channel Eq.~\eqref{eq:channel_thin}.
A sequence of steps then defines a sequence of channels and any sequence that satisfies Eqs.~\eqref{eq:step_constraints} and starts with $\mathcal{M}_1$ corresponds to a valid reduced diagram. The correlator is then given by Eq.~\eqref{eq:corr_fduk}.

\subsubsection{Thick path domain} 

This corresponds to the range $(k-1)n-(2k-4)\leq m\leq (k-1)n-2$.  To illustrate the appearance of non-skeleton diagrams in this range, consider $m=(k-1)n-(2k-4)$. After reduction using the DU$^*k$ condition, the diagram reads
\begin{align}
    &\vcenter{\hbox{\includegraphics[width=0.32\paperwidth]{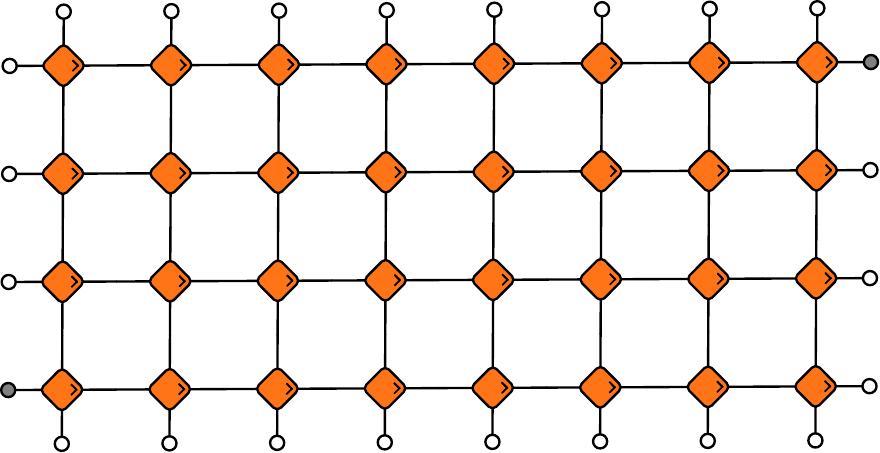}}}\,\,=\,\,\vcenter{\hbox{\includegraphics[width=0.32\paperwidth]{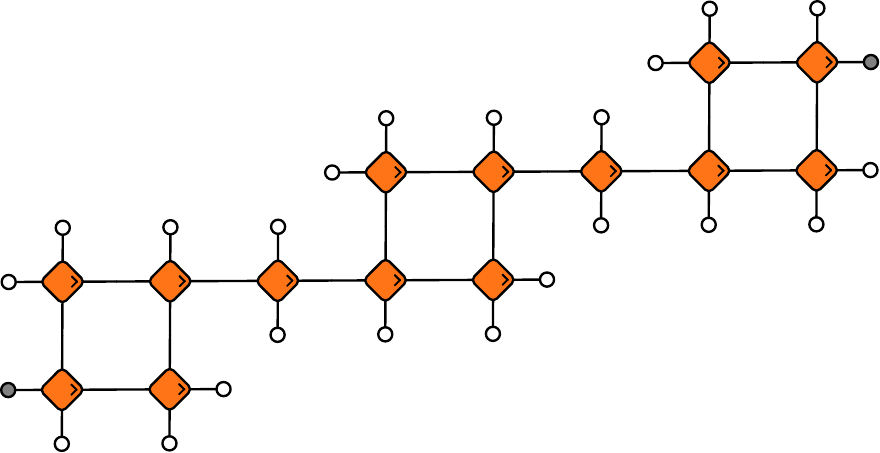}}}\,\,.
\end{align}
The bottom boundary consists of steps of maximal width $(k-1)$ with the exception of the first step only, which has width two. Note that for thin paths, the first step always has width one. Using Eq.~\eqref{eq:channel_v13_large},
the correlator can be written as
\begin{align}
    C= \bra{\sigma}\left(\mathcal{M}_2'\mathcal{M}_+^{k-3}\right)^{n-2}\mathcal{M}_2'\ket{\rho},
\end{align}
for $m=(k-1)n-(2k-4)$. This is straightforwardly generalized to $m=(k-1)n-(2k-3)+j$ with $j=1,\dots,k-2$ by noting that in these cases the bottom step has width $j+1$ while all other steps remain of maximal width. This leads to Eq.~\eqref{eq:thick_path1}.

When $j$ grows larger than $k-2$, the width of the bottom step becomes larger than $k-1$. However, now we can use the DU$k$ condition to reduce the width to $k-1$. This increases the width of the second step from the bottom above $k-1$, so it can also be reduced. In this manner, all steps are shifted to the left by $j-(k-2)$. E.g., for FDU4
\begin{align}
    &\vcenter{\hbox{\includegraphics[width=0.4\paperwidth]{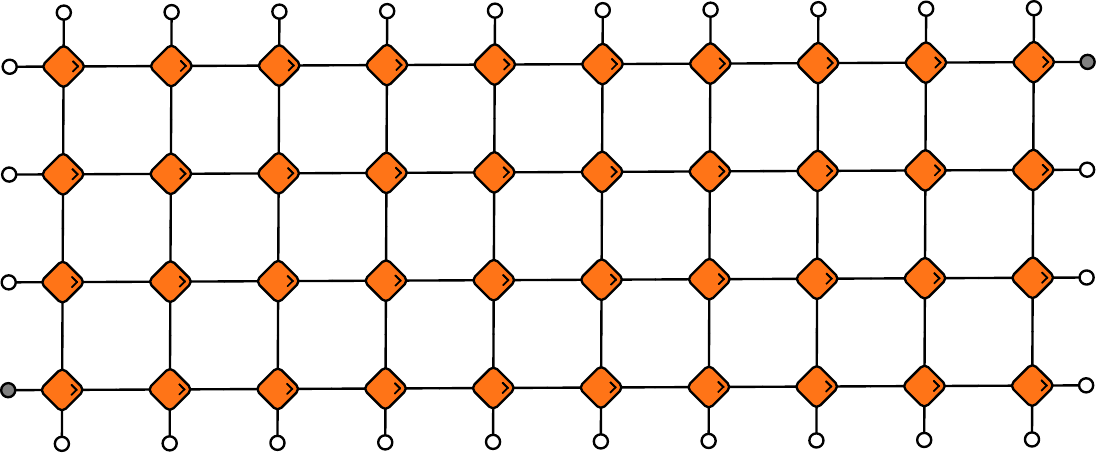}}}\,\,=\,\,\vcenter{\hbox{\includegraphics[width=0.4\paperwidth]{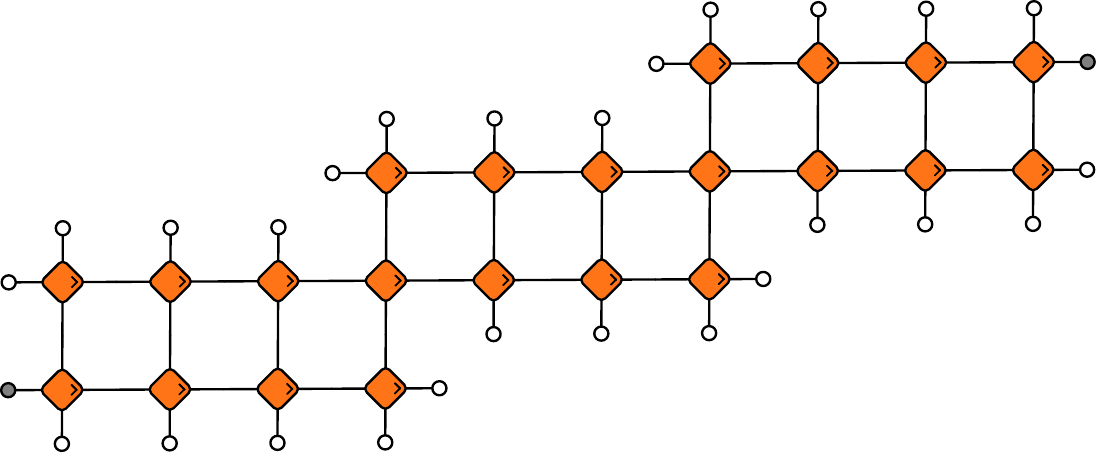}}}\,\,,\nonumber\\
    &\qquad\qquad\qquad=\,\,\vcenter{\hbox{\includegraphics[width=0.4\paperwidth]{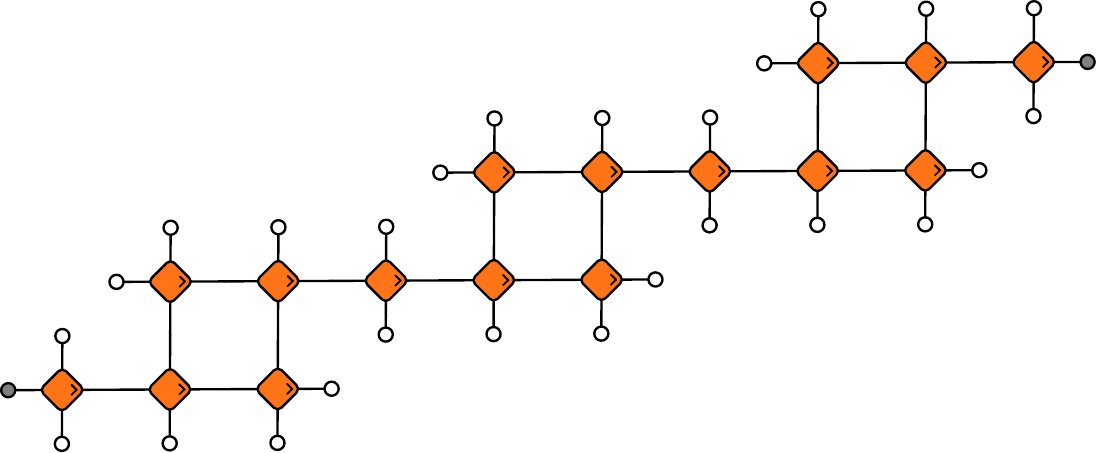}}}\,\,.
\end{align}
Generally, for $k-1\leq j\leq2k-1$ the correlator is given by Eq.~\eqref{eq:thick_path2}.

\subsection{Entanglement line tension}
\label{app:elt_higher}

In the computation of $Z_\alpha$ for FDU$k$ circuits, the $a_j$ satisfy the consistency conditions
\begin{align}
    m = \sum_{j=1}^{k-1}ja_j, \quad
    n = \sum_{j=1}^{k-1}a_j.
\end{align}
To obtain a particularly simple expression where the diagram is disconnected as much as possible we are interested in finding extremal paths. These maximize the number of steps of width $k-1$. As we only want to extract the leading behavior of $\log Z_\alpha$, it is sufficient to restrict ourselves to values of $m$ such that $(k-1)n-m=(k-2)j$ with $j\in\mathbb{N}$. In this case, a valid path is given by
\begin{equation}
    a_1 = j, \quad a_2=\dots=a_{k-2}=0, \quad a_{k-1} = n-j.
\end{equation}
It maximizes the number of steps of width $k-1$ and otherwise consists only of steps of width one. Therefore only two different channels appear in the expression
\begin{equation}
    Z_\alpha(m,n) = \bra{\medcirc}\left(\mathcal{M}^{(\alpha)}_{1}\right)^{\frac{(k-1)n-m}{k-2}}\mathcal{M}^{(\alpha)}_{+}\ket{\Box}\left(\frac{B^{(\alpha)}_{k-1}}{q^{k}}\right)^{(\alpha-1)\frac{m-n}{k-2}}.
\end{equation}
Extracting the leading order term yields Eq.~\eqref{eq:elt_fduk}.

\section{Proof of FDU$(k-1)$ for family of gates $V_k$}
\label{app:family_proofs}

In this section, we show that $V_k$ satisfies the FDU$(k-1)$ conditions. This establishes the exact solvability of the ELT and correlation functions.
First, we show that $V_k$ satisfies DU$(k-1)$. The left hand side of the condition reads
\begin{equation}
    \vcenter{\hbox{\includegraphics[width=0.9\columnwidth]{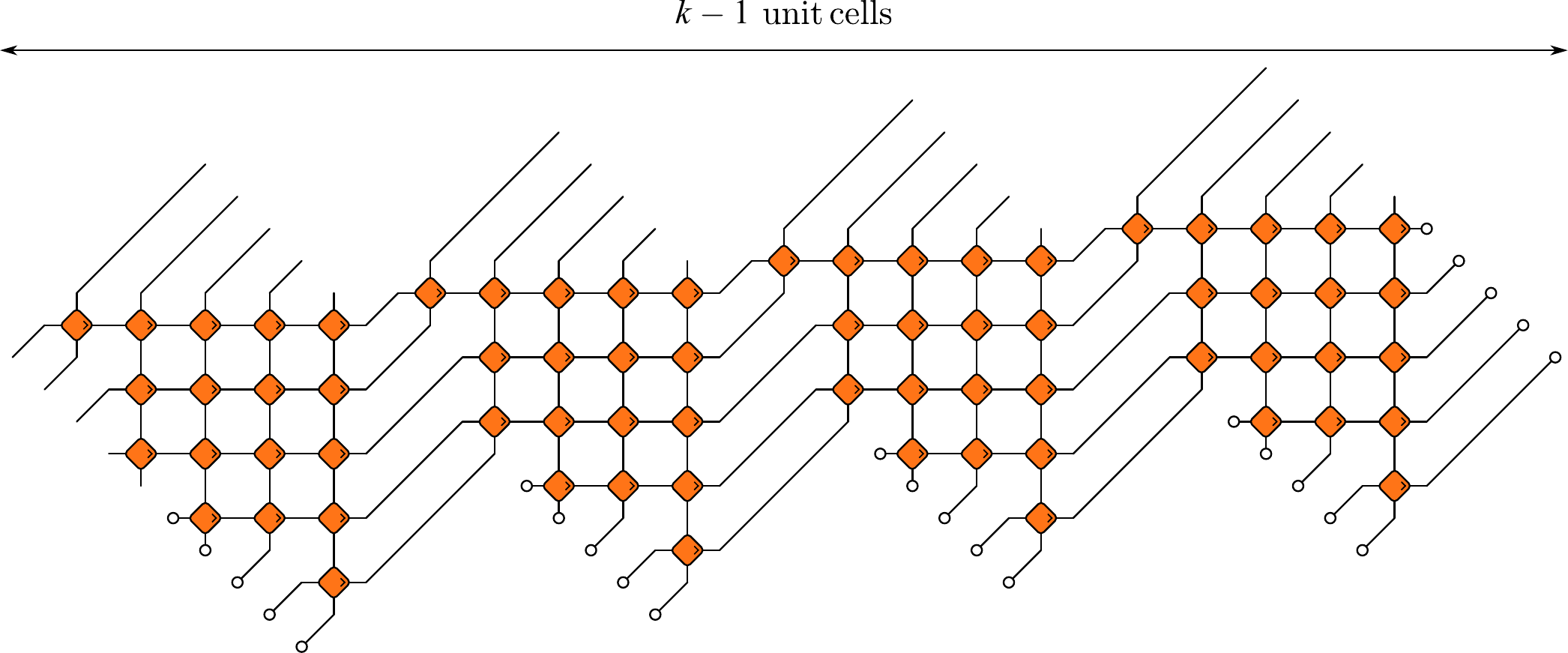}}}\,\,.
\end{equation}
After simplifying the diagram using unitarity, $k-2$ rows of gates remain in the leftmost unit cell. When going to the next unit cell to the right, the number of rows is reduced by one after the first gate in the new unit cell. This continues until the last row is reached
\begin{equation}
    \vcenter{\hbox{\includegraphics[width=0.9\columnwidth]{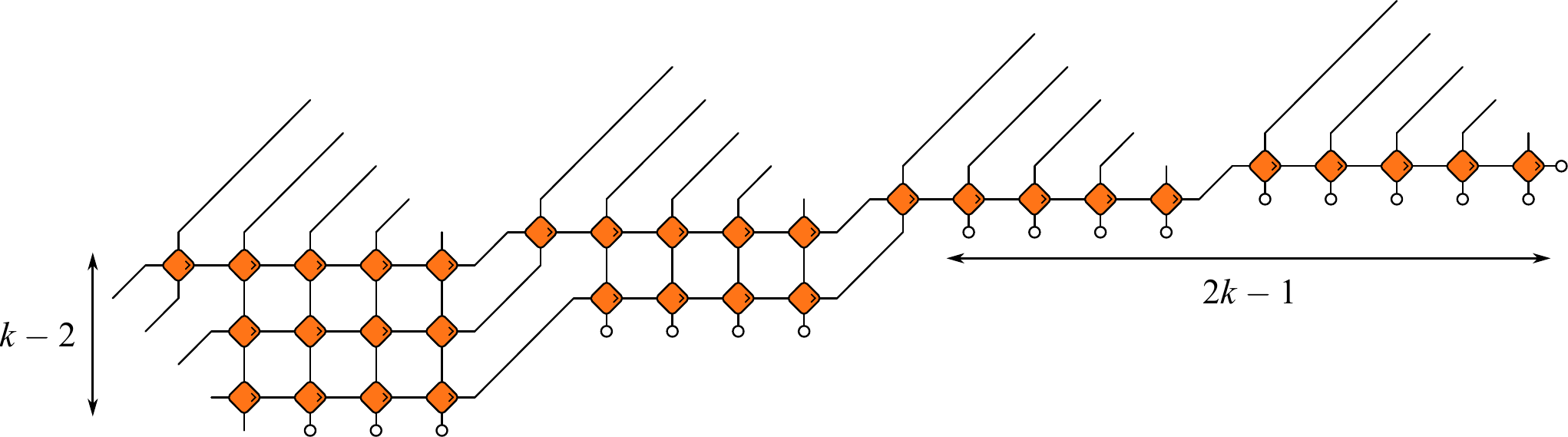}}}\,\,.
\end{equation}
For the diagram consisting of $k-1$ unit cells, this means that there are $2k-1$ gates that can be removed using dual unitarity, yielding
\begin{equation}
    \vcenter{\hbox{\includegraphics[width=0.9\columnwidth]{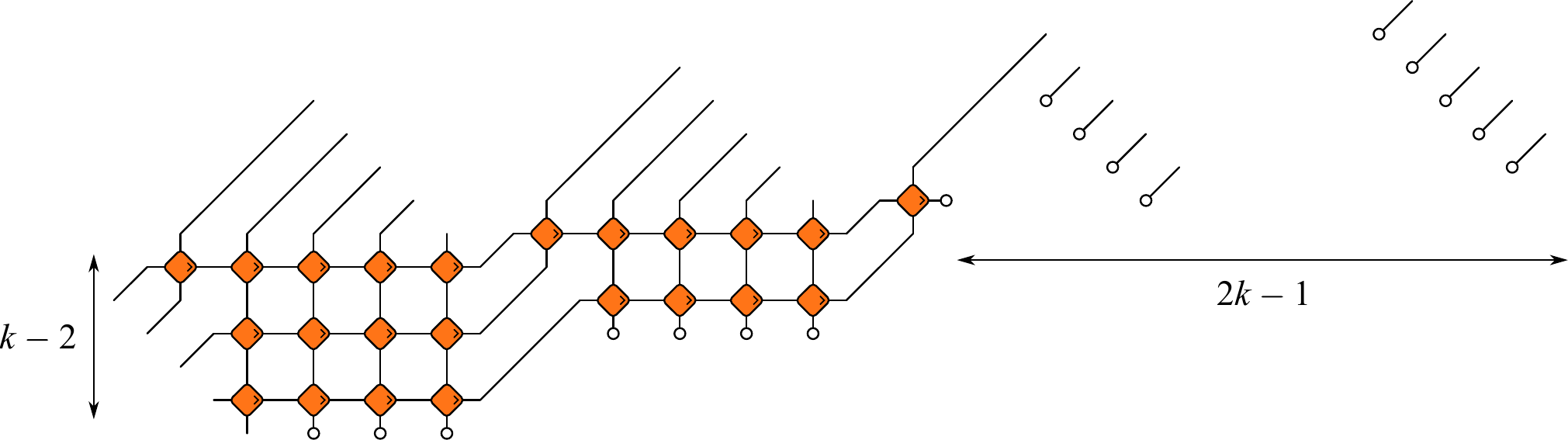}}}\,\,.
\end{equation}
In particular, the rightmost unit cell has been reduced to identities, yielding the DU$(k-1)$ condition.
Consider now the diagram on the left hand side of Eq.~\eqref{eq:cond_dustark} for $k-1$
\begin{equation}
    \vcenter{\hbox{\includegraphics[width=0.9\columnwidth]{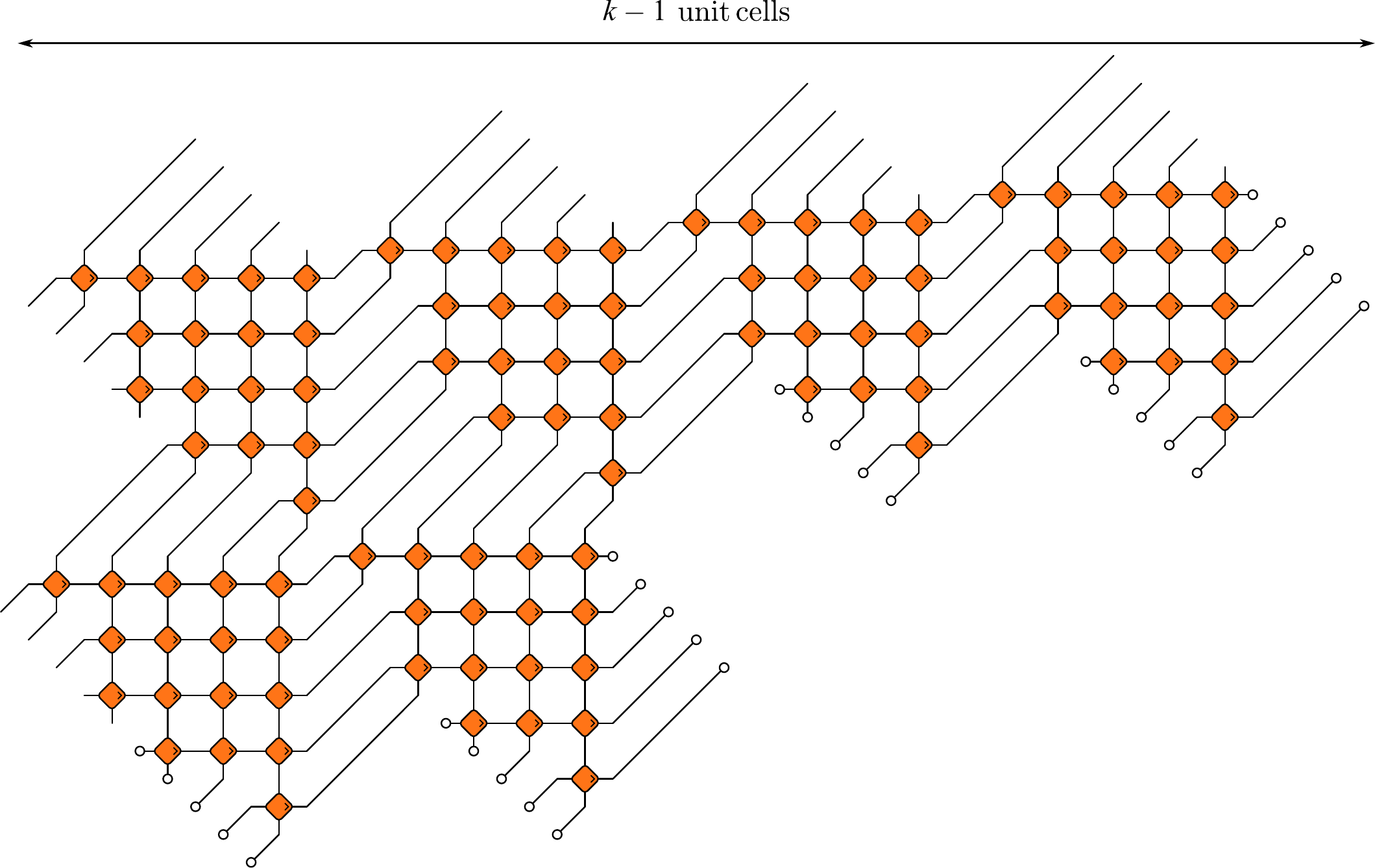}}}\,\,.
\end{equation}
Using unitarity, this can be reduced to
\begin{equation}
    \vcenter{\hbox{\includegraphics[width=0.9\columnwidth]{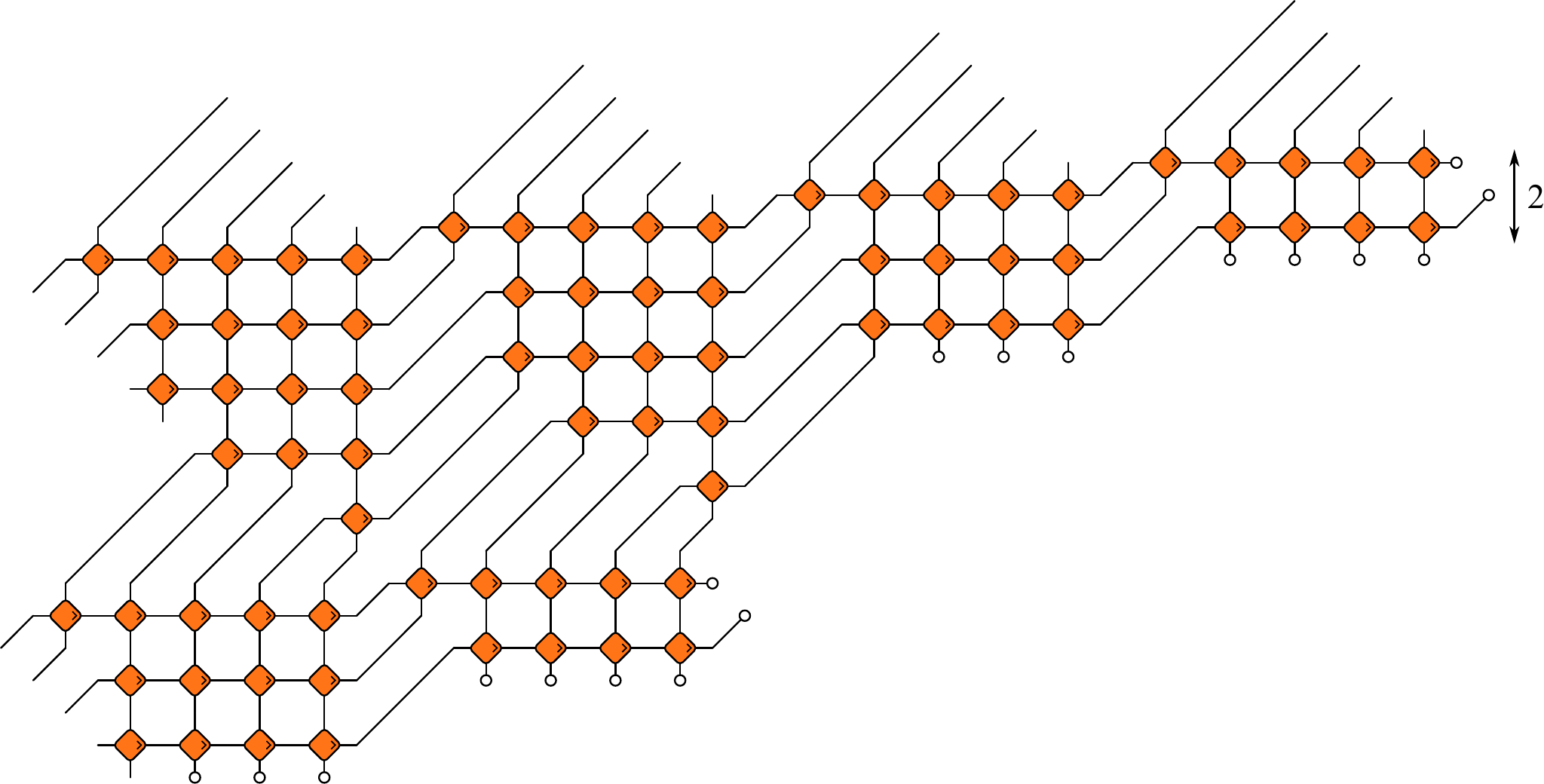}}}\,\,.
\end{equation}
After the simplification, two rows of gates remain in the rightmost unit cell in the top row of unit cells. When moving left one unit cell, the number of rows increases by one. Applying dual unitarity yields
\begin{equation}
    \vcenter{\hbox{\includegraphics[width=0.9\columnwidth]{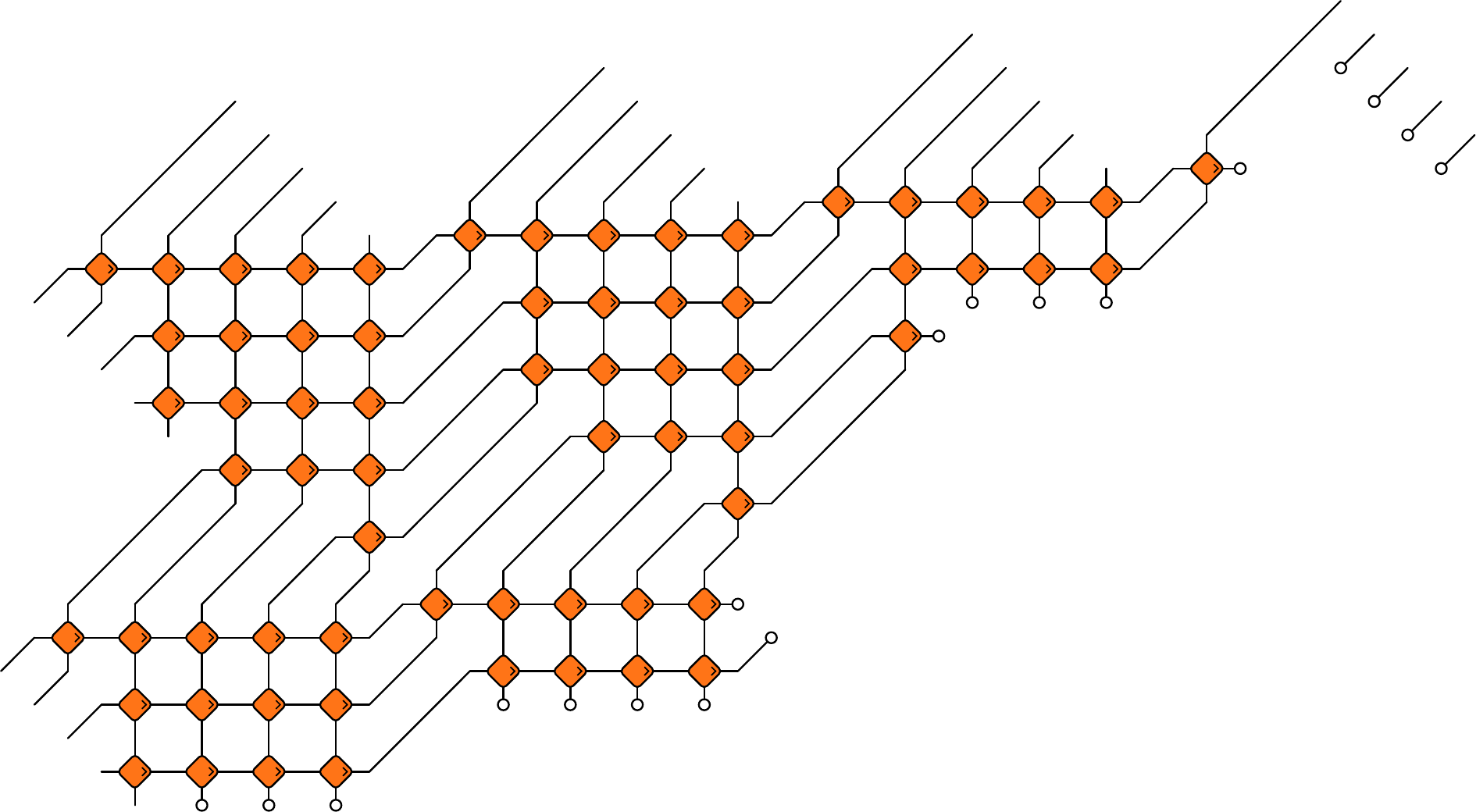}}}\,\,.
\end{equation}
Now in the last unit cell, before the bottom row of unit cells begins, a single dual-unitary gate obstructs further reduction of the $(k-1)$th row of gates. Using dual unitarity on the bottom row of unit cells, enables removing this gate from the bottom 
\begin{equation}
    \vcenter{\hbox{\includegraphics[width=0.9\columnwidth]{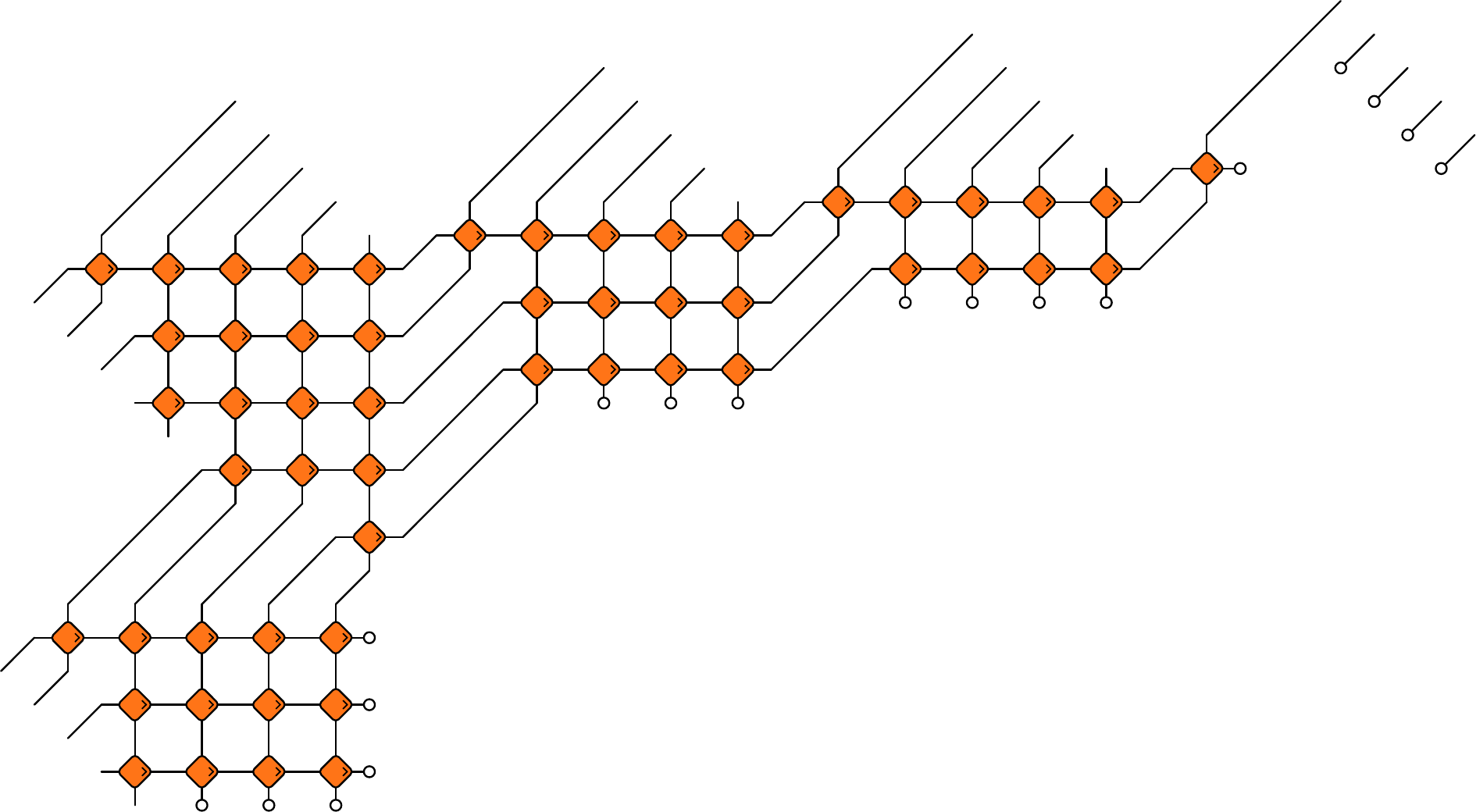}}}\,\,,
\end{equation}
and further removing all gates belonging to the bottom right unit cell. This yields one part of the DU$^*(k-1)$ condition. The proof of the remaining part is analogous.

\section{Channel description of the dynamics of $V_4$}
\label{app:channels_v4}

Let us illustrate for $V_4$ how to derive this from the theory developed in the previous sections. Considering first correlation functions, the channel $\mathcal{M}_1$ propagating correlations along $v=0$ can be simplified as follows 
\begin{align}
\mathcal{M}_1 = \,\,\vcenter{\hbox{\includegraphics[height=0.17\paperwidth]{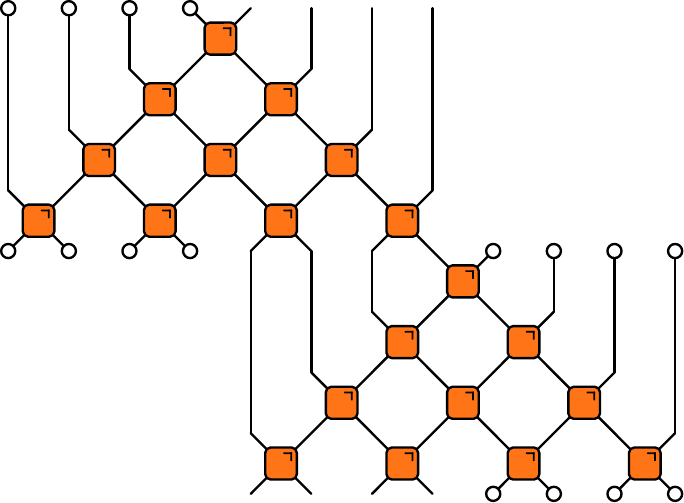}}}\,\,=\,\,\vcenter{\hbox{\includegraphics[height=0.17\paperwidth]{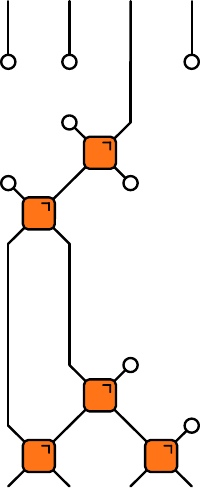}}}\,\,.
\end{align}
Any left eigenvector with non-zero eigenvalue must be of the form $\bra{\mathbbm{1}\otimes\mathbbm{1}\otimes\sigma\otimes\mathbbm{1}}=\,\,\vcenter{\hbox{\includegraphics[height=0.025\columnwidth]{figs/circle.pdf}}}\,\,\vcenter{\hbox{\includegraphics[height=0.025\columnwidth]{figs/circle.pdf}}}\,\,\vcenter{\hbox{\includegraphics[height=0.025\columnwidth]{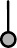}}}\,\,\vcenter{\hbox{\includegraphics[height=0.025\columnwidth]{figs/circle.pdf}}}\,$,
where the grey bullet represents an arbitrary operator $\sigma\in\mathbb{C}^{q\times q}$. Applying the channel to this ansatz yields
\begin{align}
    \bra{\mathbbm{1}\otimes\mathbbm{1}\otimes\sigma\otimes\mathbbm{1}}\mathcal{M}_1 = \,\,\vcenter{\hbox{\includegraphics[height=0.17\paperwidth]{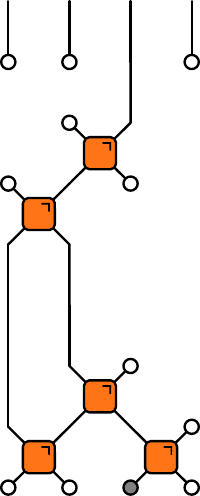}}}\,\, = \tr\left[\sigma\right]\,\,\vcenter{\hbox{\includegraphics[height=0.03\columnwidth]{figs/circle.pdf}}}\,\,\vcenter{\hbox{\includegraphics[height=0.03\columnwidth]{figs/circle.pdf}}}\,\,\vcenter{\hbox{\includegraphics[height=0.03\columnwidth]{figs/circle.pdf}}}\,\,\vcenter{\hbox{\includegraphics[height=0.03\columnwidth]{figs/circle.pdf}}}\,,
\end{align}
showing that the only such eigenvector is the product of vectorized identities. Therefore, no correlations asymptotically spread along $v=0$.
As $V_4$ is a FDU3 gate, we also consider the quantum channels propagating correlations along $v=1/3$. First, we consider $\mathcal{M}_2$
\begin{align}
\mathcal{M}_2 = \,\,\vcenter{\hbox{\includegraphics[height=0.24\paperwidth]{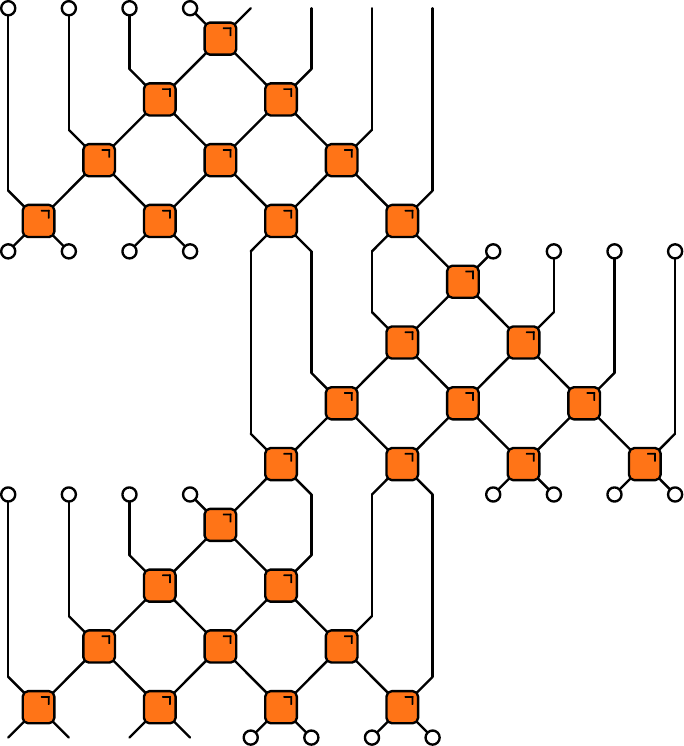}}}\,\,=\,\,\vcenter{\hbox{\includegraphics[height=0.24\paperwidth]{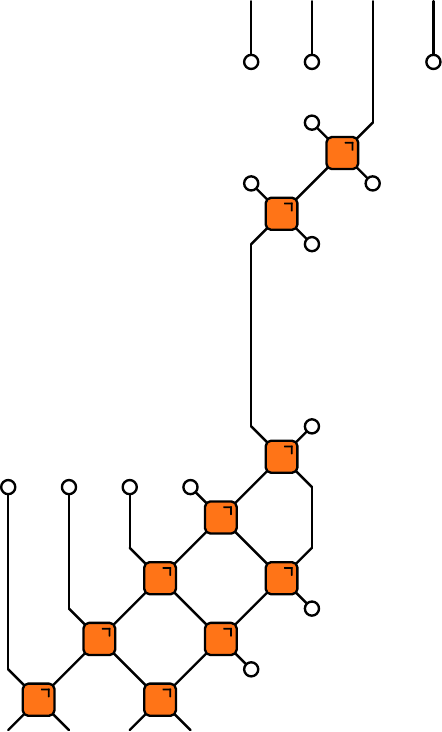}}}\,\,.
\end{align}
This channel possesses left eigenvectors of the form $\bra{\mathbbm{1}\otimes\mathbbm{1}\otimes\sigma\otimes\mathbbm{1}}$ for certain traceless operators $\sigma$ that are determined by the light-cone channels of the underlying dual-unitary gates
\begin{align}
    \bra{\mathbbm{1}\otimes\mathbbm{1}\otimes\sigma\otimes\mathbbm{1}}\mathcal{M}_2 = \,\,\vcenter{\hbox{\includegraphics[height=0.24\paperwidth]{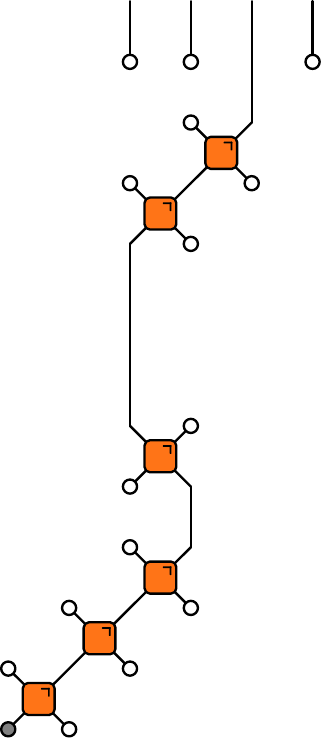}}}\,\,.
\end{align}
This implies that correlations spread along $v=1/3$. Further, $\mathcal{M}_2'$ reads
\begin{align}
\mathcal{M}_2' = \,\,\vcenter{\hbox{\includegraphics[height=0.24\paperwidth]{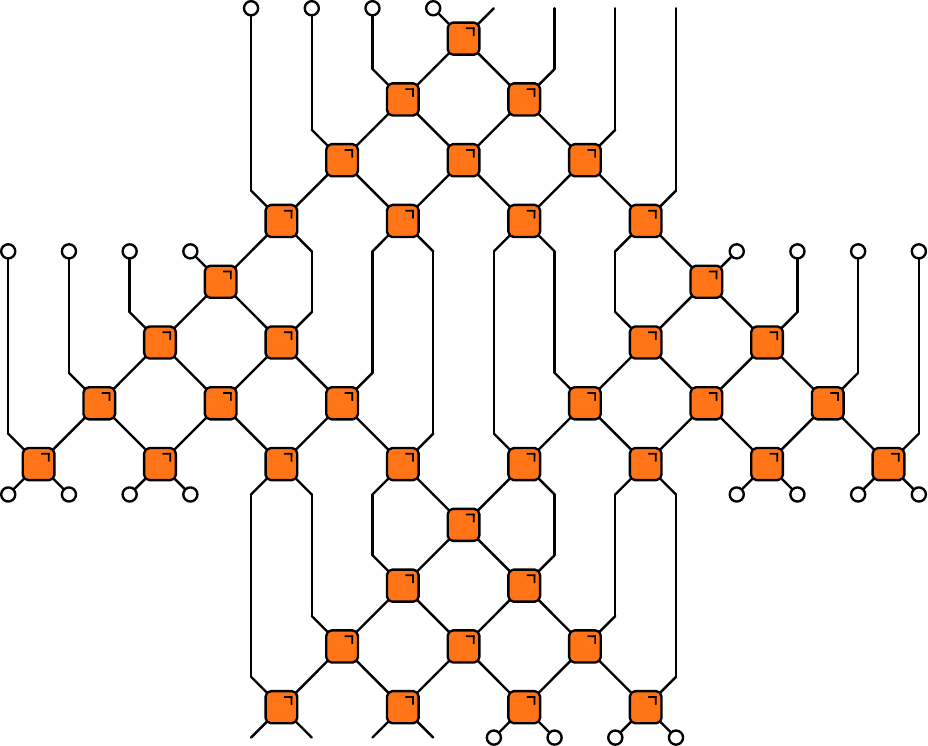}}}\,\,=\,\,\vcenter{\hbox{\includegraphics[height=0.24\paperwidth]{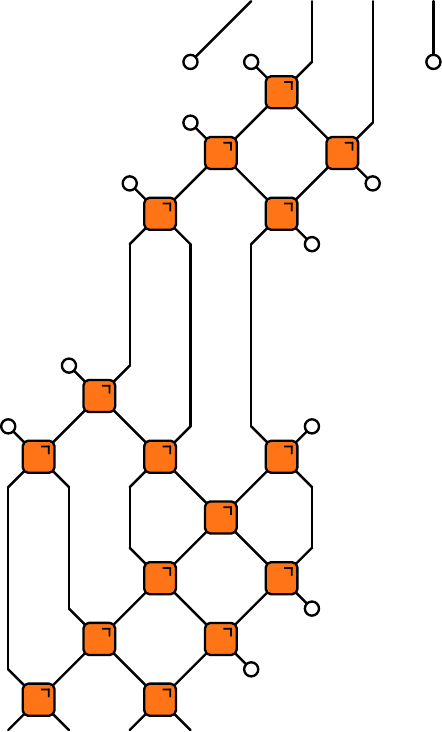}}}\,\,,
\end{align}
with left eigenvectors of the form $\bra{\mathbbm{1}\otimes\sigma\otimes\mathbbm{1}\otimes\mathbbm{1}}$ and $\bra{\mathbbm{1}\otimes\mathbbm{1}\otimes\sigma\otimes\mathbbm{1}}$.

Along $v=1$ the correlations are determined by the channel
\begin{align}
\mathcal{M}_+ = \,\,\vcenter{\hbox{\includegraphics[width=0.14\paperwidth]{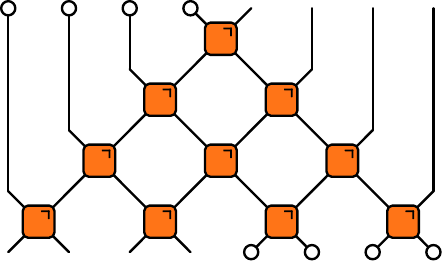}}}\,\,=\,\,\vcenter{\hbox{\includegraphics[width=0.14\paperwidth]{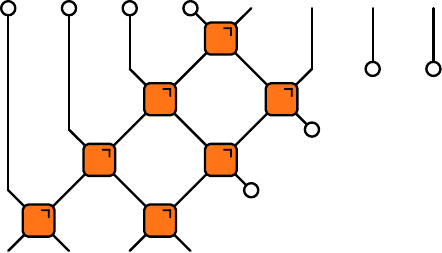}}}\,\,,
\end{align}
with left eigenvectors of the form $\bra{\sigma\otimes\mathbbm{1}\otimes\mathbbm{1}\otimes\mathbbm{1}}$.

To determine the ELT, we need the leading eigenvalue of the transfer matrix $\mathcal{M}_1^{(\alpha)}$ and the quantity $B_2^{(\alpha)}$. The transfer matrix $\mathcal{M}_1^{(\alpha)}$ can be simplified as
\begin{align}
\mathcal{M}_1^{(\alpha)} = \,\,\vcenter{\hbox{\includegraphics[height=0.2\paperwidth]{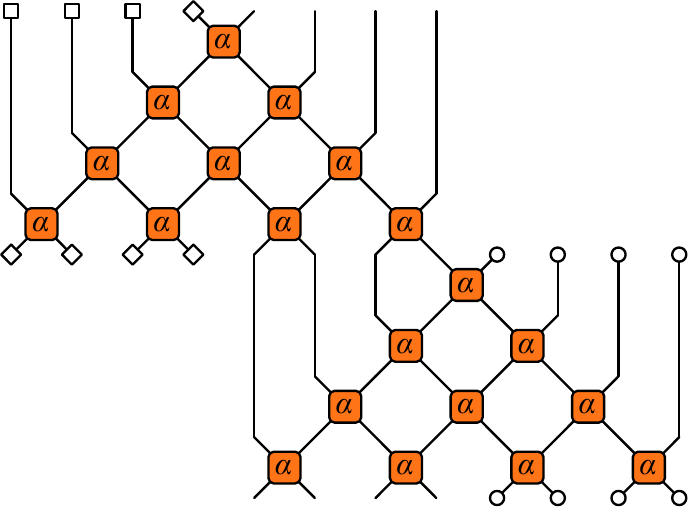}}}\,\,=\,\,\vcenter{\hbox{\includegraphics[height=0.2\paperwidth]{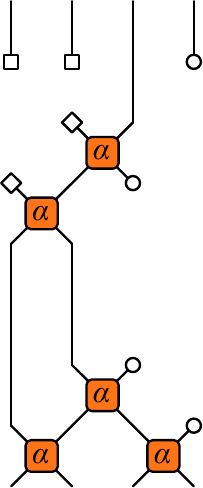}}}\,\,.
\end{align}
The only left eigenvector with non-zero eigenvalue is
$\bra{\Box\otimes\Box\otimes\Box\otimes\medcirc} = \,\,\vcenter{\hbox{\includegraphics[height=0.035\columnwidth]{figs/square.pdf}}}\,\,\vcenter{\hbox{\includegraphics[height=0.035\columnwidth]{figs/square.pdf}}}\,\,\vcenter{\hbox{\includegraphics[height=0.035\columnwidth]{figs/square.pdf}}}\,\,\vcenter{\hbox{\includegraphics[height=0.035\columnwidth]{figs/circle.pdf}}}\,$.
Application of the transfer matrix yields the eigenvalue $\lambda_1^{(\alpha)}=\left(1/d\right)^{4(\alpha-1)}=\left(1/q\right)^{\alpha-1}$
\begin{align}
    \bra{\Box\otimes\Box\otimes\Box\otimes\medcirc}\mathcal{M}_1^{(\alpha)}= \,\,\vcenter{\hbox{\includegraphics[height=0.2\paperwidth]{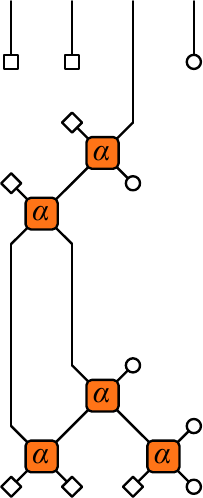}}}\quad= \quad\vcenter{\hbox{\includegraphics[height=0.2\paperwidth]{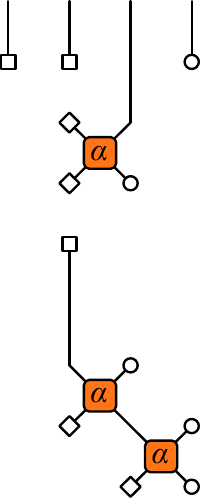}}}\,\,=\left(\frac{1}{d}\right)^{4(\alpha-1)}\bra{\Box\otimes\Box\otimes\Box\otimes\medcirc}.
\end{align}
Using Eq.~\eqref{eq:ve}, the entanglement velocity follows as $v_E=1/2$ independent of the R\'{e}nyi index. $B_2^{(\alpha)}$ is given by
\begin{align}
B_2^{(\alpha)} = q^3\left(\,\,\vcenter{\hbox{\includegraphics[height=0.25\columnwidth]{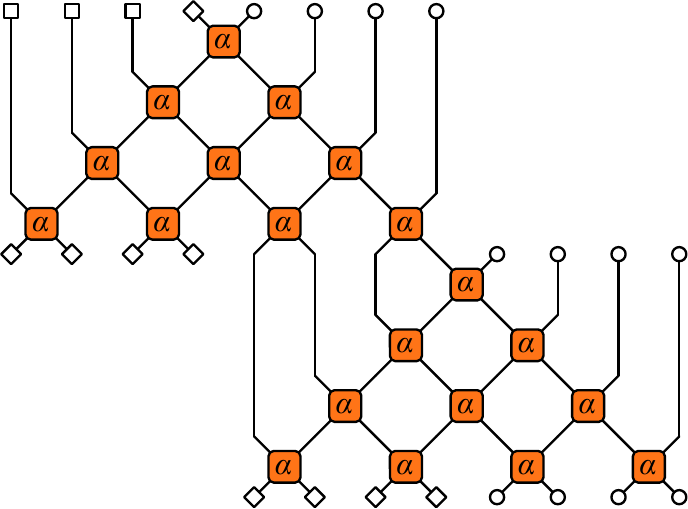}}}\,\,\right)^{\frac{1}{\alpha-1}}=d^{12}\left(\,\,\vcenter{\hbox{\includegraphics[height=0.25\columnwidth]{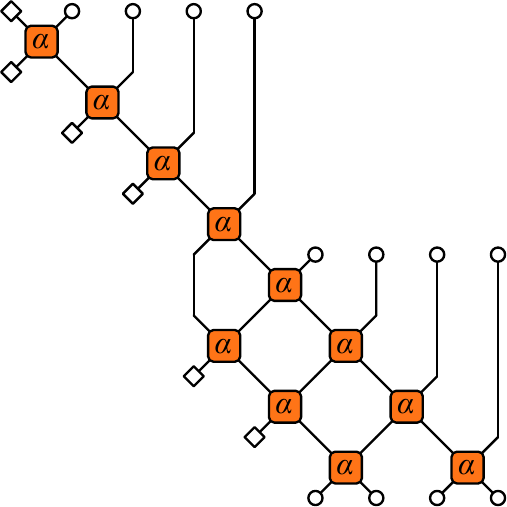}}}\,\,\right)^{\frac{1}{\alpha-1}}= d^{12}\left(\frac{1}{d}\right)^{6},
\end{align}
and is also independent of the R\'{e}nyi index. From this, we find an ELT in agreement with the result of Ref.~\cite{Rampp2025a} as
\begin{align}
    \mathcal{E}_\alpha(v) = \begin{cases}
        \frac{1}{2}, & \abs{v}\leq\frac{1}{3},\\
        \frac{1+3\abs{v}}{4}, & \frac{1}{3}<\abs{v}\leq 1.
    \end{cases} \label{eq:4pyr_elt}
\end{align}

\section{FDU3 Clifford gates in small local dimensions}
\label{app:clifford}

In this appendix we report some numerical results about the existence of FDU3 gates belonging to an important discrete subgroup of the unitary group $\mathbb{U}(q^2)$, namely, the Clifford group. While search algorithms applied to the full unitary group struggle to identify FDU3 gates, it turns out progress can be made by restricting to the Clifford group. We separately discuss the cases of gates acting on local Hilbert spaces with prime and nonprime dimensions. A similar distinction was made in Ref.~\cite{Rampp2024} for the construction of DU2 gates.

\subsection{No-go for FDU3 Clifford gates in prime dimensions} 

We first establish the absence of FDU3 gates for qubit gates.
Two-qubit Clifford gates map the two-qubit Pauli group, $\mathcal{P}_2:=\left\lbrace I,X,Y,Z\right\rbrace^{\otimes 2}$, to itself under conjugation, where $X,Y,$ and $Z$ are the $2 \times 2$ Pauli matrices \cite{Gottesman1997stabilizer}. There are 11520 two-qubit Clifford gates and numerically we found that the number of gates that satisfy one of the DU2 conditions given in Eq.~(\ref{eq:DU2}) is $11520-576=10944$. We denote the difference between the left hand side and right hand side of one of the DU2 conditions in Eq.~\eqref{eq:DU2} as $\Delta(\text{DU2})$. The number of gates with $\Delta(\text{DU2})=0$ (up to machine precision) is $10944$ and among these gates the number of gates that satisfy both the DU2 conditions is $8640$. Interestingly, the remaining $11520-10944=576$ gates that do not satisfy the DU2 conditions have $\Delta(\text{DU2})=4\neq 0$ and satisfy the DU3 conditions given in Eq.~\eqref{eq:DU3}. Although there are $24^2=576$ two-qubit Clifford gates that satisfy the DU3 condition, no two-qubit Clifford gate satisfies the complementary DU3 property, i.e. the DU$^*$3 property. As FDU3 gates require both DU3 and DU$^*$3 conditions to hold, there are no two-qubit Clifford gates that are FDU3. 

We next consider qutrit gates. For the two-qutrit case, the number of Clifford gates is prohibitively large,  $\mathcal{O}(10^6)$, preventing an exhaustive search. We instead sample $10^5$ gates uniformly from the two-qutrit Clifford group~\cite{Koenig2014efficiently}. The results are similar to the two-qubit case: we find two-qutrit Clifford gates that satisfy the DU3 condition, but these do not satisfy the DU$^*$3 condition. These results are illustrated in Table~\ref{tab:FDU3_cliff}. This suggests that Clifford FDU3 gates may not exist in prime dimensions, and we leave a thorough examination of these observations for future work. 

\begin{table}
\centering
\begin{tabular}{c|c|c|c|c}
\hline
\hline
   Local dimension $q$ & DU3 & DU$^*$3 & FDU3 & Search type \\
   \hline
                    2   &    $\frac{576}{11,520} \approx 0.05\%$      &  0  &  0 &  Exhaustive \\
                    
                    \hline
                    3    &    $\frac{2250}{10^5} \approx 0.02\%$      &  0  &  0 &  Uniform sampling  \\
                    
                    \hline
                    4    &    $\frac{572}{10^4} \approx 0.05\%$      &  $\frac{229}{10^4} \approx 0.02\%$  &   $\frac{60}{10^4} \approx 0.01\%$ &  Uniform sampling \\
                    
\hline  
\hline
\end{tabular}
\caption{A search for DU3, DU$^*$3, and FDU3 Clifford gates in small local dimensions $d\leq 4$. An exhaustive search is performed only for $q=2$; for $q=3$ and $4$ a uniform sampling is performed over the corresponding Clifford groups, taking into consideration $10^5$ and $10^4$ realizations, respectively.}
    \label{tab:FDU3_cliff}
\end{table}

\subsection{FDU3 Clifford gates in nonprime dimensions} 
For prime-power dimensions such as $q=4$, the Clifford group can be defined in different ways. For the two-ququart case ($\mathbb{C}^4 \otimes \mathbb{C}^4$) using the isomorphism $\mathbb{C}^4 \otimes \mathbb{C}^4 \equiv \mathbb{C}^2 \otimes \mathbb{C}^2 \otimes \mathbb{C}^2 \otimes \mathbb{C}^2$, we sample the Clifford gates from the four-qubit Clifford group. Unlike the smaller dimensions discussed above, for $q=4$ we are able to obtain (with a small probability) Clifford gates that satisfy DU3, DU$^*$3, and hence the FDU3 conditions. 

\begin{table}[ht]
\centering
\begin{minipage}{0.48\textwidth}
\centering
(a)
\begin{tabular}{cc||cc}
\hline
\hline
X-type generator & $C_1^\dagger X C_1$ &  Z-type generator & $C_1^\dagger Z C_1$  \\ \hline
$XIII$ & $-IIIX$ & $ZIII$ & $-ZZYZ$  \\
$IXII$ & $-ZZII$ & $IZII$ & $-IXIX$  \\
$IIXI$ & $-ZIXX$ & $IIZI$ & $+IIYI$  \\
$IIIX$ & $-YXYI$ & $IIIZ$ & $-ZIII$  \\ \hline \hline
\end{tabular}
\end{minipage}
\begin{minipage}{0.48\textwidth}
\centering
(b)
\begin{tabular}{cc||cc}
    \hline
\hline
X-type generator & $C_2^\dagger X C_2$ &  Z-type generator & $C_2^\dagger Z C_2$  \\ \hline
$XIII$ & $-YZZI$ & $ZIII$ & $-IIXY$ \\
$IXII$ & $-IYXI$ & $IZII$ & $YZIY$  \\
$IIXI$ & $YXYZ$  & $IIZI$ & $-IXYX$ \\
$IIIX$ & $-ZZYZ$ & $IIIZ$ & $XZYZ$  \\ \hline\hline
\end{tabular}
\end{minipage}
\caption{(a) An example of FDU3 Clifford gate in local dimension $q=4$. It is obtained from a search over the four-qubit Clifford group, and is shown via its action on the generators of the four-qubit Pauli group. We have used a shorthand notation for the tensor product of Pauli matrices, for example, $XIII \equiv X\otimes I\otimes I \otimes I$. (b) The four-qubit Clifford unitary $C_2$ satisfying FDU3 conditions with Schmidt rank 8 is shown via its action on the generators of the four-qubit Pauli group.
\label{table:clifford_FDU3}}
\hfill
\end{table}

A first example of a FDU3 gate is shown in Table \ref{table:clifford_FDU3}~(a) via its action on the stabilizer generators of the four-qubit Pauli group. 
The corresponding FDU3 gate has Schmidt rank of 4, with each Schmidt value equal to 2. The matrix representation of this gate in the computational basis is given by
\begin{equation}
C_1=\frac{1}{2}
\begin{pmatrix}
. & . & . & -1 & . & . & . & 1 & . & i & . & . & . & -i & . & . \\
. & -i & . & . & . & i & . & . & . & . & . & 1 & . & . & . & -1 \\
. & . & . & i & . & . & . & -i & . & -1 & . & . & . & 1 & . & . \\
. & 1 & . & . & . & -1 & . & . & . & . & . & -i & . & . & . & i \\
. & i & . & . & . & i & . & . & . & . & . & -1 & . & . & . & -1 \\
. & . & . & 1 & . & . & . & 1 & . & -i & . & . & . & -i & . & . \\
. & -1 & . & . & . & -1 & . & . & . & . & . & i & . & . & . & i \\
. & . & . & -i & . & . & . & -i & . & 1 & . & . & . & 1 & . & . \\
1 & . & . & . & 1 & . & . & . & . & . & -i & . & . & . & -i & . \\
. & . & i & . & . & . & i & . & -1 & . & . & . & -1 & . & . & . \\
i & . & . & . & i & . & . & . & . & . & -1 & . & . & . & -1 & . \\
. & . & 1 & . & . & . & 1 & . & -i & . & . & . & -i & . & . & . \\
. & . & -i & . & . & . & i & . & 1 & . & . & . & -1 & . & . & . \\
-1 & . & . & . & 1 & . & . & . & . & . & i & . & . & . & -i & . \\
. & . & -1 & . & . & . & 1 & . & i & . & . & . & -i & . & . & . \\
-i & . & . & . & i & . & . & . & . & . & 1 & . & . & . & -1 & .
\end{pmatrix},
\label{eq:FDU3_cliff_rank_4}
\end{equation}
where the non-zero entries are determined by the fourth roots of unity, $\left\lbrace \pm 1,\pm i \right\rbrace$, and `` $.$ " represents 0. The numerical search for FDU3 gates was dominated by FDU3 gates with Schmidt rank 4; however, we were also able to obtain FDU3 gates with Schmidt rank 8. An example of an FDU3 Clifford gate with Schmidt rank 8 (with each Schmidt value equal to $\sqrt{2}$) is given by
\begin{equation}
C_2 = \frac{1}{2\sqrt{2}} \begin{pmatrix}
. & -1 & . & i & 1 & . & i & . & i & . & 1 & . & . & i & . & -1 \\
1 & . & -i & . & . & -1 & . & -i & . & i & . & 1 & i & . & -1 & . \\
i & . & 1 & . & . & -i & . & 1 & . & 1 & . & -i & 1 & . & i & . \\
. & i & . & 1 & -i & . & 1 & . & -1 & . & i & . & . & -1 & . & -i \\
. & -1 & . & -i & 1 & . & -i & . & -i & . & 1 & . & . & -i & . & -1 \\
-1 & . & -i & . & . & 1 & . & -i & . & i & . & -1 & i & . & 1 & . \\
-i & . & 1 & . & . & i & . & 1 & . & 1 & . & i & 1 & . & -i & . \\
. & i & . & -1 & -i & . & -1 & . & 1 & . & i & . & . & 1 & . & -i \\
1 & . & -i & . & . & 1 & . & i & . & i & . & 1 & -i & . & 1 & . \\
. & 1 & . & -i & 1 & . & i & . & -i & . & -1 & . & . & i & . & -1 \\
. & i & . & 1 & i & . & -1 & . & -1 & . & i & . & . & 1 & . & i \\
-i & . & -1 & . & . & -i & . & 1 & . & -1 & . & i & 1 & . & i & . \\
1 & . & i & . & . & 1 & . & -i & . & -i & . & 1 & i & . & 1 & . \\
. & -1 & . & -i & -1 & . & i & . & -i & . & 1 & . & . & i & . & 1 \\
. & -i & . & 1 & -i & . & -1 & . & -1 & . & -i & . & . & 1 & . & -i \\
-i & . & 1 & . & . & -i & . & -1 & . & 1 & . & i & -1 & . & i & .
\end{pmatrix},
\label{eq:FDU3_cliff_rank_8}
\end{equation}
where again the non-zero entries are determined by the fourth roots of unity and ``$.$ " represents vanishing matrix entries. The action of the Clifford unitary $C_2$ on the generators of the four-qubit Pauli group is shown in Table \ref{table:clifford_FDU3}~(b). It is worth emphasizing that the analytical constructions of FDU3 gates discussed in the main text have a Schmidt rank 4 only; therefore, the above example lies beyond those constructions. 

\end{widetext}

\bibliography{solvability}

\begin{thebibliography}{44}%
\makeatletter
\providecommand \@ifxundefined [1]{%
 \@ifx{#1\undefined}
}%
\providecommand \@ifnum [1]{%
 \ifnum #1\expandafter \@firstoftwo
 \else \expandafter \@secondoftwo
 \fi
}%
\providecommand \@ifx [1]{%
 \ifx #1\expandafter \@firstoftwo
 \else \expandafter \@secondoftwo
 \fi
}%
\providecommand \natexlab [1]{#1}%
\providecommand \enquote  [1]{``#1''}%
\providecommand \bibnamefont  [1]{#1}%
\providecommand \bibfnamefont [1]{#1}%
\providecommand \citenamefont [1]{#1}%
\providecommand \href@noop [0]{\@secondoftwo}%
\providecommand \href [0]{\begingroup \@sanitize@url \@href}%
\providecommand \@href[1]{\@@startlink{#1}\@@href}%
\providecommand \@@href[1]{\endgroup#1\@@endlink}%
\providecommand \@sanitize@url [0]{\catcode `\\12\catcode `\$12\catcode
  `\&12\catcode `\#12\catcode `\^12\catcode `\_12\catcode `\%12\relax}%
\providecommand \@@startlink[1]{}%
\providecommand \@@endlink[0]{}%
\providecommand \url  [0]{\begingroup\@sanitize@url \@url }%
\providecommand \@url [1]{\endgroup\@href {#1}{\urlprefix }}%
\providecommand \urlprefix  [0]{URL }%
\providecommand \Eprint [0]{\href }%
\providecommand \doibase [0]{https://doi.org/}%
\providecommand \selectlanguage [0]{\@gobble}%
\providecommand \bibinfo  [0]{\@secondoftwo}%
\providecommand \bibfield  [0]{\@secondoftwo}%
\providecommand \translation [1]{[#1]}%
\providecommand \BibitemOpen [0]{}%
\providecommand \bibitemStop [0]{}%
\providecommand \bibitemNoStop [0]{.\EOS\space}%
\providecommand \EOS [0]{\spacefactor3000\relax}%
\providecommand \BibitemShut  [1]{\csname bibitem#1\endcsname}%
\let\auto@bib@innerbib\@empty
\bibitem [{\citenamefont {Akila}\ \emph {et~al.}(2016)\citenamefont {Akila},
  \citenamefont {Waltner}, \citenamefont {Gutkin},\ and\ \citenamefont
  {Guhr}}]{Akila2016}%
  \BibitemOpen
  \bibfield  {author} {\bibinfo {author} {\bibfnamefont {M.}~\bibnamefont
  {Akila}}, \bibinfo {author} {\bibfnamefont {D.}~\bibnamefont {Waltner}},
  \bibinfo {author} {\bibfnamefont {B.}~\bibnamefont {Gutkin}},\ and\ \bibinfo
  {author} {\bibfnamefont {T.}~\bibnamefont {Guhr}},\ }\bibfield  {title}
  {\bibinfo {title} {Particle-time duality in the kicked {I}sing spin chain},\
  }\href {https://doi.org/10.1088/1751-8113/49/37/375101} {\bibfield  {journal}
  {\bibinfo  {journal} {J. Phys. Math. Theor.}\ }\textbf {\bibinfo {volume}
  {49}},\ \bibinfo {pages} {375101} (\bibinfo {year} {2016})}\BibitemShut
  {NoStop}%
\bibitem [{\citenamefont {Bertini}\ \emph {et~al.}(2018)\citenamefont
  {Bertini}, \citenamefont {Kos},\ and\ \citenamefont {Prosen}}]{Bertini2018}%
  \BibitemOpen
  \bibfield  {author} {\bibinfo {author} {\bibfnamefont {B.}~\bibnamefont
  {Bertini}}, \bibinfo {author} {\bibfnamefont {P.}~\bibnamefont {Kos}},\ and\
  \bibinfo {author} {\bibfnamefont {T.}~\bibnamefont {Prosen}},\ }\bibfield
  {title} {\bibinfo {title} {Exact spectral form factor in a minimal model of
  many-body quantum chaos},\ }\href
  {https://doi.org/10.1103/physrevlett.121.264101} {\bibfield  {journal}
  {\bibinfo  {journal} {Phys. Rev. Lett.}\ }\textbf {\bibinfo {volume} {121}},\
  \bibinfo {pages} {264101} (\bibinfo {year} {2018})}\BibitemShut {NoStop}%
\bibitem [{\citenamefont {Gopalakrishnan}\ and\ \citenamefont
  {Lamacraft}(2019)}]{Gopalakrishnan2019}%
  \BibitemOpen
  \bibfield  {author} {\bibinfo {author} {\bibfnamefont {S.}~\bibnamefont
  {Gopalakrishnan}}\ and\ \bibinfo {author} {\bibfnamefont {A.}~\bibnamefont
  {Lamacraft}},\ }\bibfield  {title} {\bibinfo {title} {Unitary circuits of
  finite depth and infinite width from quantum channels},\ }\href
  {https://doi.org/10.1103/physrevb.100.064309} {\bibfield  {journal} {\bibinfo
   {journal} {Phys. Rev. B}\ }\textbf {\bibinfo {volume} {100}},\ \bibinfo
  {pages} {064309} (\bibinfo {year} {2019})}\BibitemShut {NoStop}%
\bibitem [{\citenamefont {Bertini}\ \emph
  {et~al.}(2019{\natexlab{a}})\citenamefont {Bertini}, \citenamefont {Kos},\
  and\ \citenamefont {Prosen}}]{Bertini2019}%
  \BibitemOpen
  \bibfield  {author} {\bibinfo {author} {\bibfnamefont {B.}~\bibnamefont
  {Bertini}}, \bibinfo {author} {\bibfnamefont {P.}~\bibnamefont {Kos}},\ and\
  \bibinfo {author} {\bibfnamefont {T.}~\bibnamefont {Prosen}},\ }\bibfield
  {title} {\bibinfo {title} {Exact correlation functions for dual-unitary
  lattice models in $1+1$ dimensions},\ }\href
  {https://doi.org/10.1103/physrevlett.123.210601} {\bibfield  {journal}
  {\bibinfo  {journal} {Phys. Rev. Lett.}\ }\textbf {\bibinfo {volume} {123}},\
  \bibinfo {pages} {210601} (\bibinfo {year} {2019}{\natexlab{a}})}\BibitemShut
  {NoStop}%
\bibitem [{\citenamefont {Bertini}\ \emph
  {et~al.}(2019{\natexlab{b}})\citenamefont {Bertini}, \citenamefont {Kos},\
  and\ \citenamefont {Prosen}}]{Bertini2019a}%
  \BibitemOpen
  \bibfield  {author} {\bibinfo {author} {\bibfnamefont {B.}~\bibnamefont
  {Bertini}}, \bibinfo {author} {\bibfnamefont {P.}~\bibnamefont {Kos}},\ and\
  \bibinfo {author} {\bibfnamefont {T.}~\bibnamefont {Prosen}},\ }\bibfield
  {title} {\bibinfo {title} {Entanglement spreading in a minimal model of
  maximal many-body quantum chaos},\ }\href
  {https://journals.aps.org/prx/abstract/10.1103/PhysRevX.9.021033} {\bibfield
  {journal} {\bibinfo  {journal} {Phys. Rev. X}\ }\textbf {\bibinfo {volume}
  {9}},\ \bibinfo {pages} {021033} (\bibinfo {year}
  {2019}{\natexlab{b}})}\BibitemShut {NoStop}%
\bibitem [{\citenamefont {Claeys}\ and\ \citenamefont
  {Lamacraft}(2020)}]{Claeys2020}%
  \BibitemOpen
  \bibfield  {author} {\bibinfo {author} {\bibfnamefont {P.~W.}\ \bibnamefont
  {Claeys}}\ and\ \bibinfo {author} {\bibfnamefont {A.}~\bibnamefont
  {Lamacraft}},\ }\bibfield  {title} {\bibinfo {title} {Maximum velocity
  quantum circuits},\ }\href {https://doi.org/10.1103/physrevresearch.2.033032}
  {\bibfield  {journal} {\bibinfo  {journal} {Phys. Rev. Research}\ }\textbf
  {\bibinfo {volume} {2}},\ \bibinfo {pages} {033032} (\bibinfo {year}
  {2020})}\BibitemShut {NoStop}%
\bibitem [{\citenamefont {Claeys}\ and\ \citenamefont
  {Lamacraft}(2021)}]{Claeys2021}%
  \BibitemOpen
  \bibfield  {author} {\bibinfo {author} {\bibfnamefont {P.~W.}\ \bibnamefont
  {Claeys}}\ and\ \bibinfo {author} {\bibfnamefont {A.}~\bibnamefont
  {Lamacraft}},\ }\bibfield  {title} {\bibinfo {title} {Ergodic and nonergodic
  dual-unitary quantum circuits with arbitrary local {H}ilbert space
  dimension},\ }\href {https://doi.org/10.1103/physrevlett.126.100603}
  {\bibfield  {journal} {\bibinfo  {journal} {Phys. Rev. Lett.}\ }\textbf
  {\bibinfo {volume} {126}},\ \bibinfo {pages} {100603} (\bibinfo {year}
  {2021})}\BibitemShut {NoStop}%
\bibitem [{\citenamefont {Bertini}\ \emph {et~al.}(2026)\citenamefont
  {Bertini}, \citenamefont {Claeys},\ and\ \citenamefont
  {Prosen}}]{Bertini2026}%
  \BibitemOpen
  \bibfield  {author} {\bibinfo {author} {\bibfnamefont {B.}~\bibnamefont
  {Bertini}}, \bibinfo {author} {\bibfnamefont {P.~W.}\ \bibnamefont
  {Claeys}},\ and\ \bibinfo {author} {\bibfnamefont {T.}~\bibnamefont
  {Prosen}},\ }\bibfield  {title} {\bibinfo {title} {Exactly solvable quantum
  many-body dynamics from space-time duality},\ }\href
  {https://doi.org/10.1103/yx73-dk86} {\bibfield  {journal} {\bibinfo
  {journal} {Rev. Mod. Phys.}\ }\textbf {\bibinfo {volume} {98}},\ \bibinfo
  {pages} {025001} (\bibinfo {year} {2026})}\BibitemShut {NoStop}%
\bibitem [{\citenamefont {Jonay}\ \emph {et~al.}(2021)\citenamefont {Jonay},
  \citenamefont {Khemani},\ and\ \citenamefont {Ippoliti}}]{Jonay2021}%
  \BibitemOpen
  \bibfield  {author} {\bibinfo {author} {\bibfnamefont {C.}~\bibnamefont
  {Jonay}}, \bibinfo {author} {\bibfnamefont {V.}~\bibnamefont {Khemani}},\
  and\ \bibinfo {author} {\bibfnamefont {M.}~\bibnamefont {Ippoliti}},\
  }\bibfield  {title} {\bibinfo {title} {Triunitary quantum circuits},\ }\href
  {https://doi.org/10.1103/physrevresearch.3.043046} {\bibfield  {journal}
  {\bibinfo  {journal} {Phys. Rev. Research}\ }\textbf {\bibinfo {volume}
  {3}},\ \bibinfo {pages} {043046} (\bibinfo {year} {2021})}\BibitemShut
  {NoStop}%
\bibitem [{\citenamefont {Milbradt}\ \emph {et~al.}(2023)\citenamefont
  {Milbradt}, \citenamefont {Scheller}, \citenamefont {Aßmus},\ and\
  \citenamefont {Mendl}}]{Milbradt2023}%
  \BibitemOpen
  \bibfield  {author} {\bibinfo {author} {\bibfnamefont {R.~M.}\ \bibnamefont
  {Milbradt}}, \bibinfo {author} {\bibfnamefont {L.}~\bibnamefont {Scheller}},
  \bibinfo {author} {\bibfnamefont {C.}~\bibnamefont {Aßmus}},\ and\ \bibinfo
  {author} {\bibfnamefont {C.~B.}\ \bibnamefont {Mendl}},\ }\bibfield  {title}
  {\bibinfo {title} {Ternary unitary quantum lattice models and circuits in 2+1
  dimensions},\ }\href {https://doi.org/10.1103/physrevlett.130.090601}
  {\bibfield  {journal} {\bibinfo  {journal} {Phys. Rev. Lett.}\ }\textbf
  {\bibinfo {volume} {130}},\ \bibinfo {pages} {090601} (\bibinfo {year}
  {2023})}\BibitemShut {NoStop}%
\bibitem [{\citenamefont {Sommers}\ \emph {et~al.}(2023)\citenamefont
  {Sommers}, \citenamefont {Huse},\ and\ \citenamefont
  {Gullans}}]{Sommers2023}%
  \BibitemOpen
  \bibfield  {author} {\bibinfo {author} {\bibfnamefont {G.~M.}\ \bibnamefont
  {Sommers}}, \bibinfo {author} {\bibfnamefont {D.~A.}\ \bibnamefont {Huse}},\
  and\ \bibinfo {author} {\bibfnamefont {M.~J.}\ \bibnamefont {Gullans}},\
  }\bibfield  {title} {\bibinfo {title} {Crystalline quantum circuits},\ }\href
  {https://doi.org/10.1103/prxquantum.4.030313} {\bibfield  {journal} {\bibinfo
   {journal} {PRX Quantum}\ }\textbf {\bibinfo {volume} {4}},\ \bibinfo {pages}
  {030313} (\bibinfo {year} {2023})}\BibitemShut {NoStop}%
\bibitem [{\citenamefont {Mestyán}\ \emph {et~al.}(2024)\citenamefont
  {Mestyán}, \citenamefont {Pozsgay},\ and\ \citenamefont
  {Wanless}}]{Mestyan2024}%
  \BibitemOpen
  \bibfield  {author} {\bibinfo {author} {\bibfnamefont {M.}~\bibnamefont
  {Mestyán}}, \bibinfo {author} {\bibfnamefont {B.}~\bibnamefont {Pozsgay}},\
  and\ \bibinfo {author} {\bibfnamefont {I.~M.}\ \bibnamefont {Wanless}},\
  }\bibfield  {title} {\bibinfo {title} {{Multi-directional unitarity and
  maximal entanglement in spatially symmetric quantum states}},\ }\href
  {https://doi.org/10.21468/SciPostPhys.16.1.010} {\bibfield  {journal}
  {\bibinfo  {journal} {SciPost Phys.}\ }\textbf {\bibinfo {volume} {16}},\
  \bibinfo {pages} {010} (\bibinfo {year} {2024})}\BibitemShut {NoStop}%
\bibitem [{\citenamefont {Yu}\ \emph {et~al.}(2024)\citenamefont {Yu},
  \citenamefont {Wang},\ and\ \citenamefont {Kos}}]{Yu2024}%
  \BibitemOpen
  \bibfield  {author} {\bibinfo {author} {\bibfnamefont {X.-H.}\ \bibnamefont
  {Yu}}, \bibinfo {author} {\bibfnamefont {Z.}~\bibnamefont {Wang}},\ and\
  \bibinfo {author} {\bibfnamefont {P.}~\bibnamefont {Kos}},\ }\bibfield
  {title} {\bibinfo {title} {Hierarchical generalization of dual unitarity},\
  }\href {https://doi.org/10.22331/q-2024-02-20-1260} {\bibfield  {journal}
  {\bibinfo  {journal} {Quantum}\ }\textbf {\bibinfo {volume} {8}},\ \bibinfo
  {pages} {1260} (\bibinfo {year} {2024})}\BibitemShut {NoStop}%
\bibitem [{\citenamefont {Rampp}\ \emph
  {et~al.}(2025{\natexlab{a}})\citenamefont {Rampp}, \citenamefont {Rather},\
  and\ \citenamefont {Claeys}}]{Rampp2025}%
  \BibitemOpen
  \bibfield  {author} {\bibinfo {author} {\bibfnamefont {M.~A.}\ \bibnamefont
  {Rampp}}, \bibinfo {author} {\bibfnamefont {S.~A.}\ \bibnamefont {Rather}},\
  and\ \bibinfo {author} {\bibfnamefont {P.~W.}\ \bibnamefont {Claeys}},\
  }\bibfield  {title} {\bibinfo {title} {{Geometric constructions of
  generalized dual-unitary circuits from biunitarity}},\ }\href
  {https://doi.org/10.21468/SciPostPhys.18.6.182} {\bibfield  {journal}
  {\bibinfo  {journal} {SciPost Phys.}\ }\textbf {\bibinfo {volume} {18}},\
  \bibinfo {pages} {182} (\bibinfo {year} {2025}{\natexlab{a}})}\BibitemShut
  {NoStop}%
\bibitem [{\citenamefont {Breach}\ \emph {et~al.}(2025)\citenamefont {Breach},
  \citenamefont {Placke}, \citenamefont {Claeys},\ and\ \citenamefont
  {Parameswaran}}]{Breach2025}%
  \BibitemOpen
  \bibfield  {author} {\bibinfo {author} {\bibfnamefont {O.}~\bibnamefont
  {Breach}}, \bibinfo {author} {\bibfnamefont {B.}~\bibnamefont {Placke}},
  \bibinfo {author} {\bibfnamefont {P.~W.}\ \bibnamefont {Claeys}},\ and\
  \bibinfo {author} {\bibfnamefont {S.}~\bibnamefont {Parameswaran}},\
  }\bibfield  {title} {\bibinfo {title} {Solvable quantum circuits in
  $\mathrm{Tree}+1$ dimensions},\ }\href {https://doi.org/10.1103/fbgh-rq3l}
  {\bibfield  {journal} {\bibinfo  {journal} {PRX Quantum}\ }\textbf {\bibinfo
  {volume} {6}},\ \bibinfo {pages} {040316} (\bibinfo {year}
  {2025})}\BibitemShut {NoStop}%
\bibitem [{\citenamefont {Rampp}\ \emph
  {et~al.}(2025{\natexlab{b}})\citenamefont {Rampp}, \citenamefont {Rather},\
  and\ \citenamefont {Claeys}}]{Rampp2025a}%
  \BibitemOpen
  \bibfield  {author} {\bibinfo {author} {\bibfnamefont {M.~A.}\ \bibnamefont
  {Rampp}}, \bibinfo {author} {\bibfnamefont {S.~A.}\ \bibnamefont {Rather}},\
  and\ \bibinfo {author} {\bibfnamefont {P.~W.}\ \bibnamefont {Claeys}},\
  }\bibfield  {title} {\bibinfo {title} {Solvable quantum circuits from
  spacetime lattices},\ }\href {https://doi.org/10.48550/arXiv.2512.15871}
  {\bibfield  {journal} {\bibinfo  {journal} {arXiv:2512.15871}\ } (\bibinfo
  {year} {2025}{\natexlab{b}})}\BibitemShut {NoStop}%
\bibitem [{\citenamefont {Pickering}\ and\ \citenamefont
  {Bertini}(2026)}]{Pickering2026}%
  \BibitemOpen
  \bibfield  {author} {\bibinfo {author} {\bibfnamefont {S.~H.}\ \bibnamefont
  {Pickering}}\ and\ \bibinfo {author} {\bibfnamefont {B.}~\bibnamefont
  {Bertini}},\ }\bibfield  {title} {\bibinfo {title} {Asymptotically solvable
  quantum circuits},\ }\href {https://doi.org/10.48550/ARXIV.2602.24276}
  {\bibfield  {journal} {\bibinfo  {journal} {arXiv:2602.24276}\ } (\bibinfo
  {year} {2026})}\BibitemShut {NoStop}%
\bibitem [{\citenamefont {Rampp}\ \emph {et~al.}(2024)\citenamefont {Rampp},
  \citenamefont {Rather},\ and\ \citenamefont {Claeys}}]{Rampp2024}%
  \BibitemOpen
  \bibfield  {author} {\bibinfo {author} {\bibfnamefont {M.~A.}\ \bibnamefont
  {Rampp}}, \bibinfo {author} {\bibfnamefont {S.~A.}\ \bibnamefont {Rather}},\
  and\ \bibinfo {author} {\bibfnamefont {P.~W.}\ \bibnamefont {Claeys}},\
  }\bibfield  {title} {\bibinfo {title} {Entanglement membrane in exactly
  solvable lattice models},\ }\href
  {https://doi.org/10.1103/physrevresearch.6.033271} {\bibfield  {journal}
  {\bibinfo  {journal} {Phys. Rev. Research}\ }\textbf {\bibinfo {volume}
  {6}},\ \bibinfo {pages} {033271} (\bibinfo {year} {2024})}\BibitemShut
  {NoStop}%
\bibitem [{\citenamefont {Foligno}\ \emph {et~al.}(2024)\citenamefont
  {Foligno}, \citenamefont {Kos},\ and\ \citenamefont {Bertini}}]{Foligno2024}%
  \BibitemOpen
  \bibfield  {author} {\bibinfo {author} {\bibfnamefont {A.}~\bibnamefont
  {Foligno}}, \bibinfo {author} {\bibfnamefont {P.}~\bibnamefont {Kos}},\ and\
  \bibinfo {author} {\bibfnamefont {B.}~\bibnamefont {Bertini}},\ }\bibfield
  {title} {\bibinfo {title} {Quantum information spreading in generalized
  dual-unitary circuits},\ }\href
  {https://doi.org/10.1103/physrevlett.132.250402} {\bibfield  {journal}
  {\bibinfo  {journal} {Phys. Rev. Lett.}\ }\textbf {\bibinfo {volume} {132}},\
  \bibinfo {pages} {250402} (\bibinfo {year} {2024})}\BibitemShut {NoStop}%
\bibitem [{\citenamefont {Sommers}\ \emph {et~al.}(2024)\citenamefont
  {Sommers}, \citenamefont {Gopalakrishnan}, \citenamefont {Gullans},\ and\
  \citenamefont {Huse}}]{Sommers2024}%
  \BibitemOpen
  \bibfield  {author} {\bibinfo {author} {\bibfnamefont {G.~M.}\ \bibnamefont
  {Sommers}}, \bibinfo {author} {\bibfnamefont {S.}~\bibnamefont
  {Gopalakrishnan}}, \bibinfo {author} {\bibfnamefont {M.~J.}\ \bibnamefont
  {Gullans}},\ and\ \bibinfo {author} {\bibfnamefont {D.~A.}\ \bibnamefont
  {Huse}},\ }\bibfield  {title} {\bibinfo {title} {Zero-temperature
  entanglement membranes in quantum circuits},\ }\href
  {https://doi.org/10.1103/physrevb.110.064311} {\bibfield  {journal} {\bibinfo
   {journal} {Phys. Rev. B}\ }\textbf {\bibinfo {volume} {110}},\ \bibinfo
  {pages} {064311} (\bibinfo {year} {2024})}\BibitemShut {NoStop}%
\bibitem [{\citenamefont {Liu}\ and\ \citenamefont {Ho}(2025)}]{Liu2025}%
  \BibitemOpen
  \bibfield  {author} {\bibinfo {author} {\bibfnamefont {C.}~\bibnamefont
  {Liu}}\ and\ \bibinfo {author} {\bibfnamefont {W.~W.}\ \bibnamefont {Ho}},\
  }\bibfield  {title} {\bibinfo {title} {Solvable entanglement dynamics in
  quantum circuits with generalized space-time duality},\ }\href
  {https://doi.org/10.1103/physrevresearch.7.l012011} {\bibfield  {journal}
  {\bibinfo  {journal} {Phys. Rev. Research}\ }\textbf {\bibinfo {volume}
  {7}},\ \bibinfo {pages} {l012011} (\bibinfo {year} {2025})}\BibitemShut
  {NoStop}%
\bibitem [{\citenamefont {Bertini}\ \emph {et~al.}(2024)\citenamefont
  {Bertini}, \citenamefont {De~Fazio}, \citenamefont {Garrahan},\ and\
  \citenamefont {Klobas}}]{Bertini2024a}%
  \BibitemOpen
  \bibfield  {author} {\bibinfo {author} {\bibfnamefont {B.}~\bibnamefont
  {Bertini}}, \bibinfo {author} {\bibfnamefont {C.}~\bibnamefont {De~Fazio}},
  \bibinfo {author} {\bibfnamefont {J.~P.}\ \bibnamefont {Garrahan}},\ and\
  \bibinfo {author} {\bibfnamefont {K.}~\bibnamefont {Klobas}},\ }\bibfield
  {title} {\bibinfo {title} {Exact quench dynamics of the floquet quantum east
  model at the deterministic point},\ }\href
  {https://doi.org/10.1103/physrevlett.132.120402} {\bibfield  {journal}
  {\bibinfo  {journal} {Phys. Rev. Lett.}\ }\textbf {\bibinfo {volume} {132}},\
  \bibinfo {pages} {120402} (\bibinfo {year} {2024})}\BibitemShut {NoStop}%
\bibitem [{\citenamefont {Fisher}\ \emph {et~al.}(2023)\citenamefont {Fisher},
  \citenamefont {Khemani}, \citenamefont {Nahum},\ and\ \citenamefont
  {Vijay}}]{Fisher2023}%
  \BibitemOpen
  \bibfield  {author} {\bibinfo {author} {\bibfnamefont {M.~P.}\ \bibnamefont
  {Fisher}}, \bibinfo {author} {\bibfnamefont {V.}~\bibnamefont {Khemani}},
  \bibinfo {author} {\bibfnamefont {A.}~\bibnamefont {Nahum}},\ and\ \bibinfo
  {author} {\bibfnamefont {S.}~\bibnamefont {Vijay}},\ }\bibfield  {title}
  {\bibinfo {title} {Random quantum circuits},\ }\href
  {https://doi.org/10.1146/annurev-conmatphys-031720-030658} {\bibfield
  {journal} {\bibinfo  {journal} {Annu. Rev. Condens. Matter Phys.}\ }\textbf
  {\bibinfo {volume} {14}},\ \bibinfo {pages} {335} (\bibinfo {year}
  {2023})}\BibitemShut {NoStop}%
\bibitem [{\citenamefont {Jonay}\ \emph {et~al.}(2018)\citenamefont {Jonay},
  \citenamefont {Huse},\ and\ \citenamefont {Nahum}}]{Jonay2018}%
  \BibitemOpen
  \bibfield  {author} {\bibinfo {author} {\bibfnamefont {C.}~\bibnamefont
  {Jonay}}, \bibinfo {author} {\bibfnamefont {D.~A.}\ \bibnamefont {Huse}},\
  and\ \bibinfo {author} {\bibfnamefont {A.}~\bibnamefont {Nahum}},\ }\bibfield
   {title} {\bibinfo {title} {Coarse-grained dynamics of operator and state
  entanglement},\ }\href {https://doi.org/10.48550/ARXIV.1803.00089} {\bibfield
   {journal} {\bibinfo  {journal} {arXiv:1803.00089}\ } (\bibinfo {year}
  {2018})}\BibitemShut {NoStop}%
\bibitem [{\citenamefont {Zhou}\ and\ \citenamefont {Nahum}(2019)}]{Zhou2019}%
  \BibitemOpen
  \bibfield  {author} {\bibinfo {author} {\bibfnamefont {T.}~\bibnamefont
  {Zhou}}\ and\ \bibinfo {author} {\bibfnamefont {A.}~\bibnamefont {Nahum}},\
  }\bibfield  {title} {\bibinfo {title} {Emergent statistical mechanics of
  entanglement in random unitary circuits},\ }\href
  {https://doi.org/10.1103/physrevb.99.174205} {\bibfield  {journal} {\bibinfo
  {journal} {Phys. Rev. B}\ }\textbf {\bibinfo {volume} {99}},\ \bibinfo
  {pages} {174205} (\bibinfo {year} {2019})}\BibitemShut {NoStop}%
\bibitem [{\citenamefont {Reutter}\ and\ \citenamefont
  {Vicary}(2019)}]{Reutter2019}%
  \BibitemOpen
  \bibfield  {author} {\bibinfo {author} {\bibfnamefont {D.~J.}\ \bibnamefont
  {Reutter}}\ and\ \bibinfo {author} {\bibfnamefont {J.}~\bibnamefont
  {Vicary}},\ }\bibfield  {title} {\bibinfo {title} {Biunitary constructions in
  quantum information},\ }\href
  {https://higher-structures.math.cas.cz/api/files/issues/Vol3Iss1/ReutterVicary}
  {\bibfield  {journal} {\bibinfo  {journal} {Higher Structures}\ }\textbf
  {\bibinfo {volume} {3}},\ \bibinfo {pages} {109} (\bibinfo {year}
  {2019})}\BibitemShut {NoStop}%
\bibitem [{\citenamefont {Claeys}\ \emph {et~al.}(2024)\citenamefont {Claeys},
  \citenamefont {Lamacraft},\ and\ \citenamefont {Vicary}}]{Claeys2024}%
  \BibitemOpen
  \bibfield  {author} {\bibinfo {author} {\bibfnamefont {P.~W.}\ \bibnamefont
  {Claeys}}, \bibinfo {author} {\bibfnamefont {A.}~\bibnamefont {Lamacraft}},\
  and\ \bibinfo {author} {\bibfnamefont {J.}~\bibnamefont {Vicary}},\
  }\bibfield  {title} {\bibinfo {title} {From dual-unitary to biunitary: a
  2-categorical model for exactly-solvable many-body quantum dynamics},\ }\href
  {https://doi.org/10.1088/1751-8121/ad653f} {\bibfield  {journal} {\bibinfo
  {journal} {Journal of Physics A: Mathematical and Theoretical}\ }\textbf
  {\bibinfo {volume} {57}},\ \bibinfo {pages} {335301} (\bibinfo {year}
  {2024})}\BibitemShut {NoStop}%
\bibitem [{\citenamefont {Tadej}\ and\ \citenamefont
  {Życzkowski}(2006)}]{Tadej2006}%
  \BibitemOpen
  \bibfield  {author} {\bibinfo {author} {\bibfnamefont {W.}~\bibnamefont
  {Tadej}}\ and\ \bibinfo {author} {\bibfnamefont {K.}~\bibnamefont
  {Życzkowski}},\ }\bibfield  {title} {\bibinfo {title} {A concise guide to
  complex hadamard matrices},\ }\href
  {https://doi.org/10.1007/s11080-006-8220-2} {\bibfield  {journal} {\bibinfo
  {journal} {Open Systems \& Information Dynamics}\ }\textbf {\bibinfo {volume}
  {13}},\ \bibinfo {pages} {133} (\bibinfo {year} {2006})}\BibitemShut
  {NoStop}%
\bibitem [{\citenamefont {Werner}(2001)}]{Werner2001}%
  \BibitemOpen
  \bibfield  {author} {\bibinfo {author} {\bibfnamefont {R.~F.}\ \bibnamefont
  {Werner}},\ }\bibfield  {title} {\bibinfo {title} {All teleportation and
  dense coding schemes},\ }\href {https://doi.org/10.1088/0305-4470/34/35/332}
  {\bibfield  {journal} {\bibinfo  {journal} {Journal of Physics A:
  Mathematical and General}\ }\textbf {\bibinfo {volume} {34}},\ \bibinfo
  {pages} {7081} (\bibinfo {year} {2001})}\BibitemShut {NoStop}%
\bibitem [{\citenamefont {Englert}\ and\ \citenamefont
  {Aharonov}(2001)}]{Englert2001}%
  \BibitemOpen
  \bibfield  {author} {\bibinfo {author} {\bibfnamefont {B.-G.}\ \bibnamefont
  {Englert}}\ and\ \bibinfo {author} {\bibfnamefont {Y.}~\bibnamefont
  {Aharonov}},\ }\bibfield  {title} {\bibinfo {title} {The mean king’s
  problem: prime degrees of freedom},\ }\href
  {https://doi.org/10.1016/s0375-9601(01)00271-7} {\bibfield  {journal}
  {\bibinfo  {journal} {Physics Letters A}\ }\textbf {\bibinfo {volume}
  {284}},\ \bibinfo {pages} {1} (\bibinfo {year} {2001})}\BibitemShut {NoStop}%
\bibitem [{\citenamefont {Wojcik}\ \emph {et~al.}(2003)\citenamefont {Wojcik},
  \citenamefont {Grudka},\ and\ \citenamefont {Chhajlany}}]{Wojcik2003}%
  \BibitemOpen
  \bibfield  {author} {\bibinfo {author} {\bibfnamefont {A.}~\bibnamefont
  {Wojcik}}, \bibinfo {author} {\bibfnamefont {A.}~\bibnamefont {Grudka}},\
  and\ \bibinfo {author} {\bibfnamefont {R.~W.}\ \bibnamefont {Chhajlany}},\
  }\bibfield  {title} {\bibinfo {title} {Generation of inequivalent generalized
  bell bases},\ }\href {https://arxiv.org/pdf/quant-ph/0305034} {\bibfield
  {journal} {\bibinfo  {journal} {Quantum Information Processing 2, 201}\ }
  (\bibinfo {year} {2003})},\ \Eprint {https://arxiv.org/abs/quant-ph/0305034}
  {arXiv:quant-ph/0305034 [quant-ph]} \BibitemShut {NoStop}%
\bibitem [{\citenamefont {Klappenecker}\ and\ \citenamefont
  {Rötteler}(2004)}]{Klappenecker2004}%
  \BibitemOpen
  \bibfield  {author} {\bibinfo {author} {\bibfnamefont {A.}~\bibnamefont
  {Klappenecker}}\ and\ \bibinfo {author} {\bibfnamefont {M.}~\bibnamefont
  {Rötteler}},\ }\bibinfo {title} {Constructions of mutually unbiased bases},\
  in\ \href {https://doi.org/10.1007/978-3-540-24633-6_10} {\emph {\bibinfo
  {booktitle} {Finite Fields and Applications}}}\ (\bibinfo  {publisher}
  {Springer Berlin Heidelberg},\ \bibinfo {year} {2004})\ pp.\ \bibinfo {pages}
  {137--144}\BibitemShut {NoStop}%
\bibitem [{\citenamefont {Gutkin}\ \emph {et~al.}(2020)\citenamefont {Gutkin},
  \citenamefont {Braun}, \citenamefont {Akila}, \citenamefont {Waltner},\ and\
  \citenamefont {Guhr}}]{Gutkin2020}%
  \BibitemOpen
  \bibfield  {author} {\bibinfo {author} {\bibfnamefont {B.}~\bibnamefont
  {Gutkin}}, \bibinfo {author} {\bibfnamefont {P.}~\bibnamefont {Braun}},
  \bibinfo {author} {\bibfnamefont {M.}~\bibnamefont {Akila}}, \bibinfo
  {author} {\bibfnamefont {D.}~\bibnamefont {Waltner}},\ and\ \bibinfo {author}
  {\bibfnamefont {T.}~\bibnamefont {Guhr}},\ }\bibfield  {title} {\bibinfo
  {title} {Exact local correlations in kicked chains},\ }\href
  {https://doi.org/10.1103/PhysRevB.102.174307} {\bibfield  {journal} {\bibinfo
   {journal} {Phys. Rev. B}\ }\textbf {\bibinfo {volume} {102}},\ \bibinfo
  {pages} {174307} (\bibinfo {year} {2020})}\BibitemShut {NoStop}%
\bibitem [{\citenamefont {Claeys}\ and\ \citenamefont
  {Lamacraft}(2022)}]{Claeys2022a}%
  \BibitemOpen
  \bibfield  {author} {\bibinfo {author} {\bibfnamefont {P.~W.}\ \bibnamefont
  {Claeys}}\ and\ \bibinfo {author} {\bibfnamefont {A.}~\bibnamefont
  {Lamacraft}},\ }\bibfield  {title} {\bibinfo {title} {Emergent quantum state
  designs and biunitarity in dual-unitary circuit dynamics},\ }\href
  {https://doi.org/10.22331/q-2022-06-15-738} {\bibfield  {journal} {\bibinfo
  {journal} {Quantum}\ }\textbf {\bibinfo {volume} {6}},\ \bibinfo {pages}
  {738} (\bibinfo {year} {2022})}\BibitemShut {NoStop}%
\bibitem [{\citenamefont {Prosen}(2021)}]{Prosen2021}%
  \BibitemOpen
  \bibfield  {author} {\bibinfo {author} {\bibfnamefont {T.}~\bibnamefont
  {Prosen}},\ }\bibfield  {title} {\bibinfo {title} {Many-body quantum chaos
  and dual-unitarity round-a-face},\ }\href {https://doi.org/10.1063/5.0056970}
  {\bibfield  {journal} {\bibinfo  {journal} {Chaos: An Interdisciplinary
  Journal of Nonlinear Science}\ }\textbf {\bibinfo {volume} {31}},\ \bibinfo
  {pages} {093101} (\bibinfo {year} {2021})}\BibitemShut {NoStop}%
\bibitem [{\citenamefont {Piroli}\ \emph {et~al.}(2020)\citenamefont {Piroli},
  \citenamefont {Bertini}, \citenamefont {Cirac},\ and\ \citenamefont
  {Prosen}}]{Piroli2020}%
  \BibitemOpen
  \bibfield  {author} {\bibinfo {author} {\bibfnamefont {L.}~\bibnamefont
  {Piroli}}, \bibinfo {author} {\bibfnamefont {B.}~\bibnamefont {Bertini}},
  \bibinfo {author} {\bibfnamefont {J.~I.}\ \bibnamefont {Cirac}},\ and\
  \bibinfo {author} {\bibfnamefont {T.}~\bibnamefont {Prosen}},\ }\bibfield
  {title} {\bibinfo {title} {Exact dynamics in dual-unitary quantum circuits},\
  }\href {https://doi.org/10.1103/physrevb.101.094304} {\bibfield  {journal}
  {\bibinfo  {journal} {Phys. Rev. B}\ }\textbf {\bibinfo {volume} {101}},\
  \bibinfo {pages} {094304} (\bibinfo {year} {2020})}\BibitemShut {NoStop}%
\bibitem [{\citenamefont {Haake}\ \emph {et~al.}(2018)\citenamefont {Haake},
  \citenamefont {Gnutzmann},\ and\ \citenamefont {Kuś}}]{Haake2018}%
  \BibitemOpen
  \bibfield  {author} {\bibinfo {author} {\bibfnamefont {F.}~\bibnamefont
  {Haake}}, \bibinfo {author} {\bibfnamefont {S.}~\bibnamefont {Gnutzmann}},\
  and\ \bibinfo {author} {\bibfnamefont {M.}~\bibnamefont {Kuś}},\ }\href
  {https://doi.org/10.1007/978-3-319-97580-1} {\emph {\bibinfo {title} {Quantum
  Signatures of Chaos}}}\ (\bibinfo  {publisher} {Springer International
  Publishing},\ \bibinfo {year} {2018})\BibitemShut {NoStop}%
\bibitem [{\citenamefont {Thouless}(1977)}]{Thouless1977}%
  \BibitemOpen
  \bibfield  {author} {\bibinfo {author} {\bibfnamefont {D.~J.}\ \bibnamefont
  {Thouless}},\ }\bibfield  {title} {\bibinfo {title} {Maximum metallic
  resistance in thin wires},\ }\href
  {https://doi.org/10.1103/physrevlett.39.1167} {\bibfield  {journal} {\bibinfo
   {journal} {Phys. Rev. Lett.}\ }\textbf {\bibinfo {volume} {39}},\ \bibinfo
  {pages} {1167} (\bibinfo {year} {1977})}\BibitemShut {NoStop}%
\bibitem [{\citenamefont {Altshuler}\ and\ \citenamefont
  {Shklovskii}(1986)}]{Altshuler1986}%
  \BibitemOpen
  \bibfield  {author} {\bibinfo {author} {\bibfnamefont {B.~L.}\ \bibnamefont
  {Altshuler}}\ and\ \bibinfo {author} {\bibfnamefont {B.~I.}\ \bibnamefont
  {Shklovskii}},\ }\bibfield  {title} {\bibinfo {title} {Repulsion of energy
  levels and conductivity of small metal samples},\ }\href@noop {} {\bibfield
  {journal} {\bibinfo  {journal} {Zh. Eksp. Teor. Fiz}\ }\textbf {\bibinfo
  {volume} {91}},\ \bibinfo {pages} {220} (\bibinfo {year} {1986})}\BibitemShut
  {NoStop}%
\bibitem [{\citenamefont {Ho}\ and\ \citenamefont {Choi}(2022)}]{Ho2022}%
  \BibitemOpen
  \bibfield  {author} {\bibinfo {author} {\bibfnamefont {W.~W.}\ \bibnamefont
  {Ho}}\ and\ \bibinfo {author} {\bibfnamefont {S.}~\bibnamefont {Choi}},\
  }\bibfield  {title} {\bibinfo {title} {Exact emergent quantum state designs
  from quantum chaotic dynamics},\ }\href
  {https://doi.org/10.1103/physrevlett.128.060601} {\bibfield  {journal}
  {\bibinfo  {journal} {Phys. Rev. Lett.}\ }\textbf {\bibinfo {volume} {128}},\
  \bibinfo {pages} {060601} (\bibinfo {year} {2022})}\BibitemShut {NoStop}%
\bibitem [{\citenamefont {Cotler}\ \emph {et~al.}(2023)\citenamefont {Cotler},
  \citenamefont {Mark}, \citenamefont {Huang}, \citenamefont {Hernández},
  \citenamefont {Choi}, \citenamefont {Shaw}, \citenamefont {Endres},\ and\
  \citenamefont {Choi}}]{Cotler2023}%
  \BibitemOpen
  \bibfield  {author} {\bibinfo {author} {\bibfnamefont {J.~S.}\ \bibnamefont
  {Cotler}}, \bibinfo {author} {\bibfnamefont {D.~K.}\ \bibnamefont {Mark}},
  \bibinfo {author} {\bibfnamefont {H.-Y.}\ \bibnamefont {Huang}}, \bibinfo
  {author} {\bibfnamefont {F.}~\bibnamefont {Hernández}}, \bibinfo {author}
  {\bibfnamefont {J.}~\bibnamefont {Choi}}, \bibinfo {author} {\bibfnamefont
  {A.~L.}\ \bibnamefont {Shaw}}, \bibinfo {author} {\bibfnamefont
  {M.}~\bibnamefont {Endres}},\ and\ \bibinfo {author} {\bibfnamefont
  {S.}~\bibnamefont {Choi}},\ }\bibfield  {title} {\bibinfo {title} {Emergent
  quantum state designs from individual many-body wave functions},\ }\href
  {https://doi.org/10.1103/prxquantum.4.010311} {\bibfield  {journal} {\bibinfo
   {journal} {PRX Quantum}\ }\textbf {\bibinfo {volume} {4}},\ \bibinfo {pages}
  {010311} (\bibinfo {year} {2023})}\BibitemShut {NoStop}%
\bibitem [{\citenamefont {Nielsen}\ \emph {et~al.}(2003)\citenamefont
  {Nielsen}, \citenamefont {Dawson}, \citenamefont {Dodd}, \citenamefont
  {Gilchrist}, \citenamefont {Mortimer}, \citenamefont {Osborne}, \citenamefont
  {Bremner}, \citenamefont {Harrow},\ and\ \citenamefont
  {Hines}}]{Nielsen2003}%
  \BibitemOpen
  \bibfield  {author} {\bibinfo {author} {\bibfnamefont {M.~A.}\ \bibnamefont
  {Nielsen}}, \bibinfo {author} {\bibfnamefont {C.~M.}\ \bibnamefont {Dawson}},
  \bibinfo {author} {\bibfnamefont {J.~L.}\ \bibnamefont {Dodd}}, \bibinfo
  {author} {\bibfnamefont {A.}~\bibnamefont {Gilchrist}}, \bibinfo {author}
  {\bibfnamefont {D.}~\bibnamefont {Mortimer}}, \bibinfo {author}
  {\bibfnamefont {T.~J.}\ \bibnamefont {Osborne}}, \bibinfo {author}
  {\bibfnamefont {M.~J.}\ \bibnamefont {Bremner}}, \bibinfo {author}
  {\bibfnamefont {A.~W.}\ \bibnamefont {Harrow}},\ and\ \bibinfo {author}
  {\bibfnamefont {A.}~\bibnamefont {Hines}},\ }\bibfield  {title} {\bibinfo
  {title} {Quantum dynamics as a physical resource},\ }\href
  {https://doi.org/10.1103/physreva.67.052301} {\bibfield  {journal} {\bibinfo
  {journal} {Phys. Rev. A}\ }\textbf {\bibinfo {volume} {67}},\ \bibinfo
  {pages} {052301} (\bibinfo {year} {2003})}\BibitemShut {NoStop}%
\bibitem [{\citenamefont {Gottesman}(1997)}]{Gottesman1997stabilizer}%
  \BibitemOpen
  \bibfield  {author} {\bibinfo {author} {\bibfnamefont {D.}~\bibnamefont
  {Gottesman}},\ }\href {https://arxiv.org/abs/quant-ph/9705052} {\bibinfo
  {title} {Stabilizer codes and quantum error correction}} (\bibinfo {year}
  {1997}),\ \Eprint {https://arxiv.org/abs/quant-ph/9705052}
  {arXiv:quant-ph/9705052 [quant-ph]} \BibitemShut {NoStop}%
\bibitem [{\citenamefont {Koenig}\ and\ \citenamefont
  {Smolin}(2014)}]{Koenig2014efficiently}%
  \BibitemOpen
  \bibfield  {author} {\bibinfo {author} {\bibfnamefont {R.}~\bibnamefont
  {Koenig}}\ and\ \bibinfo {author} {\bibfnamefont {J.~A.}\ \bibnamefont
  {Smolin}},\ }\bibfield  {title} {\bibinfo {title} {How to efficiently select
  an arbitrary clifford group element},\ }\href
  {https://doi.org/10.1063/1.4903507} {\bibfield  {journal} {\bibinfo
  {journal} {Journal of Mathematical Physics}\ }\textbf {\bibinfo {volume}
  {55}},\ \bibinfo {pages} {122202} (\bibinfo {year} {2014})}\BibitemShut
  {NoStop}%
\end{thebibliography}%
\end{document}